\documentclass[11pt,fleqn]{article}

\usepackage{a4}
\usepackage{amstext}
\usepackage{amsfonts}
\usepackage{amssymb}
\usepackage{amsmath}
\usepackage{color}
\usepackage{subfig}
\usepackage{epsfig}
\usepackage{booktabs}
\usepackage{caption}
\usepackage{multirow}
\usepackage{footnote}
\usepackage{dsfont}
\usepackage{longtable}
\usepackage{xcolor,colortbl}
\usepackage[normalem]{ulem}
\usepackage{soul} 
\usepackage{float}
\usepackage{bbm}
\usepackage{ifthen}
\usepackage{geometry}
\usepackage{marginnote}

\newcommand{\D}[5]{
G^{#1}({\bf #2},#3;{\bf #4},#5)
}

\newcommand{\sptV}[0]{
V_s
}

\newcommand{\Op}[1]{
\ifthenelse{\equal{#1}{1}}{\text{$q\bar q$}}{}
\ifthenelse{\equal{#1}{2}}{\text{$K\bar K$, point}}{}
\ifthenelse{\equal{#1}{3}}{\text{$\eta_s \pi$, point} }{}
\ifthenelse{\equal{#1}{4}}{\text{$Q\bar Q$}}{}
\ifthenelse{\equal{#1}{5}}{\text{$K\bar K$, 2part}}{}
\ifthenelse{\equal{#1}{6}}{\text{$\eta_s\pi$, 2part}}{}
}

\begin{document}

% ********************
% ********************
% ********************
% ********************
% ********************

\begin{center}

{\Large {\bf{} Investigating efficient methods for computing four-quark correlation functions}}

\vspace{0.5cm}

\textbf{Abdou Abdel-Rehim${}^{1,2}$, Constantia Alexandrou${}^{1,3}$, Joshua Berlin${}^{4}$, \\ Mattia Dalla Brida${}^{5}$, Jacob Finkenrath${}^{1}$, Marc Wagner${}^{4}$}

${}^{1}$Computation-based Science and Technology Research Center, The Cyprus Institute, \\ 20 Kavafi Street, 2121 Nicosia, Cyprus

${}^{2}$Department of Engineering, Science and Mathematics SUNY Polytechnic Institute, Utica, New York 13502, USA

${}^{3}$Department of Physics, University of Cyprus, P.O.\ Box 20537, 1678 Nicosia, Cyprus

${}^{4}$Goethe-Universit\"at Frankfurt am Main, Institut f\"ur Theoretische Physik, Max-von-Laue-Stra{\ss}e 1, D-60438 Frankfurt am Main, Germany

${}^{5}$ Dipartimento di Fisica, Universit\`a di Milano-Bicocca 
	 \& INFN, sezione di Milano-Bicocca, Piazza della Scienza 3,
	 I-20126 Milano, Italy\\ 

\vspace{0.4cm}

January 25, 2017

\end{center}

\vspace{0.1cm}

\begin{tabular*}{16cm}{l@{\extracolsep{\fill}}r} \hline \end{tabular*}

\vspace{-0.40cm}
\begin{center} \textbf{Abstract} \end{center}
\vspace{-0.40cm}

We discuss and compare the efficiency of various methods, combinations of point-to-all propagators, stochastic timeslice-to-all propagators, the one-end trick and sequential propagators, to compute two-point correlation functions of two-quark and four-quark interpolating operators of different structure including quark-antiquark type, mesonic molecule type, diquark-antidiquark type and two-meson type. Although we illustrate our methods in the context of the $a_0(980)$, they can be applied for other multi-quark systems, where similar diagrams appear. Thus our results could provide helpful guidelines on the choice of methods for correlation function computation for future lattice QCD studies of meson-meson scattering and possibly existing tetraquark states.

\begin{tabular*}{16cm}{l@{\extracolsep{\fill}}r} \hline \end{tabular*}

\thispagestyle{empty}

% ********************
% ********************
% ********************
% ********************
% ********************

\newpage

\setcounter{page}{1}

\section{Introduction}

In recent years the study of four-quark systems has revived a lot of interest within the field of lattice QCD. These systems allow to study meson-meson scattering, as well as to investigate the existence and structure of possibly existing four-quark states, e.g.\ of mesonic molecule type or of diquark-antidiquark type. In particular the latter is motivated by experimental results providing strong indications for the existence of tetraquarks (e.g.\ the recently observed charged $Z_c$ and $Z_b$ states \cite{Ablikim:2013mio,Liu:2013dau,Belle:2011aa} or the mass ordering of the nonet of light scalar mesons, which is inverted compared to the expectation from a standard quark-antiquark picture \cite{pdg:sep2014}).

In order to perform this study within lattice QCD we need to compute two-point correlation functions involving both two-quark and four-quark interpolating operators. Depending on the details of these operators diagrams involving up to four quark propagators with a non-trivial spacetime structure are present. In contrast to simple quark-antiquark correlation functions, which can be computed in a straightforward way, e.g.\ by using standard point-to-all propagators, these four-quark correlation functions require in many cases all-to-all propagators (propagators from any point in space on a timeslice to any other point in space on another timeslice, what makes them considerably expensive) and, hence, more advanced techniques.

One possibility to compute them is the distillation method \cite{Peardon:2009gh}, which provides all-to-all propagators between specifically smeared quark field operators (Laplacian Heaviside smearing). Distillation has recently been applied to such four-quark correlation functions (cf.\ e.g.\ \cite{Lang:2015sba,Dudek:2016cru,Moir:2016srx}). The distillation method however comes with an expensive overhead and only pays if a much larger number of interpolating fields is used. Moreover, for large volume lattices it is impractical and one has to apply its stochastic variant, which introduces additional stochastic noise terms for each quark propagator.

In this work we explore a different strategy, namely combining several traditional techniques to compute quark propagators and four-quark correlation functions: (A)~point-to-all propagators (cf.\ e.g.\ \cite{DeGrand:2006zz,Gattringer:2010zz}); (B)~stochastic timeslice-to-all propagators (cf.\ e.g.\ \cite{Bernardson:1993he,Dong:1993pk}); (C)~ the one-end trick (cf.\ e.g.\ \cite{Foster:1998vw,McNeile:2006bz}); (D)~sequential propagators (cf.\ e.g.\ \cite{Martinelli:1988rr}). There are many different types of diagrams and for each type we discuss several methods (combinations of the above mentioned techniques (A) to (D)) and determine numerically the most efficient method. We study the $a_0(980)$ channel (quantum numbers $I(J^P) = 1(0^+)$) at a lattice spacing of $a\approx 0.09$fm and spacetime volume of $(32a)^3\times64a$ with a variety of interpolating operators including quark-antiquark type, mesonic molecule type, diquark-antidiquark type and two-meson type (the latter describes two independent mesons with total zero momentum). % Even though other systems and lattice setups might lead to somehow different quantitative results regarding the efficiency comparisons, we expect that the qualitative picture will be the same.
In other words our work is intended to provide guidelines for future lattice QCD work concerned with arbitrary four-quark correlation functions, in particular guidelines for a quick and hence time-saving decision, which diagrams of a given correlation matrix to compute with which combinations of the above listed techniques.

Parts of this work have already been presented at recent conferences \cite{Wagner:2012ay,
Wagner:2013nta,Wagner:2013jda,Wagner:2013vaa,Abdel-Rehim:2014zwa,Berlin:2015faa}.

The paper is structured as follows. In section~\ref{sec:opbasis} we introduce the six interpolating operators we investigate and the resulting $6 \times 6$ correlation matrix. In section~\ref{sec:propcomp} we recapitulate the above mentioned four techniques for propagator and correlation function computation. Section~\ref{seq:compofdiagrams} is the main section of this work, where various methods are discussed for each element and diagram of the correlation matrix, with a numerical comparison of their efficiency. In section~\ref{SEC588} we conclude, in particular we summarize a few general rules regarding the choice of an efficient method, which seem to hold for most correlation matrix elements investigated. Section~\ref{SEC876} outlines briefly the used lattice setup and gives detailed examples for the calculation of certain correlation functions.

% ********************
% ********************
% ********************
% ********************
% ********************

\newpage

\section{\label{sec:opbasis}Interpolating operators and correlation matrix}

To study the $a_0(980)$ meson, we consider several interpolating operators $\mathcal{O}^j$, which create states with quantum numbers $I(J^P) = 1(0^+)$, when applied to the vaccuum:
\begin{align}
\label{eq:operatorone}   & \mathcal{O}^1  & &= \mathcal{O}^{\Op{1}} & &= \frac{1}{\sqrt{V_s}} & &\sum_{\bf{x}} \Big({\bar d}({\bf x}) u({\bf x})\Big) \\
\label{eq:operatortwo}   & \mathcal{O}^2  & &= \mathcal{O}^{\Op{2}} & &= \frac{1}{\sqrt{V_s}} & &\sum_{\bf{x}} \Big({\bar s}({\bf x}) \gamma_5 u({\bf x})\Big) \Big({\bar d}({\bf x}) \gamma_5 s({\bf x})\Big) \\
\label{eq:operatorthree} & \mathcal{O}^3  & &= \mathcal{O}^{\Op{3}} & &= \frac{1}{\sqrt{V_s}} & &\sum_{\bf{x}} \Big({\bar s}({\bf x}) \gamma_5 s({\bf x})\Big) \Big({\bar d}({\bf x}) \gamma_5 u({\bf x})\Big) \\
\label{eq:operatorfour}  & \mathcal{O}^4  & &= \mathcal{O}^{\Op{4}} & &= \frac{1}{\sqrt{V_s}} & &\sum_{\bf{x}} \epsilon_{a b c} \Big({\bar s}_b({\bf x}) (C \gamma_5) {\bar d}_c^T({\bf x})\Big) \epsilon_{a d e} \Big(u_d^T({\bf x}) (C \gamma_5) s_e({\bf x})\Big) \\
\label{eq:operatorfive}  & \mathcal{O}^5  & &= \mathcal{O}^{\Op{5}} & &= \;\,\frac{1}{V_s}        & &\sum_{{\bf x},{\bf y}} \Big({\bar s}({\bf x}) \gamma_5 u({\bf x})\Big) \Big({\bar d}({\bf y}) \gamma_5 s({\bf y})\Big) \\
\label{eq:operatorsix}   & \mathcal{O}^6  & &= \mathcal{O}^{\Op{6}} & &= \;\,\frac{1}{V_s}        & &\sum_{{\bf x},{\bf y}} \Big({\bar s}({\bf x}) \gamma_5 s({\bf x})\Big) \Big({\bar d}({\bf y}) \gamma_5 u({\bf y})\Big)
\end{align}
($V_s$ is the spatial volume, $C$ is the charge conjugation matrix). These interpolating operators are of different structure. $\mathcal{O}^{q\bar q}$ generates a quark-antiquark pair, while the other operators (\ref{eq:operatortwo}) to (\ref{eq:operatorsix}) generate two quarks and two antiquarks. $\mathcal{O}^{\Op{2}}$ and $\mathcal{O}^{\Op{3}}$ are of mesonic molecule type, i.e.\ resemble a $K \bar{K}$ pair or $\eta_s \pi$ pair\footnote{${\bar s}_{\bf x} \gamma_5 {s}_{\bf x}$ excites a meson-like structure composed of an $s \bar{s}$ pair, which is expected to have significant overlap to both $\eta$ and $\eta'$ and, hence, is denoted by $\eta_s$.} centered around the same spatial point $\mathbf{x}$. $\mathcal{O}^{\Op{4}}$ corresponds to a diquark-antidiquark pair\footnote{We consider only the lightest (anti-)diquarks, which have spin structure $C \gamma_5$ \cite{Jaffe:2004ph,Alexandrou:2006cq,Wagner:2011fs}.}. These three operators are candidates to model the structure of a possibly existing bound four-quark state, i.e.\ of a tetraquark. The remaining two operators $\mathcal{O}^{\Op{5}}$ and $\mathcal{O}^{\Op{6}}$ also generate meson pairs ($K \bar{K}$ and $\eta_s \pi$), this time, however, at independent spatial points $\mathbf{x}$ and $\mathbf{y}$. They should be suited to resolve low-lying two-meson states within the $I(J^P) = 1(0^+)$ sector.

The interpolating operators (\ref{eq:operatorone}) to (\ref{eq:operatorsix}) enter a $6 \times 6$ correlation matrix,
\begin{align}
\label{eq:corrmatrix} C_{j k}(t) = \Big\langle \mathcal{O}^j(t_2) \mathcal{O}^{k \dag}(t_1) \Big\rangle \quad , \quad t = t_2 - t_1 > 0 ,
\end{align}
where $\langle \ldots \rangle$ denotes the lattice QCD path integral expectation value. Computing such expectations values at affordable numerical costs and with small statistical errors is a highly non-trivial task, in particular, when four-quark operators like (\ref{eq:operatortwo}) to (\ref{eq:operatorsix}) are involved. One possible strategy, which we follow throughout this work, is to suitably combine standard techniques for propagator and correlator computation (point-to-all propagators, stochastic propagators, the one-end trick, sequential propagators). One of the main goals of this work is to discuss, which combinations of techniques are possible in principle for each of the matrix elements $C_{j k}$ and, in a second step, to determine numerically which combination is most efficient, i.e.\ results in the smallest statistical error at comparable computational cost.

Even though we focus on the $a_0(980)$ meson with quantum numbers $I(J^P) = 1(0^+)$, our findings are of general interest regarding the study of tetraquark systems and unstable mesonic resonances. The study of such systems usually requires the computation of correlation matrices of identical or similar structure as (\ref{eq:corrmatrix}) with interpolating operators (\ref{eq:operatorone}) to (\ref{eq:operatorsix}). For example, after replacing the quark flavors according to $(u,d,s) \rightarrow (c,s,u/d)$, one would obtain a matrix suited to study the $J^P = 0^+$ $D_{s0}^\ast(2317)$ meson, and probe whether two- or for-quark structures are dominating. Similarly, $(u,d,s) \rightarrow (u,d,c)$ would allow to study certain charmonium states, for which are tetraquark structure is frequently discussed, e.g. the $Z(4430)$ meson. 

\begin{figure}[htb]
\begin{center}
${}\hspace{-2.97cm}{}$\includegraphics[scale=.7,page=1]{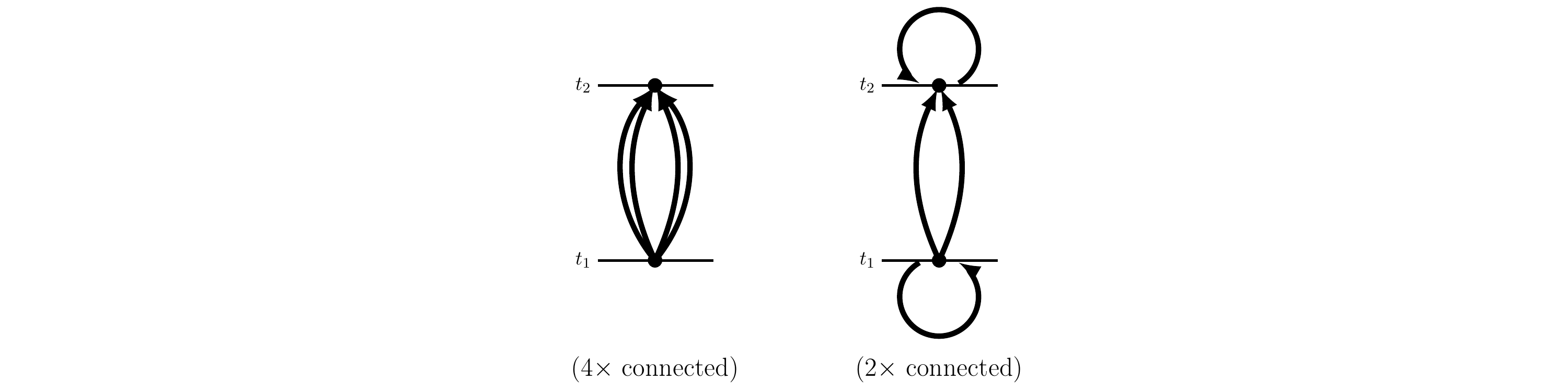}
\caption{\label{fig:examplecorrdiagram}Diagrammatic representation of $C_{2 2}$.}
\end{center}
\end{figure}

\begin{figure}[htb]
\begin{center}
\includegraphics[scale=.138]{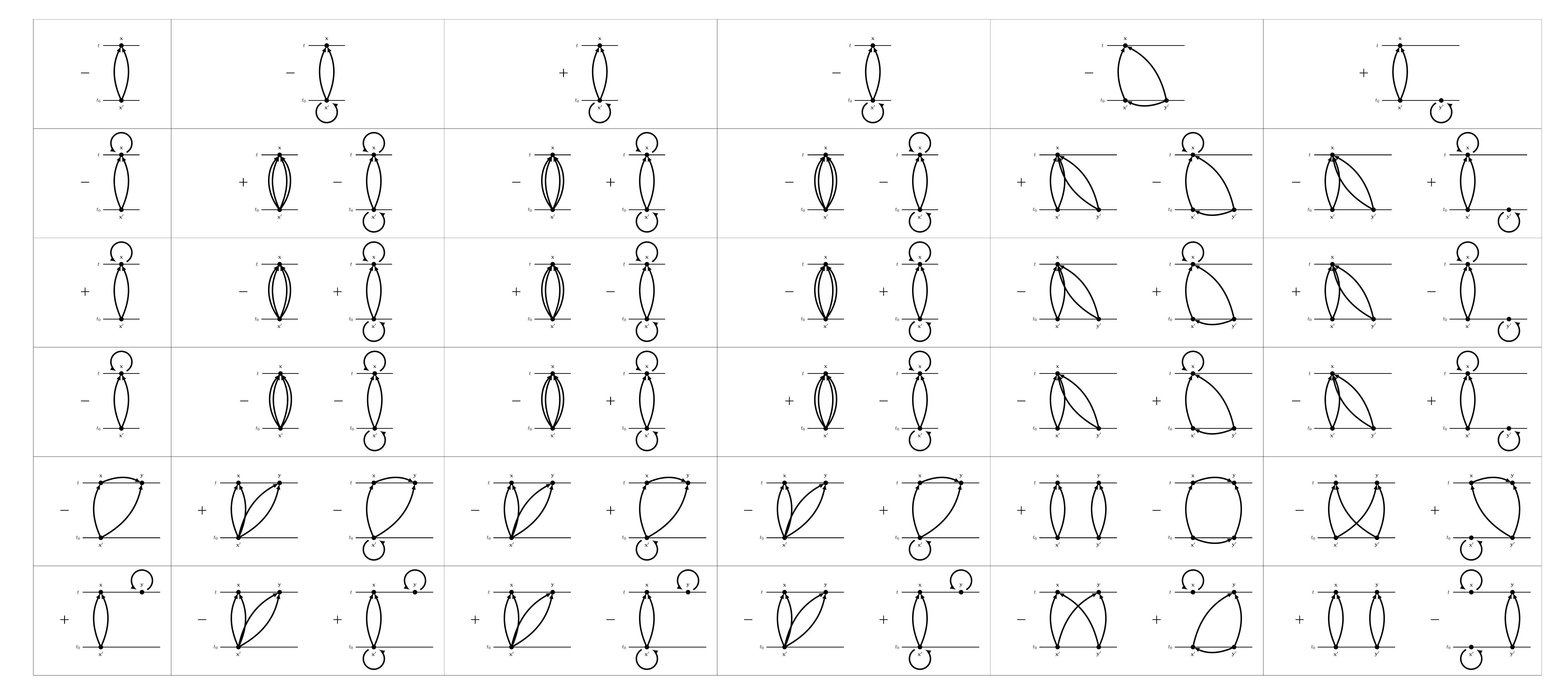}
\caption{\label{fig:diagrammatrix}Diagrammatic representation of the $6 \times 6$ correlation matrix $C_{j k}$ (eq.\ (\ref{eq:corrmatrix})).}
\end{center}
\end{figure}

To ease notation, we will often picture correlation matrix elements in a diagrammatic way, where quark propagators are represented by arrows. These diagrams do not exhibit information about color and spin, but clearly display the spacetime structure. In particular one can read off, which combinations of methods is suited to compute a diagram. In our case, each correlation matrix element (\ref{eq:corrmatrix}) corresponds to either a single diagram or a sum of two diagrams. Due to the flavor structure of the interpolating operators (either $u \bar{d}$ or $u \bar{d} s \bar{s}$) there are within each diagram either two or four quark propagators connecting the timeslices at $t_1$ and $t_2$. The diagrams are, hence, denoted as ``$2 \times$ connected'' and ``$4 \times$ connected'', respectively (for an example cf.\ Figure~\ref{fig:examplecorrdiagram}). The diagrammatic representation of the full $6 \times 6$ correlation matrix is shown in Figure~\ref{fig:diagrammatrix}. Note that certain correlation matrix elements have identical diagrammatic representations, since they only differ in their color and spin structure.

\newpage

$\quad$

% ********************
% ********************
% ********************
% ********************
% ********************

\newpage

\section{\label{sec:propcomp}Techniques for propagator computation}

The lattice action for quarks is bilinear in the quark fields $q$ and $\bar{q}$,
\begin{align}
S^{(q)} = \sum_{x,y} \bar{q}_{a,A}(x) D_{a,A;b,B}^{(q)}(x;y) q_{b,B}(y) ,
\end{align}
where $D^{(q)}$ denotes the Dirac operator for quark flavor $q$ and indices $a,b,\ldots$, $A,B,\ldots$ and $x,y,\ldots$ label color, spin and spacetime, respectively. The propagator $G^{(q)}$ is the inverse of the Dirac operator, i.e.\ the solution of the linear system
\begin{align}
\sum_y D_{a,A;b,B}^{(q)}(x;y) G_{b,B;c,C}^{(q)}(y;z) = \delta_{a;c} \delta_{A;C} \delta(x;z) .
\end{align}

In the following subsections we discuss several standard techniques for propagator computation which are well-know in the literature, namely: (1) point-to-all propagators, (2) stochastic propagators, (3) the one-end trick and (4) sequential propagators. We illustrate some of these techniques in the context of a simple example, the correlation function of the interpolating operator $\mathcal{O}^1 = \mathcal{O}^{\Op{1}} = (1 / \sqrt{\sptV}) \sum_{\bf x} \bar d({\bf x}) u({\bf x})$ (eq.\ (\ref{eq:operatorone})), i.e.\ correlation matrix element $C_{1 1}$ in (\ref{eq:corrmatrix}) and Figure~\ref{fig:diagrammatrix}. Integrating over the Grassmann valued quark fields allows us to express the correlation function in terms of quark propagators,
\begin{align}
\nonumber & C_{1 1}(t) = -\frac{1}{\sptV} \sum_{{\bf x},{\bf y}} \Big\langle \textrm{Tr}\Big(\D{(d)}{x}{t_1}{y}{t_2} \D{(u)}{y}{t_2}{x}{t_1}\Big) \Big\rangle_U \ \ = \\
\label{eq:examplecorrfunc} & \hspace{0.7cm} = -\frac{1}{\sptV} \sum_{{\bf x},{\bf y}} \Big\langle \textrm{Tr}\Big(\gamma_5 \Big(\D{(d)}{y}{t_2}{x}{t_1}\Big)^\dag \gamma_5 \D{(u)}{y}{t_2}{x}{t_1} \Big) \Big\rangle_U ,
\end{align}
where $\gamma_5$ hermiticity has been used, $\textrm{Tr}(\ldots)$ denotes the trace in spin and color space and $\langle \ldots \rangle_U$ is the average over gauge link configurations distributed proportionally to $e^{-S_\textrm{eff}} = e^{-(S_\textrm{gauge} - \ln(\det(Q)))}$. 

% ********************

\subsection{\label{sec:pointtoallprops}Point-to-all propagators}

The exact computation of a propagator $G^{(q)}(x;y)$ from any point in spacetime $x$ to any other point $y$ is numerically not feasible, because for typical lattices with e.g.\ $32^4$ lattice sites both $D^{(q)}$ and $G^{(q)}$ are matrices with $\mathcal{O}(10^7)$ entries. However, using translation invariance it is often possible to simply compute propagators from a single spacetime point $x$ to any other point $y$. For example in (\ref{eq:examplecorrfunc}), $\sum_\textbf{x}$ can be replaced by the spatial volume $\sptV$,
\begin{align}
\label{eq:examplecorrfuncreplacedsum} C_{1 1}(t) = -\bigg\langle \sum_\mathbf{y} \textrm{Tr}\Big(\gamma_5 \left(\D{(d)}{y}{t_2}{x}{t_1}\right)^\dag \gamma_5 \D{(u)}{y}{t_2}{x}{t_1}\Big) \bigg\rangle_U ,
\end{align}
where $\mathbf{x}$ denotes an arbitrary but fixed point in space. The spacetime column index of both propagators is now $(\mathbf{x},t_1)$, i.e.\ it is not anymore necessary to compute the full matrix $G^{(u/d)}$, but only 12 columns of this matrix ($\textrm{3 color} \times \textrm{4 spin}$), which is numerically feasible.

Explicitly, one has to solve 12 linear systems,
\begin{align}
\label{eq:pointtoalllinearsystem} \sum_y D_{a,A;b,B}^{(q)}(x;y) \phi_{b,B}^{(q)}(y)[c,C,z] = \xi_{a,A}(x)[c,C,z] \quad , \quad \xi_{a,A}(x)[c,C,z] = \delta_{a;c} \delta_{A;C} \delta(x;z) ,
\end{align}
where $c=1,2,3$ and $C=1,2,3,4$ label 12 different point sources. (Each solution $\phi$ corresponds to a single column of the inverse of the Dirac matrix $D$; $\phi$ are, therefore, also commonly called ``inversions''). The propagator ending at spacetime point $x$, a so-called point-to-all propagator (cf.\ e.g.\ \cite{DeGrand:2006zz,Gattringer:2010zz}), is then
\begin{align}
G_{b,B;a,A}^{(q)}(y;x) = \phi_{b,B}^{(q)}(y)[a,A,x] ,
\end{align}
The example correlation function \eqref{eq:examplecorrfunc} expressed in terms of such point-to-all propagators is
\begin{align}
\label{EQN643} C_{1 1}(t) = -\bigg\langle (\gamma_5)_{A;B} \bigg(\sum_{\bf y} \phi^{(d)}(\mathbf{y},t_2)[a,B,{\bf x},t_1]^\dag \gamma_5 \phi^{(u)}(\mathbf{y},t_2)[a,A,{\bf x},t_1]\bigg) \bigg\rangle_U .
\end{align}

Of course, for each diagram, translation invariance allows one to replace only a single spatial sum $\sum_\mathbf{x}$ by an arbitrary fixed $\mathbf{x}$. Diagrams, where all propagators either start or end at the same spacetime point can, hence, be expressed exclusively in terms of point-to-all propagators (e.g.\ $C_{1 1}$, $C_{1 2}$, left diagram of $C_{2 2}$). However, since this is not the case for the majority of diagrams, additional methods to compute propagators are necessary.

After replacing a spatial sum $\sum_\mathbf{x}$ by a fixed $\mathbf{x}$, spatial averaging to reduce statistical errors is not implemented anymore. For each set of 12 inversions (\ref{eq:pointtoalllinearsystem}) and each diagram only a single sample is computed. Of course, it is possible to compute additional samples by choosing different {\bf x} values. This however requires additional sets of 12 inversions for each different $\mathbf{x}$, and quickly becomes expensive.

% ********************

\subsection{\label{sec:stochtimeslicetoallprops}Stochastic timeslice-to-all propagators}

While it is not possible in practice to compute the propagator from any spacetime point $x$ to any other spacetime point $y$ exactly, one can at least estimate it stochastically. Quite common are so-called stochastic timeslice-to-all propagators (cf.\ e.g.\ \cite{Bernardson:1993he,Dong:1993pk}): stochastically estimated propagators from any space point $\mathbf{x}$ in a given time-slice $t_0$, to any other spacetime point $y$.

Again linear systems have to be solved labeled now by $n = 1,\ldots,N$,
\begin{align}
\label{eq:stochasticinversion} \sum_y D_{a,A;b,B}^{(q)}(x;y) \phi_{b,B}^{(q)}(y)[t_0,n] = \xi_{a,A}(x)[t_0,n] \quad , \quad \xi_{a,A}(x)[t_0,n]  = \delta(x_0,t_0) \Xi_{a,A}(\mathbf{x})[n] ,
\end{align} 
where $\Xi_{a,A}(\mathbf{x})[n]$ are random numbers satisfying
\begin{align}
\label{EQN010} \frac{1}{N} \sum_{n=1}^N \Xi^*_{a,A}(\mathbf{x})[n] \Xi_{b,B}(\mathbf{y})[n] 
= \delta_{a;b} \delta_{A;B} \delta(\mathbf{x};\mathbf{y}) + 
\textrm{unbiased noise}.
\end{align}
A convenient choice is $\Xi_{a,A}(\mathbf{x})[n] \in\mathbb{Z}(2)\times\mathbb{Z}(2)$ which results in an unbiased noise proportional to ${\rm O}\Big(\frac{1}{\sqrt{N}}\Big)$. As usual since the noise average and the average over the gauge field commute, in practice one can take a fairly small number $N$ of noise sources per gauge configuration, but not smaller than the number of propagators in the diagram. 

Using (\ref{eq:stochasticinversion}) and (\ref{EQN010}) it is straightforward to show
\begin{align}
\label{eq:timeslicetoallprop} G^{(q)}(y;\mathbf{x},t_0) \ \ = \ \ \frac{1}{N} \sum_{n=1}^N \phi^{(q)}(y)[t_0,n] \xi(\mathbf{x},t_0)[t_0,n]^\dagger + \textrm{unbiased noise}.
\end{align}

The example correlation function \eqref{eq:examplecorrfunc} expressed in terms of stochastic timeslice-to-all propagators is
\begin{align}
\nonumber & C_{1 1}(t) = -\frac{1}{N (N-1)} \sum_{n \neq \tilde{n}} \frac{1}{V_s} \\
\nonumber & \hspace{0.7cm} \bigg\langle \bigg(\sum_\mathbf{y} \phi^{(d)}(\mathbf{y},t_2)[t_1,\tilde{n}]^\dagger \gamma_5 \phi^{(u)}(\mathbf{y},t_2)[t_1,n]\Bigg)
\bigg(\sum_\mathbf{x} \xi(\mathbf{x},t_1)[t_1,n]^\dagger \gamma_5 \xi(\mathbf{x},t_1)[t_1,\tilde{n}]\bigg) \bigg\rangle_U . \\
\label{eq:stochasticpropsonexamplecorrfunc} &
\end{align}
Note that each propagator needs to be estimated by a different pair of stochastic sources $\xi[n]$ and corresponding inversions $\phi[n]$ (guaranteed here by $\sum_{n \neq \tilde{n}}$).

Stochastic timeslice-to-all propagators are most flexible, i.e.\ replacing a spatial sum $\sum_\mathbf{x}$ by a fixed $\mathbf{x}$ as in the case of point-to-all propagators is not necessary. In principle all diagrams of the correlation matrix (\ref{eq:corrmatrix}) can be computed using exclusively stochastic timeslice-to-all propagators. A severe drawback of these propagators is, however, that they introduce additional stochastic noise. The number of stochastic noise terms is $\approx V_s^M \times (\textrm{number of signal terms})$, where $M$ is the number of stochastic timeslice-to-all propagators in a diagram\footnote{For example the number of signal terms in (\ref{eq:stochasticpropsonexamplecorrfunc}) is $\propto V_s^2$ (all $(\mathbf{x},t_1)$ connected by a pair of propagators with all $(\mathbf{y},t_2)$), while the number of stochastic noise terms is $\propto V_s^4$.}. While using a single stochastic propagator, i.e.\ $M = 1$, typically leads to acceptable signal-to-noise ratios, the noise grows quite rapidly with the number of stochastic propagators. Already for $M > 2$,
the signal can easily be lost in the noise if these techniques are applied naively. Therefore, a promising strategy might be to combine a single stochastic timeslice-to-all propagator with several point-to-all propagators, as we shall see later on.

% ********************

\subsection{\label{sec:oneendtrick}The one-end trick}

The one-end trick is an efficient technique to estimate the product of two propagators stochastically (cf.\ e.g.\ \cite{Foster:1998vw,McNeile:2006bz}). The product must be of the form
\begin{align}
\label{eq:oneendtrickconfig} \sum_\mathbf{y} G^{(q_1)}(x;\mathbf{y},t) \Gamma G^{(q_2)}(\mathbf{y},t;z) ,
\end{align}
i.e.\ the propagators are connected at spacetime point $(\mathbf{y},t)$ with a sum over $\mathbf{y}$, but no further propagators starting or ending at $(\mathbf{y},t)$. The one-end trick is thus particularly suited to compute correlation matrix elements, where at least one of the two interpolating operators is either a $q \bar{q}$ or a two-meson operator, i.e.\ $\mathcal{O}^{\Op{1}}$, $\mathcal{O}^{\Op{5}}$ or $\mathcal{O}^{\Op{6}}$ defined in section~\ref{sec:opbasis}.

One has to solve $2 N$ linear systems labeled by $n = 1,\ldots,N$,
\begin{align}
\label{EQN313} & \sum_y D_{a,A;b,B}^{(q_1)}(x;y) \phi_{b,B}^{(q_1)}(y)[t_0,n] = \xi_{a,A}(x)[t_0,n] \\
 & \sum_y D_{a,A;b,B}^{(q_2)}(x;y) \tilde{\phi}_{b,B}^{(q_2)}(y)[t_0,\Gamma,n] = (\gamma_5 \Gamma^\dagger \xi)_{a,A}(x)[t_0,n] ,
\end{align} 
where $\xi$ is a stochastic timeslice source defined in (\ref{eq:stochasticinversion}). With the resulting $\phi$ and $\tilde{\phi}$ the product of propagators can be estimated as,
\begin{align}
\sum_\mathbf{y} G^{(q_1)}(x;\mathbf{y},t) \Gamma G^{(q_2)}(\mathbf{y},t;z) = \frac{1}{N} \sum_{n=1}^N \phi^{(q_1)}(x)[t,n] \tilde{\phi}^{(q_2)}(z)[t,\Gamma,n]^\dagger \gamma_5 + \textrm{unbiased noise}.
\end{align}

Applying the one-end trick to the example correlation function \eqref{eq:examplecorrfunc} results in
\begin{align}
\label{eq:oneendonexamplecorrfunc} C_{1 1}(t) = -\frac{1}{N} \sum_{n=1}^N \frac{1}{\sptV} \bigg\langle \sum_\mathbf{y} \tilde{\phi}^{(d)}(\mathbf{y},t_2)[t_1,\mathbbm{1},n]^\dagger \gamma_5 \phi^{(u)}(\mathbf{y},t_2)[t_1,n] \bigg\rangle_U .
\end{align}

Note that there is a big advantage in efficiency when using the one-end trick, compared to using two ordinary stochastic timeslice-to-all propagators discussed in the previous subsection. The one-end trick introduces $\approx V_s \times (\textrm{number of signal terms})$ stochastic noise terms, i.e.\ for the example correlation function (\ref{eq:oneendonexamplecorrfunc}) $\propto V_s^3$ stochastic noise terms. In contrast to that using two stochastic timeslice-to-all propagators (\ref{eq:stochasticpropsonexamplecorrfunc}) will generate $\propto V_s^4$ stochastic noise terms.

% ********************

\subsection{\label{SEC498}Sequential propagators}

Quite similar to the one-end trick the technique of sequential propagators (cf.\ e.g.\ \cite{Martinelli:1988rr}) is applicable when two propagators are connected at spacetime point $(\mathbf{y},t)$ with a sum over $\mathbf{y}$, but no further propagators start or end at $(\mathbf{y},t)$, i.e.\ as in (\ref{eq:oneendtrickconfig}),
\begin{align}
\label{eq:oneendtrickconfig_} \sum_\mathbf{y} G^{(q_1)}(x;\mathbf{y},t) \Gamma G^{(q_2)}(\mathbf{y},t;z) .
\end{align}

In practice, sequential propagators need to be combined with the methods discussed in the previous subsections, point-to-all propagators, stochastic timeslice-to-all propagators and/or the one-end trick. In the following we explain the basic idea of using sequential propagators in combination with point-to-all propagators. Corresponding equations for stochastic timeslice-to-all propagators or the one-end trick are straightforward to derive and, therefore, not presented.

In a first step a point-to-all propagator for flavor $q_2$ from a fixed spacetime point $z$ to any other spacetime point $y$ is computed by solving 12 linear systems as explained in section~\ref{sec:pointtoallprops}. This propagator $\phi_{a,A}^{(q_2)}(x)[b,B,y] = G_{a,A;b,B}^{(q_2)}(x;y)$ is then used as the right hand side of further sets of 12 linear systems,
\begin{align}
\label{EQN311} \sum_y D_{a,A;b,B}^{(q_1)}({\bf x},x_0;y) \psi_{b,B}^{(q_1;q_2)}(y)[t,\Gamma;c,C,z] = (\Gamma \phi)_{a,A}^{(q_2)}({\bf x},x_0)[c,C,z] \delta(x_0;t)
\end{align} 
(labeled by indices $c = 1,2,3$, $C = 1,2,3,4$ and $t$), which have to be solved with respect to $\psi$. Then
\begin{align}
\label{EQN866} \bigg(\sum_\mathbf{y} G^{(q_1)}(x;\mathbf{y},t) \Gamma G^{(q_2)}(\mathbf{y},t;z)\bigg)_{A,B}^{a,b} = \psi_{a,A}^{(q_1;q_2)}(x)[t,\Gamma;b,B,z] .
\end{align}

The use of sequential propagators is quite efficient when $t = z_0$, i.e.\ when $q_2$ propagates within timeslice $t = z_0$. Then the total number of inversions is limited to 24
(12 to obtain $\phi$, then 12 more to obtain the final result $\psi$). The result is then exact. If one is, however, interested in several $t \neq z_0$, one has to solve 12 linear systems for each value of $t$ to obtain the corresponding $\psi$, which can be a rather computer time consuming task.

Even though sequential propagators have a similar application as the one-end trick (cf.\ (\ref{eq:oneendtrickconfig}) and (\ref{eq:oneendtrickconfig_})) with the additional practical limitation to $t = z_0$, it is essential to use this technique, when computing the correlation matrix (\ref{eq:corrmatrix}). In contrast to the one-end trick, which is a self contained technique, sequential propagators always need to be combined with other techniques and, hence, offer a lot of flexibility. For example the triangular diagram appearing e.g.\ in $C_{1 5}$ can be computed in a very efficient way, when combining sequential propagators and the one-end trick (for details cf.\ section~\ref{SEC459}).

% ********************

\subsection{\label{sec:coloreddiagrams}Diagrammatic representation of propagators and correlation functions}

In the following we introduce a diagrammatic representation of propagators and correlation functions, which will be particularly useful in section~\ref{seq:compofdiagrams}. There we discuss how to compute the elements of the correlation matrix~\ref{eq:corrmatrix} using the previously introduced numerical techniques.

Propagators are represented by arrows, which point from the sources (denoted by $\xi$ in sections~\ref{sec:pointtoallprops} to \ref{SEC498}) to the corresponding solutions of the linear systems (denoted by $\phi$). Point sources (\ref{eq:pointtoalllinearsystem}) (which are restricted to a specific spacetime point) are represented by black boxes, while stochastic timeslice sources (\ref{eq:stochasticinversion}) and the solutions of the linear systems (which have non-zero values for any space or spacetime point, respectively) are represented by black circles. Point-to-all propagators are colored in blue, stochastic timeslice-to-all propagators in red and a one-end trick combination of two propagators in green. Since sequential propagators are always used in combination with one of the three techniques mentioned in the previous subsection, they are colored accordingly in blue, red or green. Two propagators expressed in terms of a sequential propagator can easily be identified by a big circle at their junction.

After correlation functions are expressed in terms of sources $\xi$ and solutions $\phi$ using the techniques discussed in sections~\ref{sec:pointtoallprops} to \ref{SEC498}, there are typically one or more spatial sums. These sums are indicated in the diagrams by $\sum \longleftrightarrow$ in black. When such sums are implicitly realized by either the one-end trick or by sequential propagators, they are indicated by $\sum \longleftrightarrow$ in gray.

\begin{figure}[htb]
\begin{center}
${}\hspace{-2.97cm}{}$\includegraphics[scale=.7,page=1]{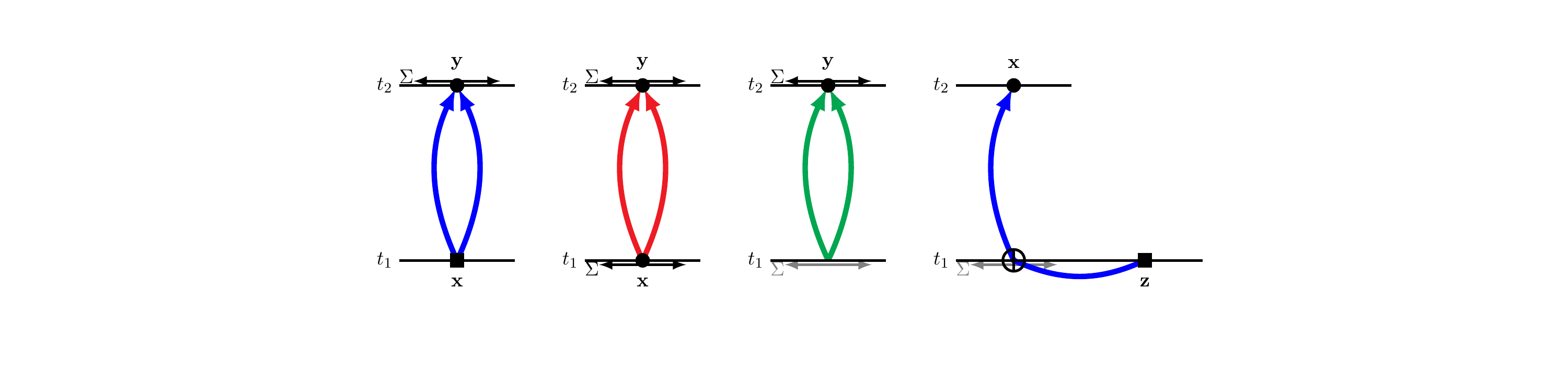}
\caption{\label{fig:techcolorscheme}From left to right: the correlation function $C_{1 1}$ expressed in terms of point-to-all propagators (\ref{EQN643}), stochastic timeslice-to-all propagators (\ref{eq:stochasticpropsonexamplecorrfunc}) and the one-end trick (\ref{eq:oneendonexamplecorrfunc}) and the sequential propagator (\ref{EQN866}).}
\end{center}
\end{figure}

To illustrate this diagrammatic representation, we show a couple of examples in Figure~\ref{fig:techcolorscheme}: the correlation function $C_{1 1}$ expressed in terms of point-to-all propagators (\ref{EQN643}), stochastic timeslice-to-all propagators (\ref{eq:stochasticpropsonexamplecorrfunc}) and the one-end trick (\ref{eq:oneendonexamplecorrfunc}) as well as the sequential propagator (\ref{EQN866}).

% ********************
% ********************
% ********************
% ********************
% ********************

\newpage

\section{\label{seq:compofdiagrams}Computation of the correlation matrix}

Each diagram of the correlation matrix \eqref{eq:corrmatrix} (cf.\ also Figure~\ref{fig:diagrammatrix}) can be computed in several ways (denoted as methods in the following) by combining the techniques for propagator computation discussed in section~\ref{sec:propcomp}. An efficient method satisfies the following three criteria,
\begin{itemize}
\item[(A)] requires only a small number of inversions,

\item[(B)] averages the diagram over space (to reduce statistical fluctuations originating from the gauge links),

\item[(C)] introduces no or only a moderate number of stochastic noise terms (due to stochastic timeslice-to-all propagators or the one-end trick).
\end{itemize}
In practice ideal methods do not exist. In particular (B) and (C) exclude each other: to avoid additional stochastic noise terms, one has to use exclusively point-to-all-propagators, which do not average the diagram over space. Therefore, the best compromise has to be found for each diagram. We do this in this section by comparing the efficiency of different methods numerically. 

To compare two methods $(a)$ and $(b)$ quantitatively, we define the quality ratio
\begin{align}
\label{EQN670} R^{(a),(b)}(t) = \frac{\Delta C^{(a)}(t) \cdot \sqrt{\tau^{(a)}}}{\Delta C^{(b)}(t) \cdot \sqrt{\tau^{(b)}}} ,
\end{align}
where $\Delta C^{(x)}(t)$ denotes the statistical error of the diagram at temporal separation $t$ obtained with method $(x)$ and $\tau^{(x)}$ the corresponding computing time. $R^{(a),(b)}(t) < 1$ indicates that method $(b)$ is inferior to method $(a)$, i.e.\ that the statistical error obtained with method $(a)$ at comparable computing time is smaller, while $R^{(a),(b)}(t) > 1$ indicates the opposite. In the following we label the investigated combinations of techniques for each diagram by $(a)$, $(b)$, $(c)$, ... according to decreasing quality, i.e.\ method $(a)$ is always the best. We compare the remaining methods $(b)$, $(c)$, ... exclusively to method $(a)$, hence, it is convenient to use the notation $R^{(b)}(t) = R^{(a),(b)}(t)$.

The numerical comparisons presented in the following are based on gauge link configurations generated by the PACS-CS collaboration \cite{Aoki:2008sm}. Details regarding these configurations are collected in appendix~\ref{SEC876}. When using point-to-all propagators, we did a single set of 12 inversions per gauge link configuration corresponding to a set of 12 point sources at the same randomly chosen point in space. For stochastic timeslice-to-all propagators, we used 15 independently chosen stochastic sources per gauge link configuration, i.e.\ $N = 15$ in section~\ref{sec:stochtimeslicetoallprops}. When using the one-end trick, we inverted 3 independently chosen stochastic sources per gauge link configuration, i.e.\ $N = 3$ in section~\ref{sec:oneendtrick}. The resulting quality ratios $R^{(x)}$ are only weakly dependent on $N$, since larger $N$ lead to smaller statistical errors $\Delta C^{(x)}$ (suppressed by approximately $1 / \sqrt{N}$)), but also to larger computing times $\tau^{(x)}$ (increased by approximately $N$), which has been taken into account in the definition (\ref{EQN670}).

For some techniques the number of inversions and, hence, the computing time is proportional to the number of temporal separations $t$, for which the correlation matrix element is computed. In such cases we limited our computations to the range $0 \leq t/a \leq 14$, which are typical temporal separations where we have signal in our correlators. Since these larger numbers of inversions are always done to compute the strange quark propagator, and are thus relatively cheap, the quality ratios $R^{(x)}$ are only weakly dependent on variations of the computed temporal range.

We do not compare all possible methods, but focus on promising methods, i.e.\ methods, which fulfill (A) to (C), the criteria listed above, at least to some extent. In particular, we do not consider methods making excessive use of stochastic techniques and, hence, introduce a large number of stochastic noise terms. The maximum we allow is either a stochastic timeslice-to-all propagator in combination with the one-end trick or twice the one-end trick\footnote{The number of stochastic noise terms is $\approx V_s^M \times (\textrm{number of signal terms})$, where $M$ counts the number of stochastic timeslice-to-all propagators and applications of the one-end trick (cf.\ sections \ref{sec:stochtimeslicetoallprops} and \ref{sec:oneendtrick}). We set the maximum to $M = 2$.}. Moreover, whenever possible, we combine the techniques for propagator computation in such a way that additional inversions for each considered temporal separation $t$ are not necessary (cf.\ the previous paragraph). In particular, to avoid a very large number of inversions, we exclude methods where additional inversions for each considered temporal separation $t$ are done for point-to-all propagators (point-to-all propagators require 12 inversions per separation).

The following subsections focus on the results of the comparisons of different methods. Technical details regarding the implementation of some of the diagrams can be found in appendix~\ref{APP002}.

% ********************

\subsection{\label{sec:2q2q}Two-quark -- two-quark correlation function} 

% ********************
% ********************
% ********************

\subsubsection{\label{SEC489}$C_{1 1}$ ($1 \equiv q \bar{q}$)}

Promising methods for numerical computation are the following (cf.\ also section \ref{sec:propcomp}, where this diagram has extensively been discussed, to illustrate various techniques for propagator computation).
\begin{figure}[H]
\hspace{-8.4cm}\includegraphics[scale=.7, page=1]{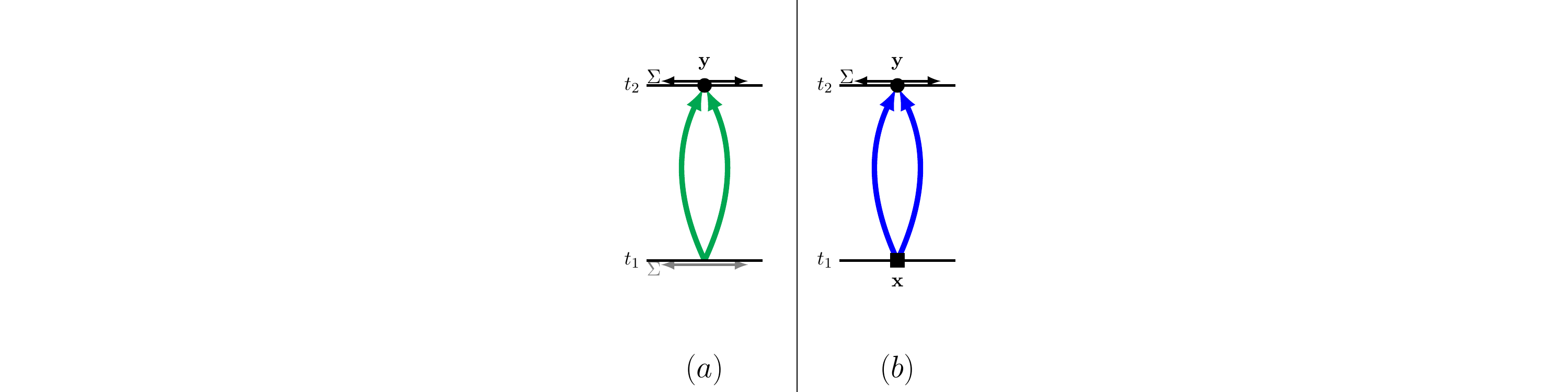}
\end{figure}
\begin{itemize}
\item[$(a)$] \textbf{one-end trick}: \\
2 $u/d$ inversions 
\item[$(b)$] \textbf{2 point-to-all propagators}: \\ 
12 $u/d$ inversions
\end{itemize}

\begin{figure}[htb]
\begin{center}
\includegraphics[scale=0.39,page=01]{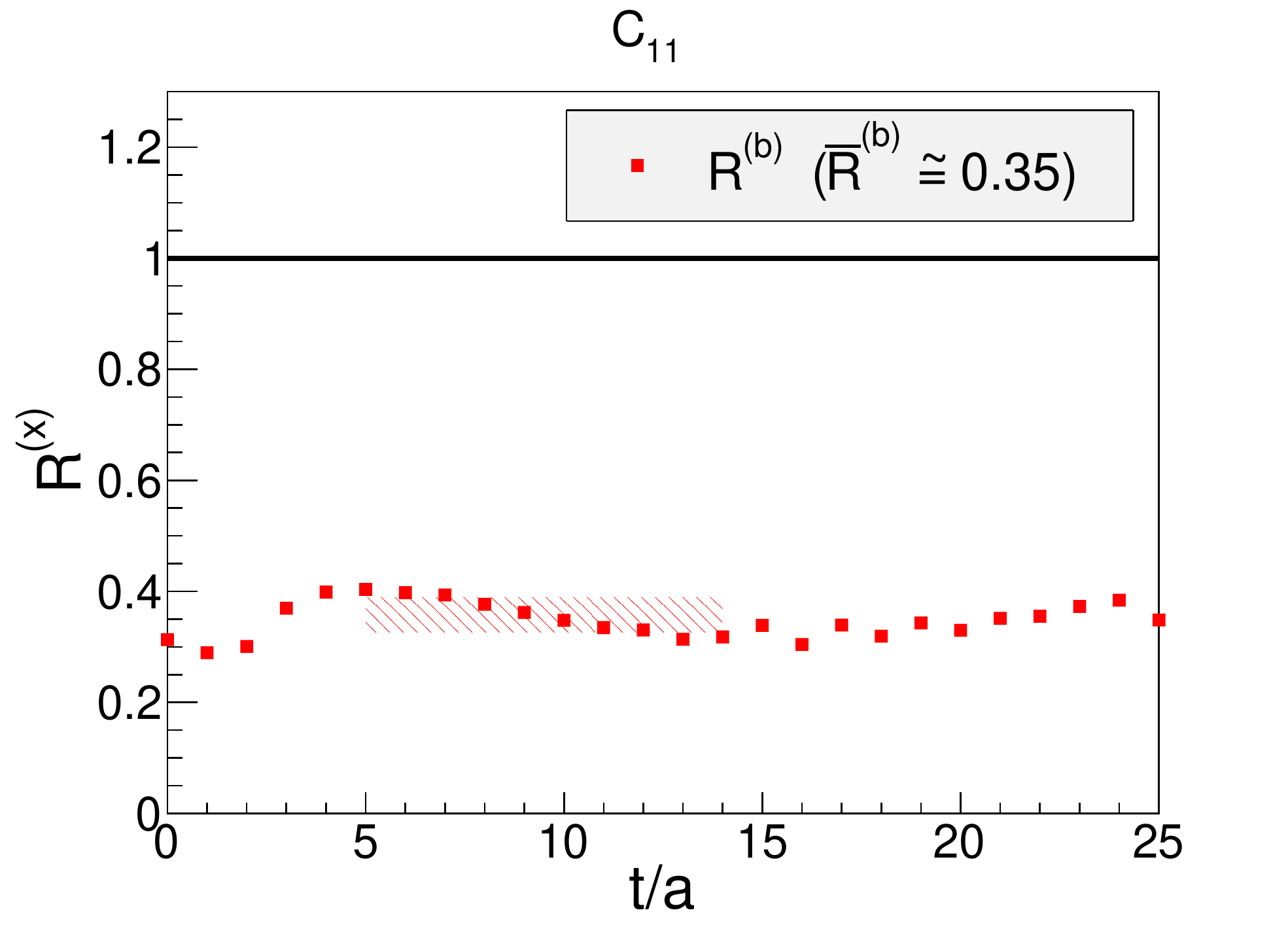}
\caption{\label{FIG755}Efficiency of different methods for $C_{1 1}$.}
\end{center}
\end{figure}

The one-end trick averages the diagram over space (reflected by $\sum \longleftrightarrow$ both at timeslice $t_1$ and at timeslice $t_2$ in $(a)$), but at the same time introduces stochastic noise terms. A numerical comparison using the lattice setup described in appendix~\ref{SEC876} results in $R^{(b)}(t) \approx 0.3 \ldots 0.4$ and $\bar{R}^{(b)} = 0.35$, where $\bar{R}^{(b)}$ is the average of $R^{(b)}(t)$ in the range $5 \leq t/a \leq 14$, which are typical temporal separations to extract energy levels from our correlation functions (cf.\ Figure~\ref{FIG755}). Consequently, the one-end trick leads to statistical errors smaller by a factor around $2$ compared to point-to-all propagators, when investing similar amounts of computing time. Hence, this is our preferred method to compute $C_{1 1}$. It should, however, be noted that the efficiency might depend to some extent on the lattice setup, in particular could be somewhat different, when another $u/d$ quark mass or spatial volume is used\footnote{For example in \cite{Wiese2012} it has been found that the efficiency of the two methods is quite similar, when the $u/d$ quark mass is replaced by the much heavier charm quark mass. Similarly, counting the number of signal terms and the number of stochastic noise terms in (\ref{eq:oneendonexamplecorrfunc}) ($\propto V_s^2$ and $\propto V_s^3$, respectively) suggests that the one-end trick becomes even more efficient for larger spatial volumes $V_s$.}.

% ********************
% ********************
% ********************

\subsection{\label{sec:2q4q}Two-quark -- four-quark correlation functions} 

% ********************

\subsubsection{\label{SEC490}$C_{1 2}$, $C_{1 3}$, $C_{1 4}$ ($1 \equiv q \bar{q}$, $2 \equiv K \bar{K}\textrm{, point}$, $3 \equiv \eta_s \pi\textrm{, point}$, $4 \equiv Q \bar{Q}$)}

Promising methods for numerical computation are the following (cf.\ also appendix~\ref{APP003}, where certain technical aspects are discussed).
\begin{figure}[H]
\hspace{-7.1cm}\includegraphics[scale=.7, page=2]{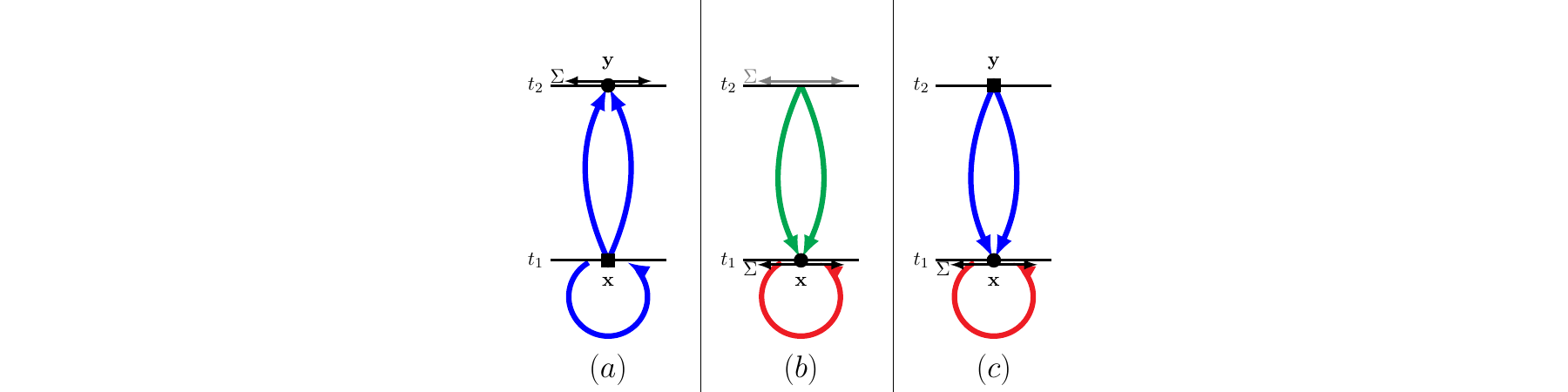}
\end{figure}
\begin{itemize}
\item[$(a)$] \textbf{3 point-to-all propagators}: \\
12 $u/d$ inversions, 12 $s$ inversions
\item[$(b)$] \textbf{one-end trick}: \\
2 $u/d$ inversions \\
\textbf{stochastic timeslice-to-all propagator}: \\
$\# t$ $s$ inversions ($\# t$ denotes the number of temporal separations computed)
\item[$(c)$] \textbf{2 point-to-all propagators}: \\
12 $u/d$ inversions \\
\textbf{stochastic timeslice-to-all propagator}: \\
$\# t$ $s$ inversions
\end{itemize}

\begin{figure}[htb]
\begin{center}
\includegraphics[scale=0.39,page=02]{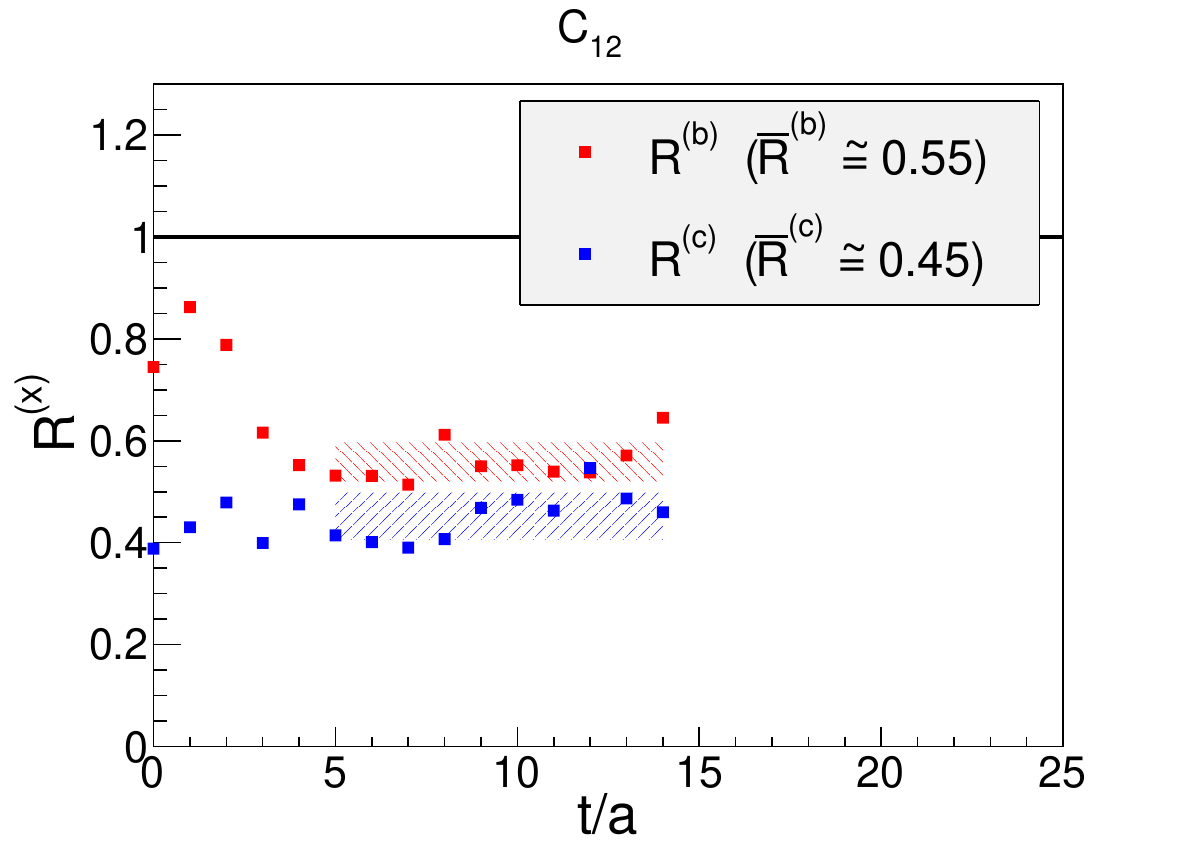}
\includegraphics[scale=0.39,page=03]{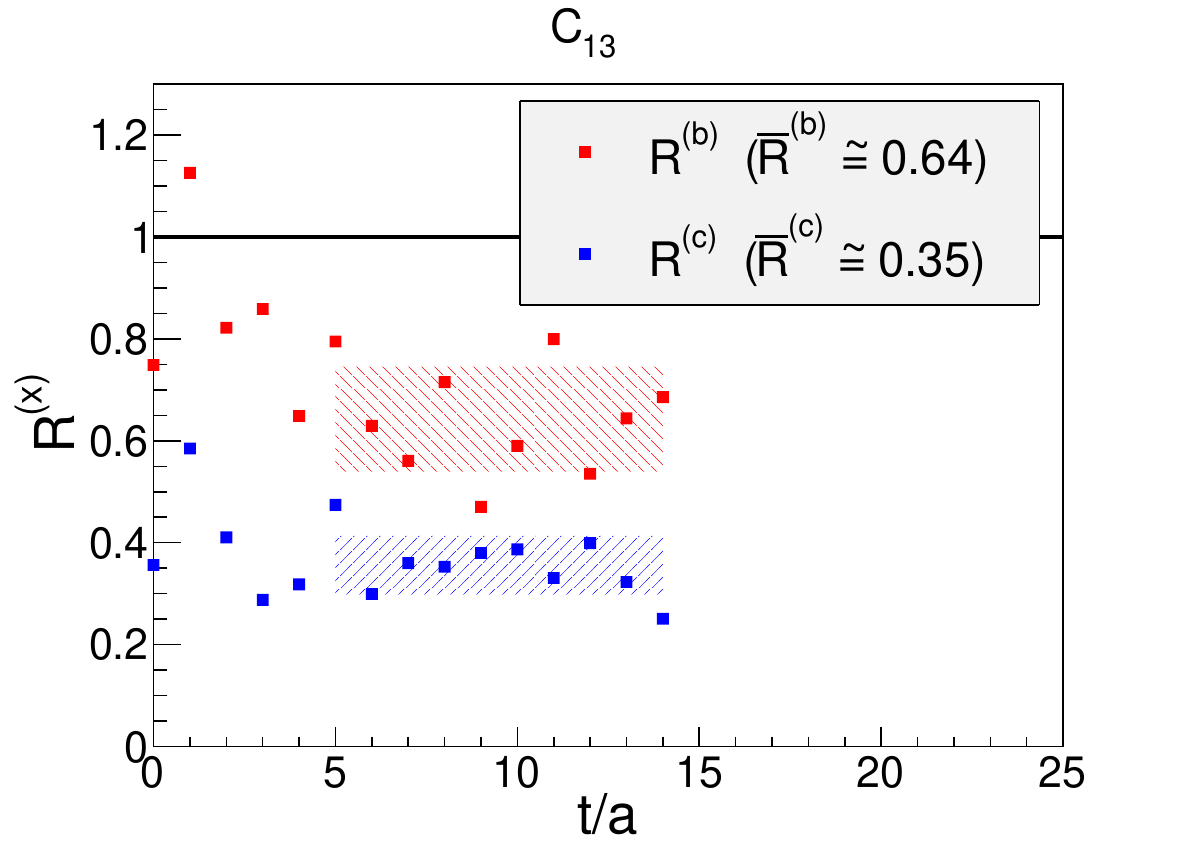}
\includegraphics[scale=0.39,page=04]{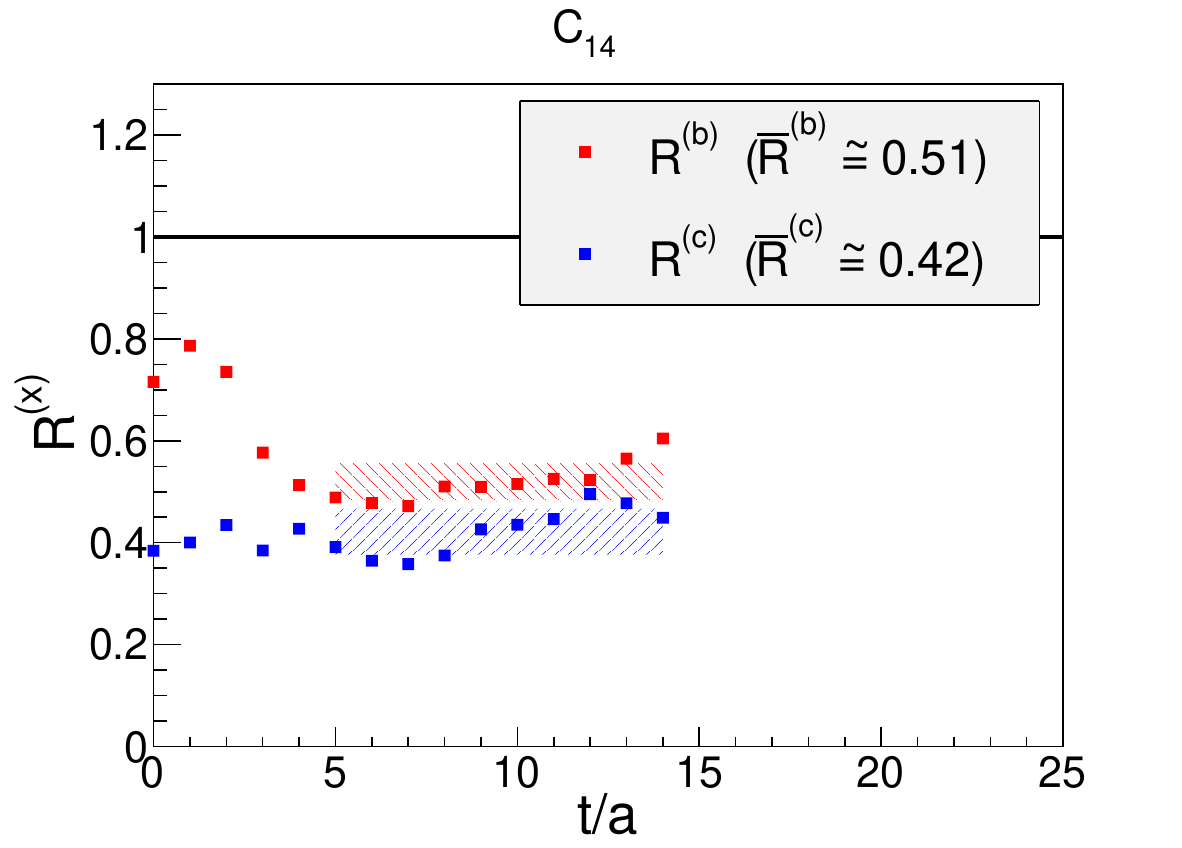}
\caption{\label{FIG756}Efficiency of different methods for $C_{1 2}, C_{1 3}, C_{1 4}$.}
\end{center}
\end{figure}

Method~$(b)$ averages the diagram over space (reflected by $\sum \longleftrightarrow$ both at timeslice $t_1$ and at timeslice $t_2$), but at the same time introduces a rather large number of stochastic noise terms. Methods $(a)$ and $(c)$ on the other hand compute only a single sample. For all three diagrams $C_{1 2}$, $C_{1 3}$ and $C_{1 4}$ method~$(a)$, which does not introduce any stochastic noise terms, is clearly more efficient, as shown in Figure~\ref{FIG756}), with corresponding quality ratios $\bar{R}^{(b)} , \bar{R}^{(c)} \approx 0.35 \ldots 0.63$.

% ********************
% ********************
% ********************

\subsection{\label{sec:2q2m}Two-quark -- two-meson correlation functions} 

% ********************

\subsubsection{\label{SEC459}$C_{1 5}$ ($1 \equiv q \bar{q}$, $5 \equiv K \bar{K} \textrm{, 2part}$)}

Since the $s$ quark propagates within timeslice $t_1$, $C_{1 5}$ can be computed efficiently using a sequential propagator (the $s$ quark propagator together with either the $u$ or the $d$ quark propagator). Promising methods are the combination of a sequential propagator with the one-end trick or with point-to-all propagators as shown in the figure below (cf.\ also appendix~\ref{APP004}, where certain technical aspects are discussed).
\begin{figure}[H]
\hspace{-4.5cm}\includegraphics[scale=.7, page=6]{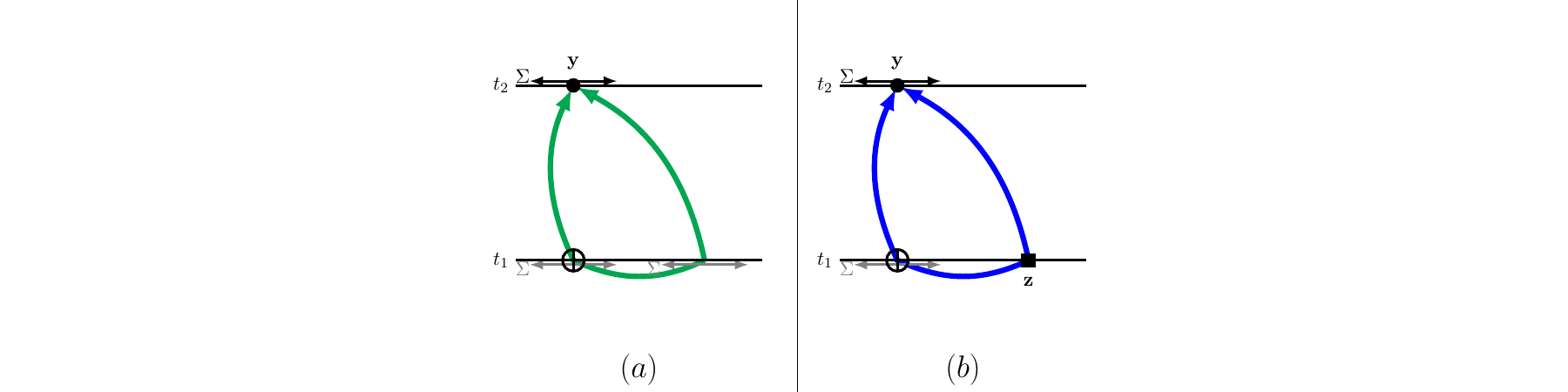}
\end{figure}
\begin{itemize}
\item[$(a)$] \textbf{one-end trick (with 1 sequential propagator)}: \\
2 $u/d$ inversions, 1 $s$ inversion
\item[$(b)$] \textbf{2 point-to-all propagators (1 is a sequential propagator)}: \\
24 $u/d$ inversions, 12 $s$ inversions
\end{itemize}

\begin{figure}[htb]
\begin{center}
\includegraphics[scale=0.39,page=05]{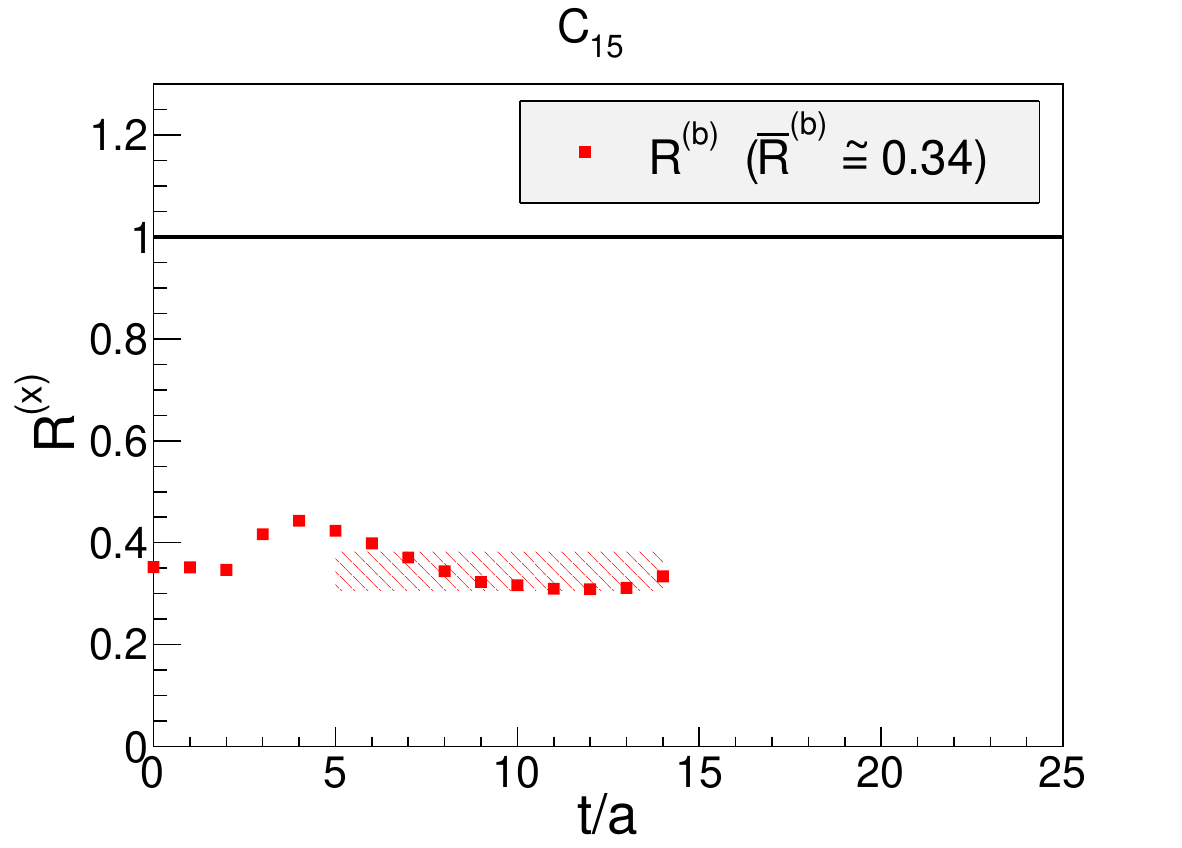}
\caption{\label{FIG757}Efficiency of different methods for $C_{1 5}$.}
\end{center}
\end{figure}

As for the matrix element $C_{1 1}$ discussed in section~\ref{SEC489}, the one-end trick is more efficient than point-to-all propagators (cf.\ Figure~\ref{FIG757}).

% ********************

\subsubsection{\label{SEC298}$C_{1 6}$ ($1 \equiv q \bar{q}$, $6 \equiv \eta_s \pi \textrm{, 2part}$)}

$C_{1 6}$ is a product of two disconnected parts.
%
% \begin{align}
% \label{EQN076} C_{1 6}(t) =  +\frac{1}{\sptV^{3/2}} \sum_{{\bf x},{\bf y},{\bf z}} \Big\langle \textrm{Tr}\Big(\D{(d)}{x}{t_1}{y}{t_2} \D{(u)}{y}{t_2}{x}{t_1} \gamma_5\Big) \textrm{Tr}\Big(\D{(s)}{z}{t_1}{z}{t_1} \gamma_5\Big) \Big\rangle_U .
% \end{align}
%
The quark loop can either be computed using a point-to-all or a stochastic timeslice-to-all propagator. Promising methods to compute this are the following.
\begin{figure}[H]
\hspace{-4.5cm}\includegraphics[scale=.7, page=5]{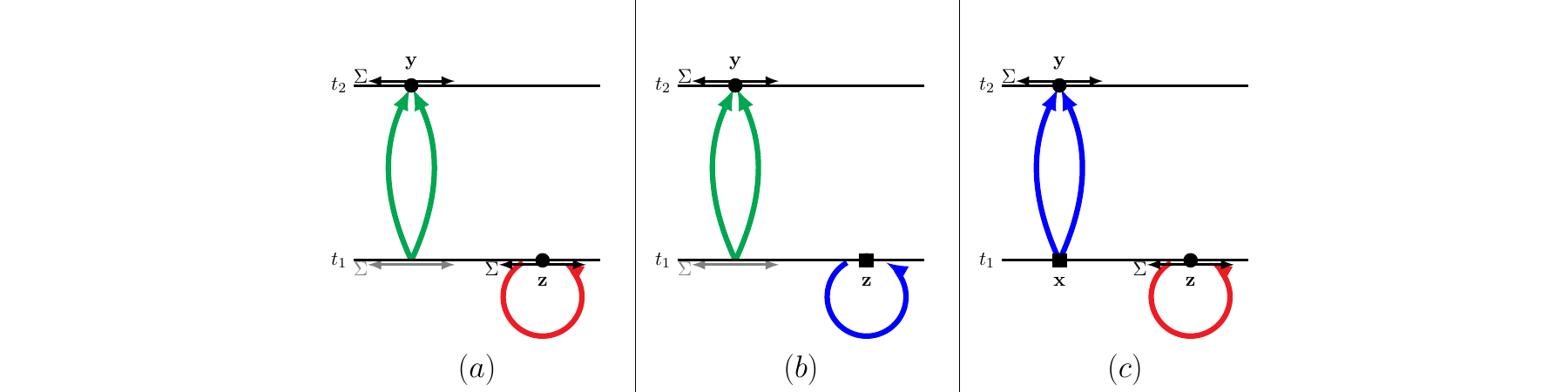}
\end{figure}
\begin{itemize}
\item[$(a)$] \textbf{one-end trick}: \\
1 $u/d$ inversion \\
\textbf{stochastic timeslice-to-all propagator}: \\
1 $s$ inversion
\item[$(b)$] \textbf{one-end trick}: \\
1 $u/d$ inversion \\
\textbf{point-to-all propagator}: \\
12 $s$ inversions
\item[$(c)$] \textbf{2 point-to-all-propagators}: \\
12 $u/d$ inversions \\
\textbf{stochastic timeslice-to-all propagator}: \\
1 $s$ inversions
\end{itemize}
%
% From now on we do not anymore discuss in detail, how to implement these different methods numerically. Equations like (\ref{EQN610}) to (\ref{EQN609}), (\ref{EQN611}) and (\ref{EQN612}) can be derived in a straightforward way starting from (\ref{EQN076}) and the shown diagrams and using the theoretical basics from section~\ref{sec:propcomp}.

\begin{figure}[htb]
\begin{center}
\includegraphics[scale=0.39,page=06]{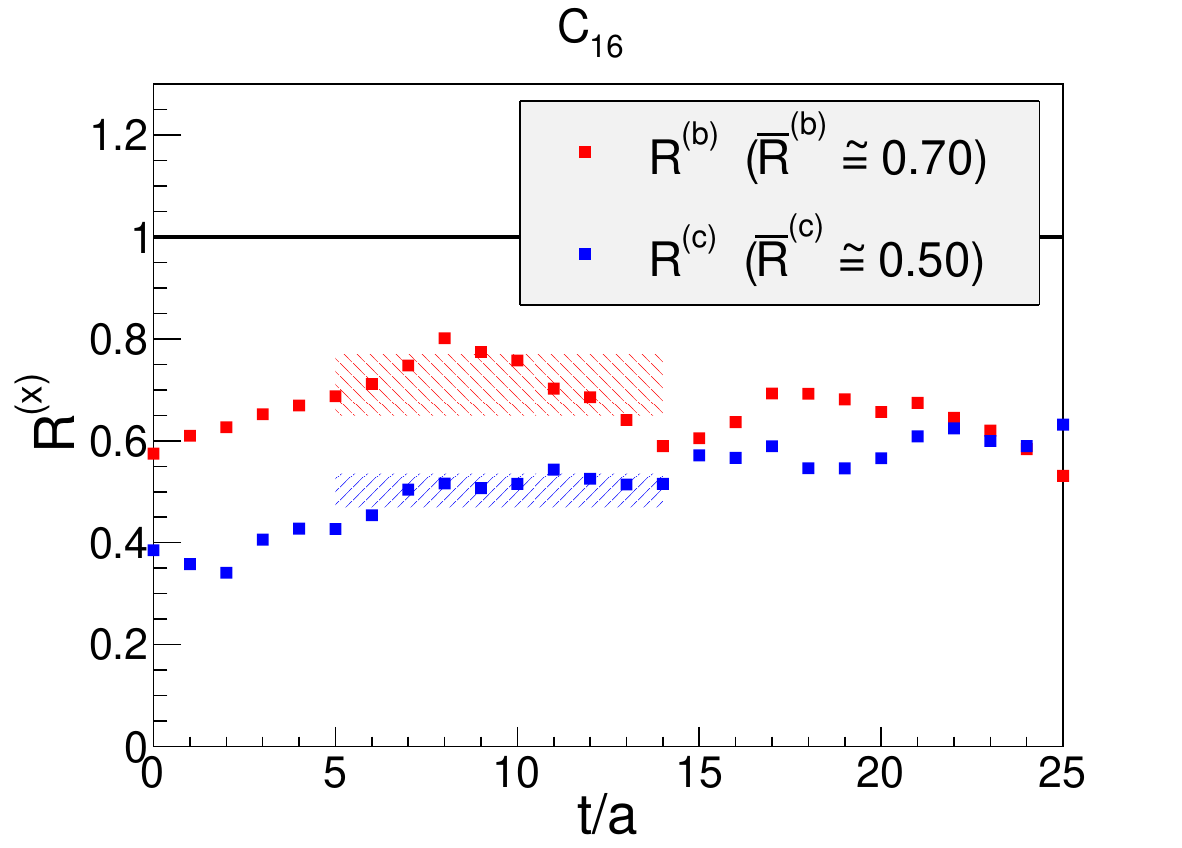}
\caption{\label{FIG758}Efficiency of different methods for $C_{1 6}$.}
\end{center}
\end{figure}

Method~$(a)$ introduces a rather large number of stochastic noise terms. However, in contrast to method~$(b)$ and method~$(c)$, which also introduce stochastic noise terms, method~$(a)$ averages the diagram over space and results in being more efficient (cf.\ Figure~\ref{FIG758}).

% ********************
% ********************
% ********************

\subsection{\label{sec:4q4q}Four-quark -- four-quark correlation functions}

% ********************

\subsubsection{$4 \times$ connected $C_{2 2}$, $C_{2 3}$, $C_{2 4}$, $C_{3 3}$, $C_{3 4}$, $C_{4 4}$ ($2 \equiv K \bar{K}\textrm{, point}$, $3 \equiv \eta_s \pi\textrm{, point}$, $4 \equiv Q \bar{Q}$)}

The correlation matrix elements $C_{2 2}$, $C_{2 3}$, $C_{2 4}$, $C_{3 3}$, $C_{3 4}$ and $C_{4 4}$ have different spin and color structures, but are identical with respect to spacetime. Therefore, it is appropriate to discuss them together. Since both at timeslice $t_1$ and at timeslice $t_2$ four quarks are located at the same point in space, there is only a single possibility to compute this type of diagram efficiently.
\begin{figure}[H]
\hspace{-9.8cm}\includegraphics[scale=.7, page=3]{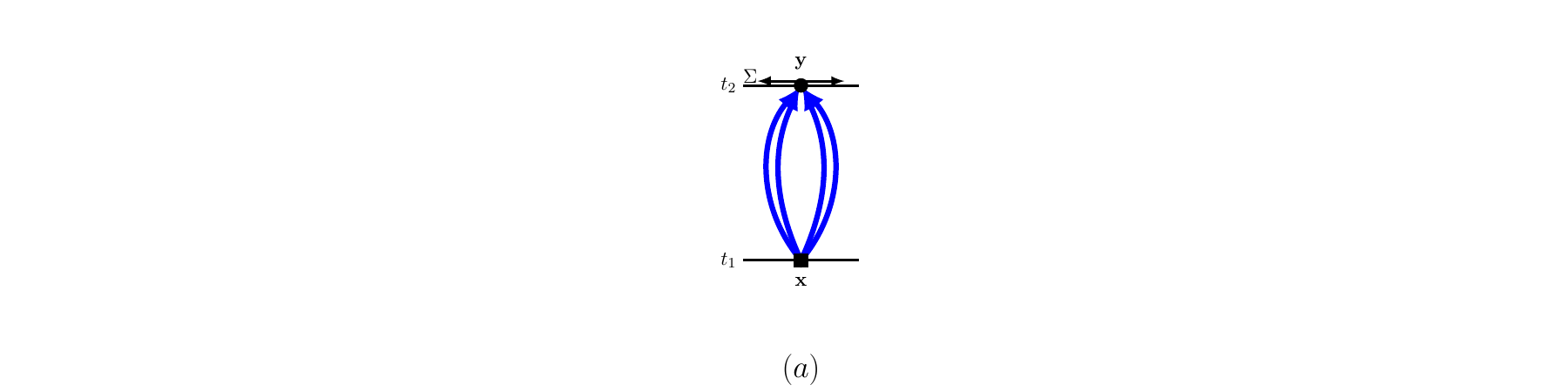}
\end{figure}
\begin{itemize}
\item[$(a)$] \textbf{4 point-to-all-propagators}: \\
12 $u/d$ inversions, 12 $s$ inversions
\end{itemize}

% ********************

\subsubsection{\label{SEC457}$2 \times$ connected $C_{2 2}$, $C_{2 3}$, $C_{2 4}$, $C_{3 3}$, $C_{3 4}$, $C_{4 4}$ ($2 \equiv K \bar{K}\textrm{, point}$, $3 \equiv \eta_s \pi\textrm{, point}$, $4 \equiv Q \bar{Q}$)}

Again there is only a single possibility to compute this type of diagram efficiently.
\begin{figure}[H]
\hspace{-9.8cm}\includegraphics[scale=.7, page=4]{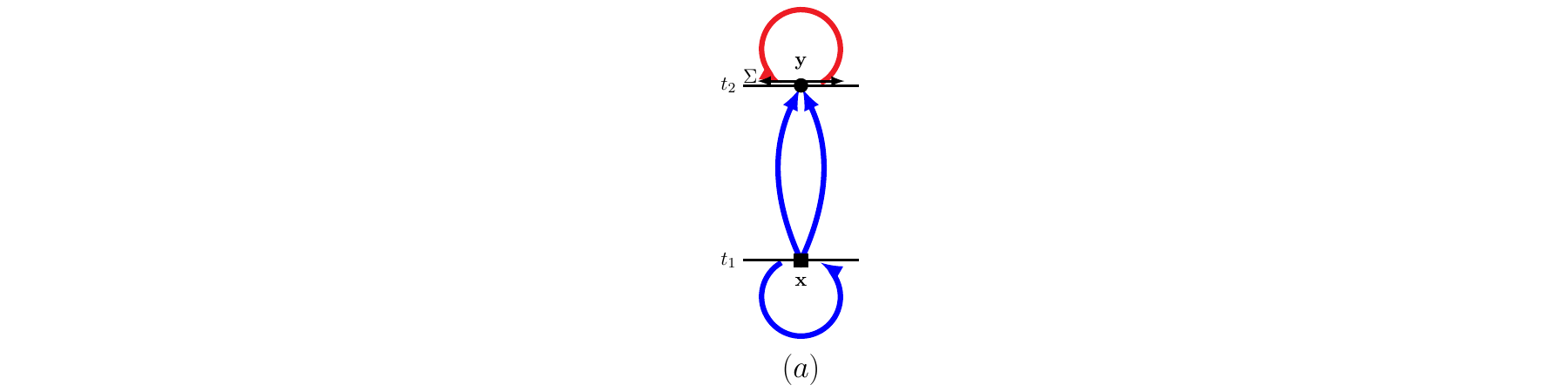}
\end{figure}
\begin{itemize}
\item[$(a)$] \textbf{3 point-to-all-propagators}: \\
12 $u/d$ inversions, 12 $s$ inversions \\
\textbf{stochastic timeslice-to-all propagator}: \\
$\# t$ $s$ inversions
\end{itemize}

% ********************

\subsubsection{\label{SEC369}Comparison of $4 \times$ connected and $2 \times$ connected diagrams}

The statistical error of each of the six correlation matrix elements $C_{2 2}$, $C_{2 3}$, $C_{2 4}$, $C_{3 3}$, $C_{3 4}$ and $C_{4 4}$ is a combination of the statistical errors of the corresponding $4 \times$ connected and $2 \times$ connected diagrams and will be dominated by the larger of the two errors. Therefore, it is interesting to compare the statistical errors of the $4 \times$ connected and the $2 \times$ connected diagrams. For that purpose we use again the quality ratio defined in (\ref{EQN670}), $R^{2 \times \textrm{con}}(t) \equiv R^{4 \times \textrm{con},2 \times \textrm{con}}(t)$.

\begin{figure}[htb]
\begin{center}
\includegraphics[scale=0.39,page=01]{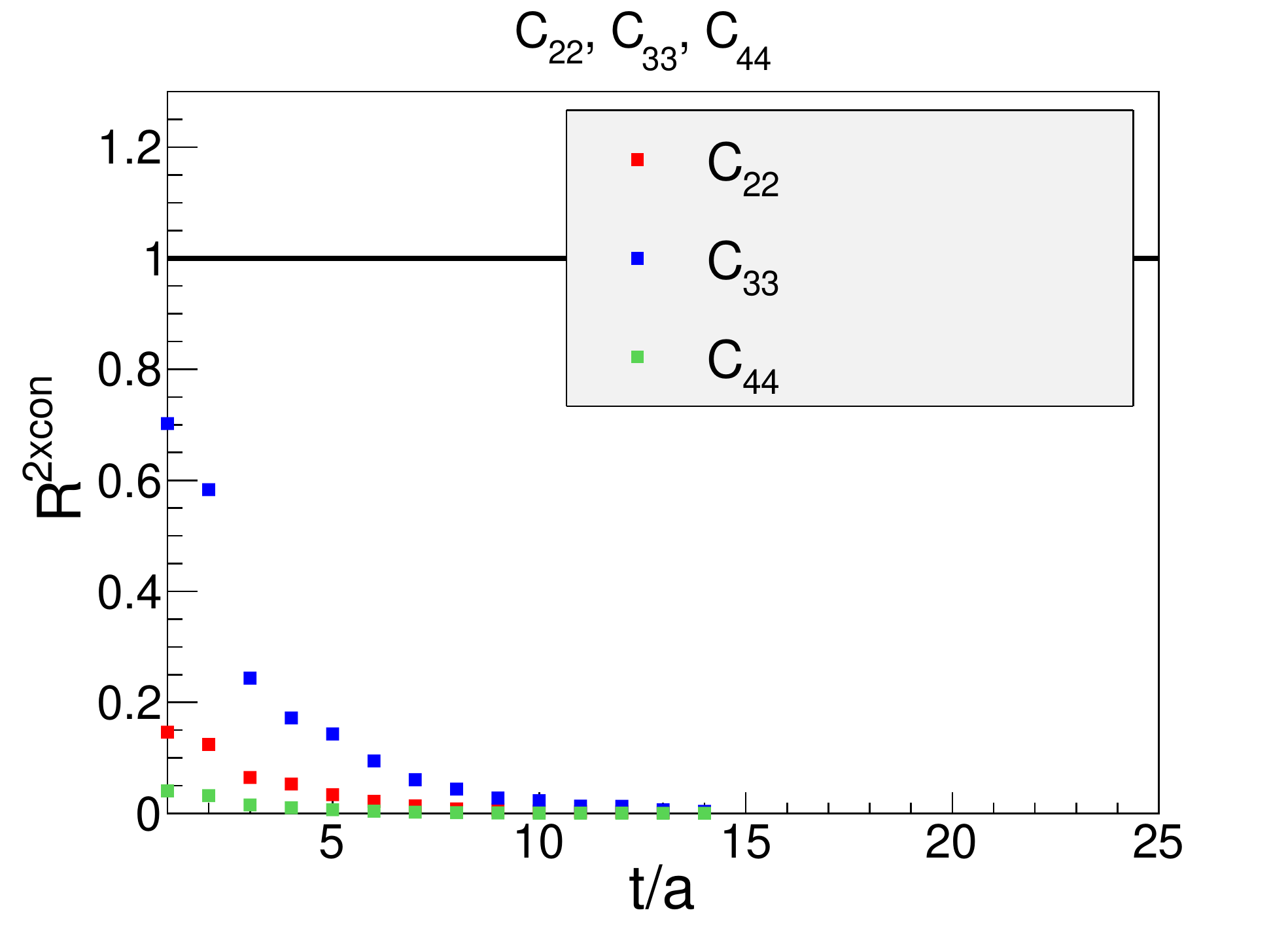}
\includegraphics[scale=0.39,page=03]{Compare_4x2x.pdf}
\includegraphics[scale=0.39,page=02]{Compare_4x2x.pdf}
\includegraphics[scale=0.39,page=04]{Compare_4x2x.pdf}
\caption{\label{FIG201}Comparison of $4 \times$ connected and $2 \times$ connected for $C_{2 2}$, $C_{2 3}$, $C_{2 4}$, $C_{3 3}$, $C_{3 4}$, $C_{4 4}$.}
\end{center}
\end{figure}

In the top row of Figure~\ref{FIG201} these quality ratios are shown for all six correlation matrix elements $C_{2 2}$, $C_{2 3}$, $C_{2 4}$, $C_{3 3}$, $C_{3 4}$ and $C_{4 4}$. They rapidly decrease with the temporal separation $t$ and are mostly below $0.1$ or even significantly smaller for $5 \leq t \leq 14$, where effective mass plateaus are typically read off. Consequently, the statistical errors of the correlation matrix elements will be dominated by the $2 \times$ connected diagrams. Thus, it is much more important to determine the optimal method of computation for the $2 \times$ connected than for the $4 \times$ connected diagrams.

To investigate this drastic difference in the statistical errors in more detail, we plot in the bottom row of Figure~\ref{FIG201} a related quantity, $\log(R^{2 \times \textrm{con}}(t) / R^{2 \times \textrm{con}}(t+a))$. Assuming an exponential behavior $R^{2 \times \textrm{con}}(t) \propto e^{-\alpha t}$, the plotted quantity $\log(\ldots)$ corresponds to the exponent $\alpha$. For all six cases $\log(\ldots)$ fluctuates around the same constant, $\alpha \approx 0.4$, which indeed shows that the statistical errors of the $2 \times$ connected diagrams increase exponentially in comparison to the $4 \times$ connected diagrams, i.e. proportionally to $e^{\alpha t}$.

\begin{figure}[htb]
\begin{center}
\includegraphics[scale=0.7,page=01]{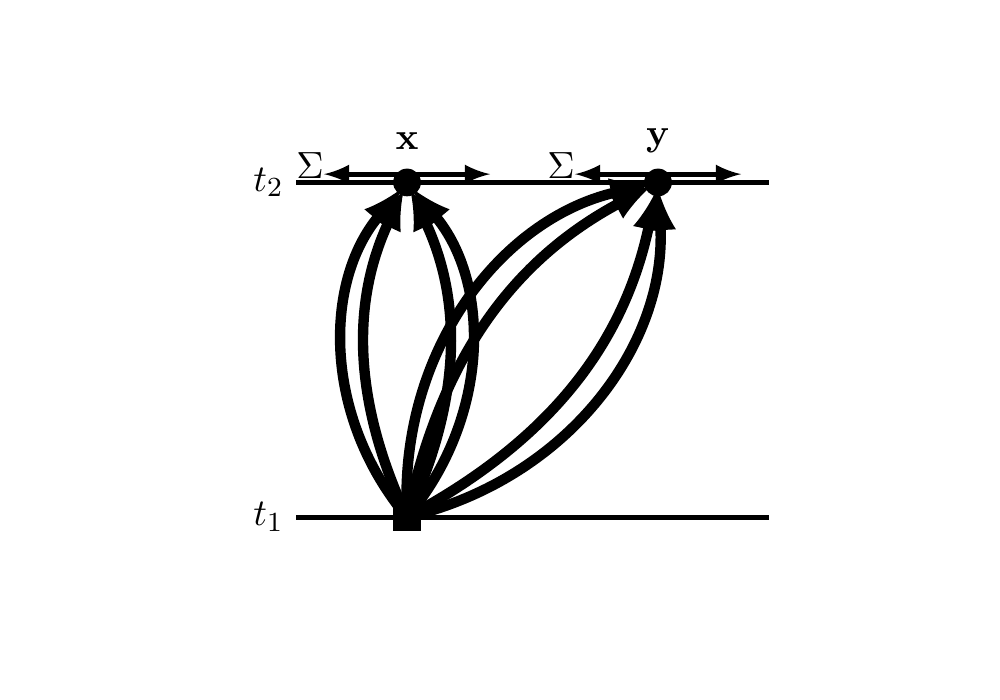}
\includegraphics[scale=0.7,page=02]{error_4x_2x.pdf}
\caption{\label{FIG731}Squared diagrams, which are proportional to the statistical errors of the $4 \times$ connected diagram (left) and the $2 \times$ connected diagram (right) of $C_{2 2}$.}
\end{center}
\end{figure}

This behavior can be understood in the following way. The squared statistical error of a diagram is proportional to the squared diagram and, hence, can be expressed as the correlation function of two eight-quark operators with appropriately chosen spin, color and spacetime structure. As an example consider,
\begin{align}
\label{EQN731} & \Big(\Delta C_{2 2}^{4 \times \textrm{con}}(t)\Big)^2 \propto \Big\langle \mathcal{O}[s^{(1)},s^{(2)},s^{(3)},s^{(4)}](t_2) \tilde{\mathcal{O}}[s^{(1)},s^{(2)},s^{(3)},s^{(4)}]^\dag(t_1) \Big\rangle \\
\label{EQN732} & \Big(\Delta C_{2 2}^{2 \times \textrm{con}}(t)\Big)^2 \propto \Big\langle \mathcal{O}[s^{(1)},s^{(1)},s^{(2)},s^{(2)}](t_2) \tilde{\mathcal{O}}[s^{(3)},s^{(3)},s^{(4)},s^{(4)}]^\dag(t_1) \Big\rangle
\end{align}
with
\begin{align}
 & \mathcal{O}[s_1,s_2,s_3,s_4] = \bigg(\sum_{\bf{x}} \Big({\bar s}_1({\bf x}) \gamma_5 u({\bf x})\Big) \Big({\bar d}({\bf x}) \gamma_5 s_2({\bf x})\Big)\bigg) \bigg(\sum_{\bf{y}} \Big({\bar s}_3({\bf y}) \gamma_5 u({\bf y})\Big) \Big({\bar d}({\bf y}) \gamma_5 s_4({\bf y})\Big)\bigg) \\
 & \tilde{\mathcal{O}}[s_1,s_2,s_3,s_4] = \sum_{\bf{x}} \Big({\bar s}_1({\bf x}) \gamma_5 u({\bf x})\Big) \Big({\bar d}({\bf x}) \gamma_5 s_2({\bf x})\Big) \Big({\bar s}_3({\bf x}) \gamma_5 u({\bf x})\Big) \Big({\bar d}({\bf x}) \gamma_5 s_4({\bf x})\Big) .
\end{align}
Note that four different degenerate strange quarks $s^{(1)},\ldots,s^{(4)}$ have been introduced in such a way that the correlation functions (\ref{EQN731}) and (\ref{EQN732}) reproduce exactly the squared diagrams of interest shown in Figure~\ref{FIG731}. Since both $\mathcal{O}$ and $\tilde{\mathcal{O}}$ with flavor structure $[s^{(1)},s^{(2)},s^{(3)},s^{(4)}]$ generate quantum numbers $I(J^P) = 2(0^+)$ and strangeness $S^{(1,3)} = +1$ and $S^{(2,4)} = -1$ for the four strange flavors, the correlation function (\ref{EQN731}) will decay asymptotically according to $e^{-4 m_K t}$. Similarly, both $\mathcal{O}$ and $\tilde{\mathcal{O}}$ with flavor structures $[s^{(1)},s^{(1)},s^{(2)},s^{(2)}]$ and $[s^{(3)},s^{(3)},s^{(4)},s^{(4)}]$ will generate quantum numbers $I(J^P) = 2(0^+)$, but strangeness $S^{(1)} = S^{(2)} = S^{(3)} = S^{(4)} = 0$ for all four strange flavors and, hence, the correlation function (\ref{EQN732}) will decay asymptotically according to $e^{-2 m_\pi t}$. Consequently, one has

\begin{eqnarray}
\frac{\Delta C_{2 2}^{4 \times \textrm{connected}}(t)}{\Delta C_{2 2}^{2 \times \textrm{connected}}(t)} \ \ \propto \ \ e^{-(2 m_K - m_\pi) t} .
\end{eqnarray}
Inserting the masses $a m_K \approx 0.274$ and $a m_\pi \approx 0.137$ yields $\alpha = 0.411$ in agreement with the numerical findings from Figure~\ref{FIG201}.

In conclusion the relative exponential increase of the statistical errors of the $2 \times$ connected diagrams is not associated with the method of computation, but rather it is an intrinsic property of these diagrams.

% ********************
% ********************
% ********************

\subsection{\label{sec:4q4q}Four-quark -- two-meson correlation functions}

% ********************

\subsubsection{\label{SEC972}$4 \times$ connected $C_{2 5}$, $C_{2 6}$, $C_{3 5}$, $C_{3 6}$, $C_{4 5}$, $C_{4 6}$ ($2 \equiv K \bar{K}\textrm{, point}$, $3 \equiv \eta_s \pi\textrm{, point}$, $4 \equiv Q \bar{Q}$, $5 \equiv K \bar{K} \textrm{, 2part}$, $6 \equiv \eta_s \pi \textrm{, 2part}$)}

The correlation matrix elements $C_{2 5}$, $C_{2 6}$, $C_{3 5}$, $C_{3 6}$, $C_{4 5}$ and $C_{4 6}$ have different spin and color structures, but their $4 \times$ connected diagrams are identical with respect to spacetime and, hence, can be discussed together. Promising methods for numerical computation are the following.
\begin{figure}[H]
\hspace{-4.5cm}\includegraphics[scale=.7, page=7]{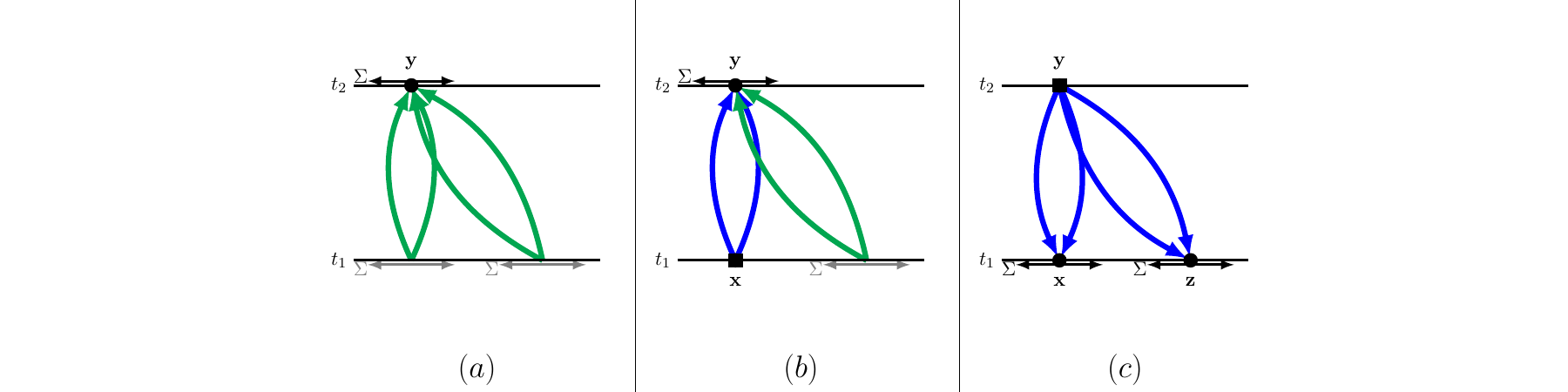}
\end{figure}
\begin{itemize}
\item $C_{2 5}$, $C_{3 5}$, $C_{4 5}$:
\begin{itemize}
\item[$(a)$] \textbf{2}$\times$\textbf{ one-end trick}: \\
2 $u/d$ inversions, 2 $s$ inversions
\item[$(b)$] \textbf{2 point-to-all-propagators}: \\
12 $u/d$ inversions, 12 $s$ inversions \\
\textbf{one-end trick}: \\
1 $u/d$ inversion, 1 $s$ inversion
\item[$(c)$] \textbf{4 point-to-all-propagators}: \\
12 $u/d$ inversions, 12 $s$ inversions
\end{itemize}
Note that there are two variants of method~$(b)$, when the operator $\mathcal{O}^6$ is considered, either applying the one-end trick to the $\pi$ meson (method $(b_1)$) or to the $\eta_s$ meson (method $(b_2)$).
\item $C_{2 6}$, $C_{3 6}$, $C_{4 6}$:
\begin{itemize}
\item[$(a)$] \textbf{2}$\times$\textbf{ one-end trick}: \\
1 $u/d$ inversions, 1 $s$ inversions
\item[$(b_1)$] \textbf{one-end trick}: \\
1 $u/d$ inversion \\
\textbf{2 point-to-all-propagators}: \\
12 $s$ inversions
\item[$(b_2)$] \textbf{2 point-to-all-propagators}: \\
12 $u/d$ inversions \\
\textbf{one-end trick}: \\
1 $s$ inversion
\item[$(c)$] \textbf{4 point-to-all-propagators}: \\
12 $u/d$ inversions, 12 $s$ inversions
\end{itemize}
\end{itemize}

\begin{figure}[p]
\begin{center}
\includegraphics[scale=0.39,page=07]{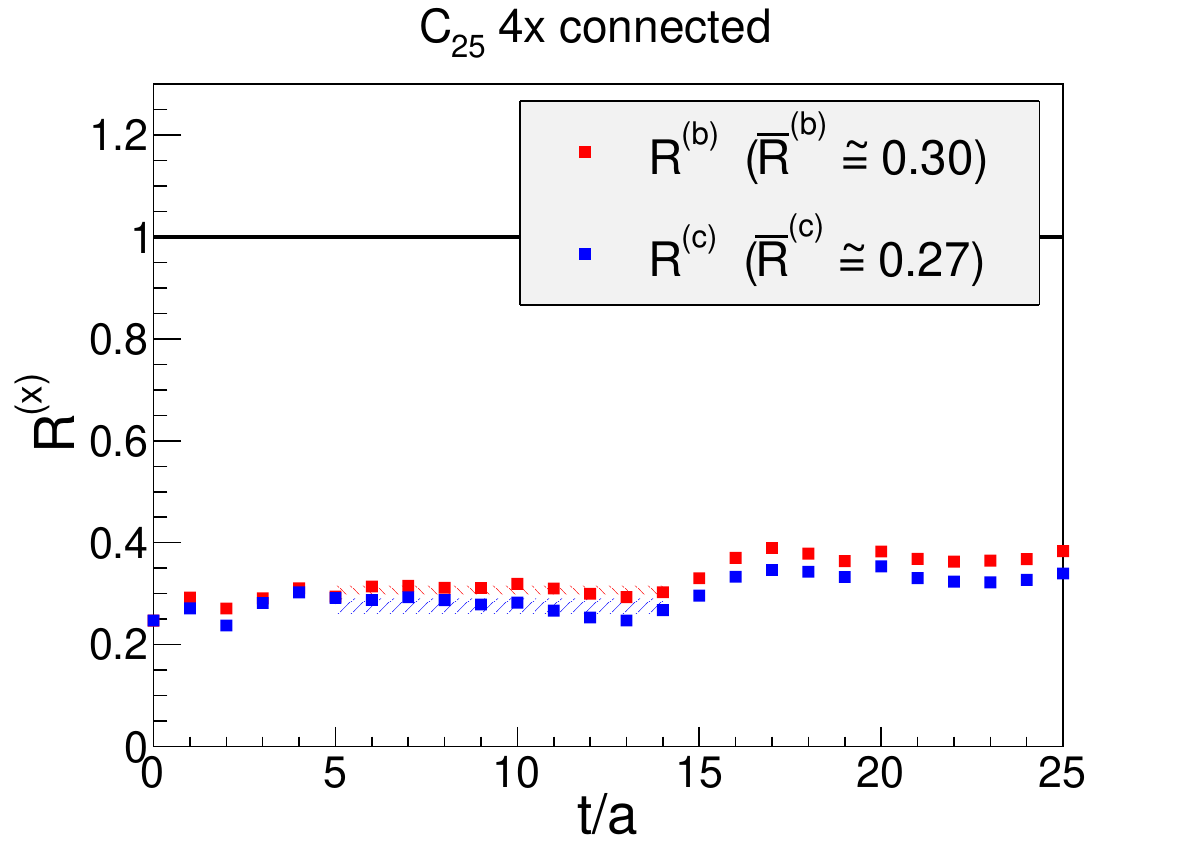}
\includegraphics[scale=0.39,page=08]{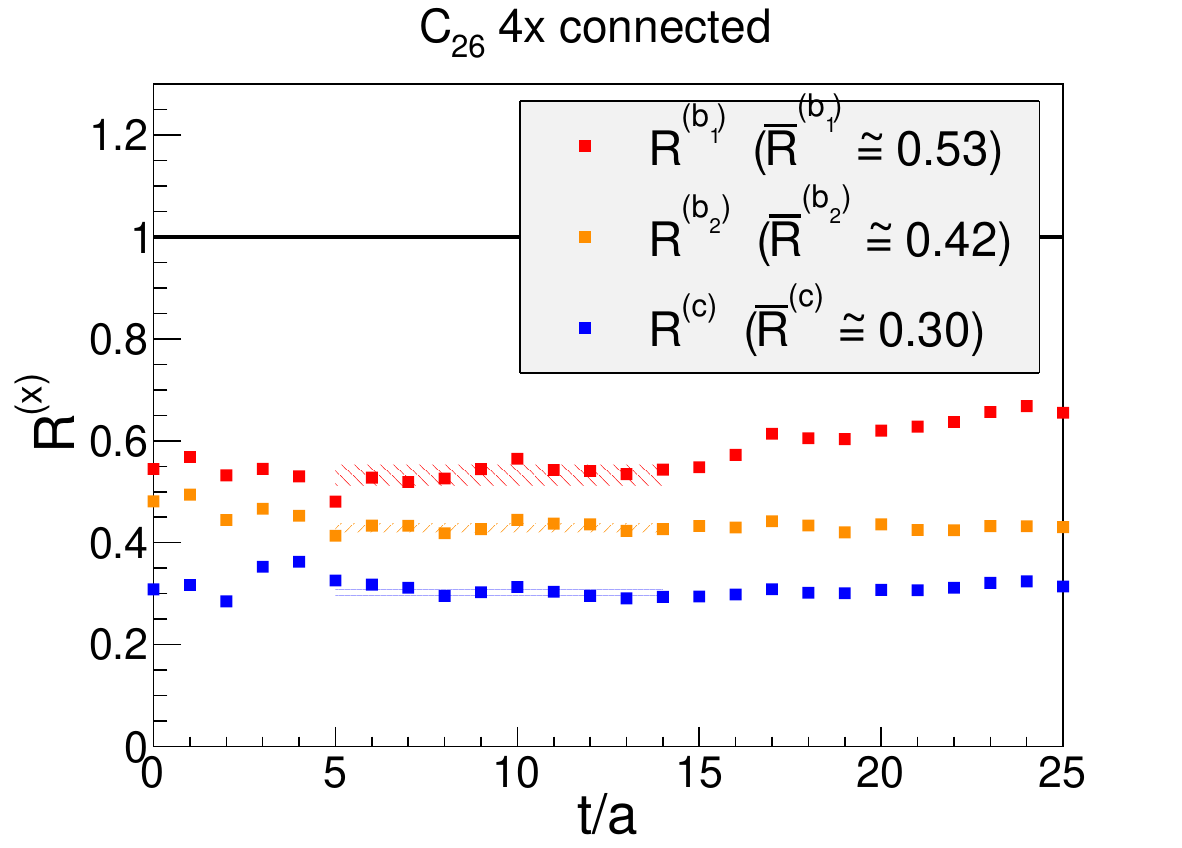}
\includegraphics[scale=0.39,page=09]{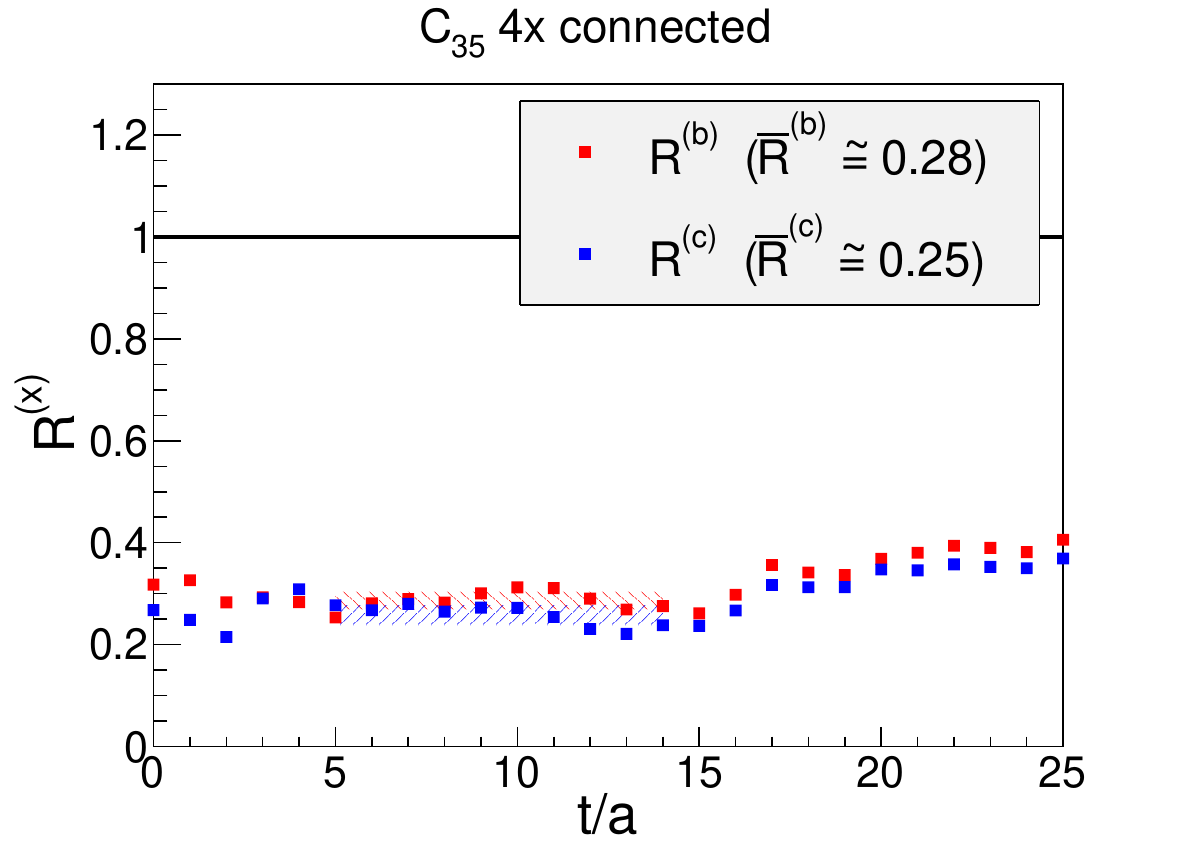}
\includegraphics[scale=0.39,page=10]{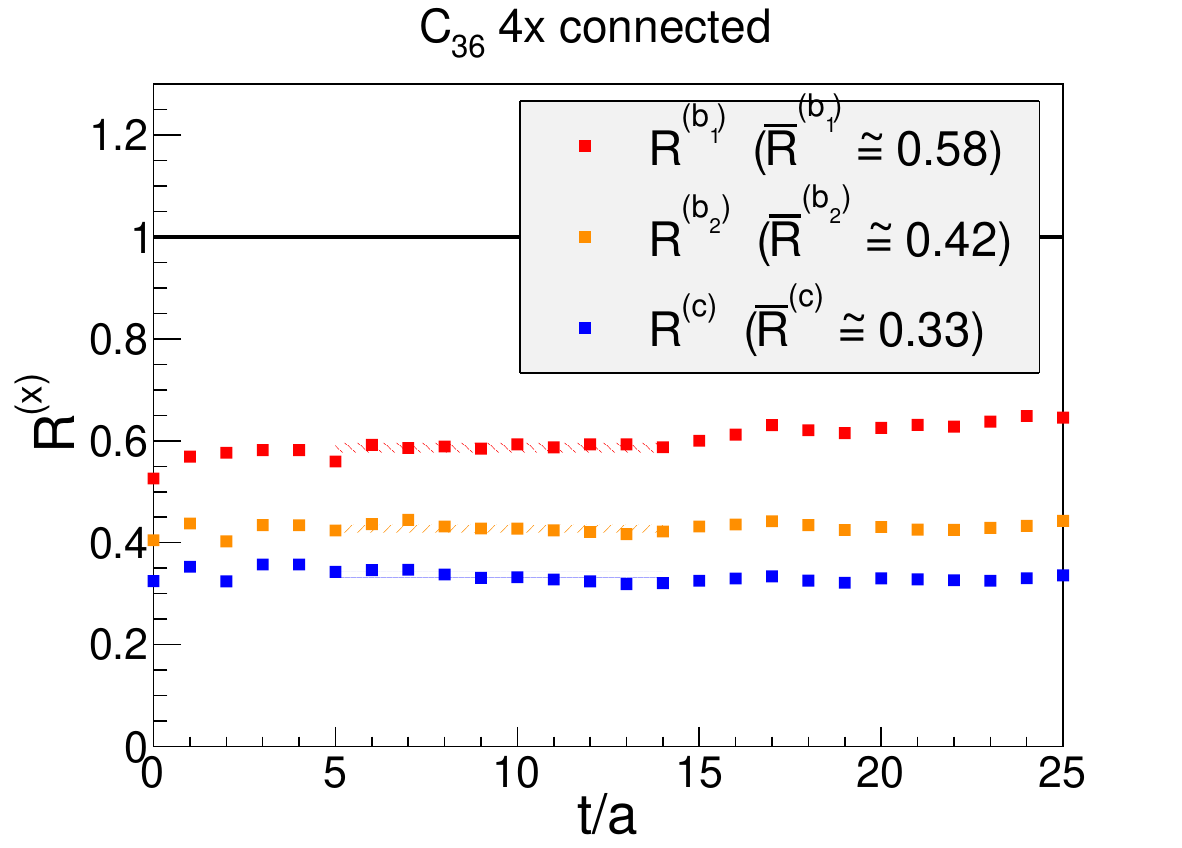}
\includegraphics[scale=0.39,page=11]{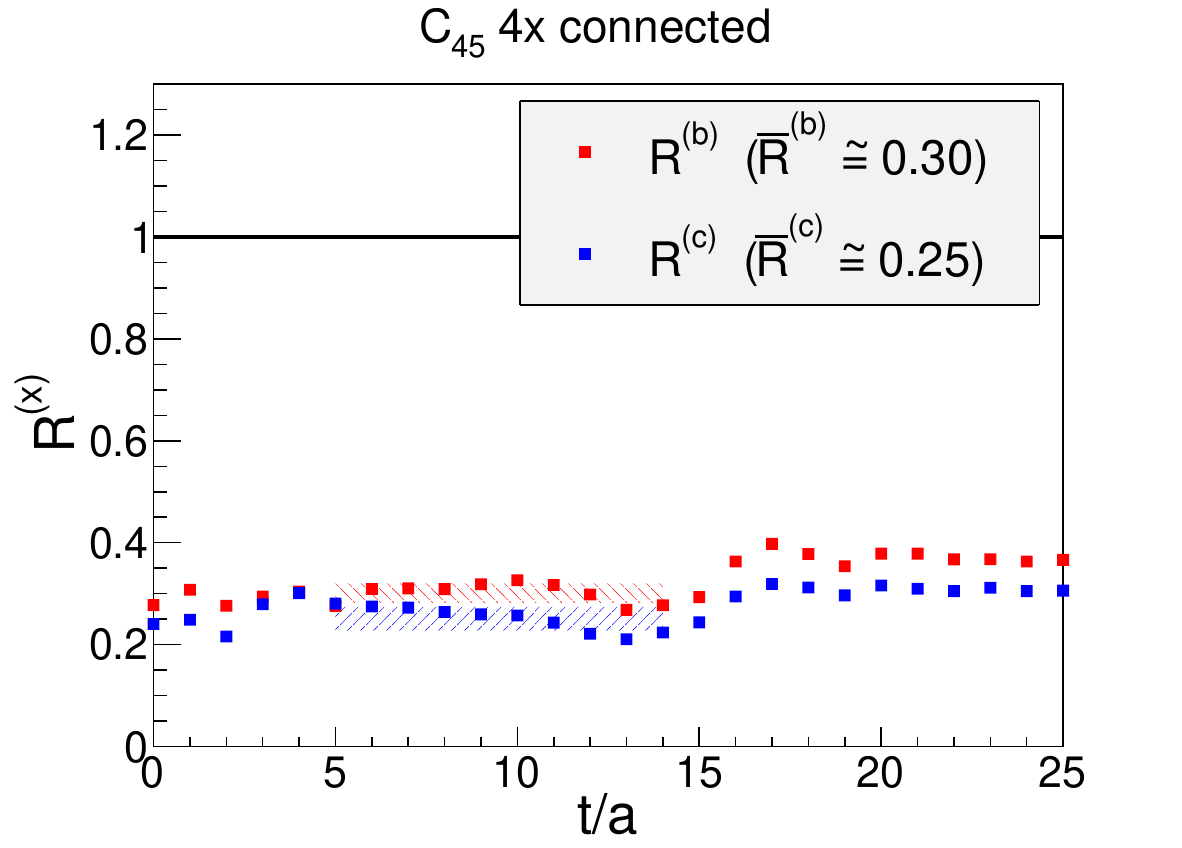}
\includegraphics[scale=0.39,page=12]{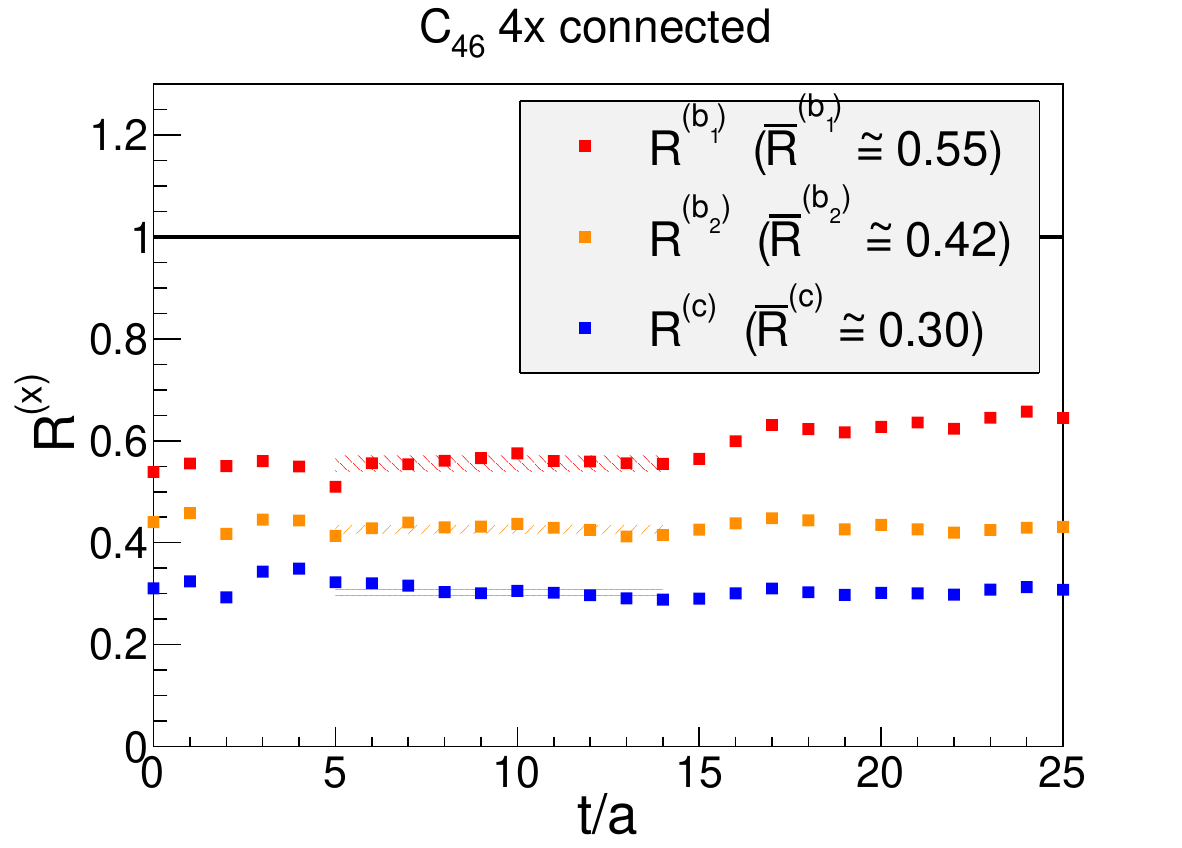}
\caption{\label{FIG759}Efficiency of different methods for $4 \times$ connected $C_{2 5}$, $C_{2 6}$, $C_{3 5}$, $C_{3 6}$, $C_{4 5}$, $C_{4 6}$.}
\end{center}
\end{figure}

A numerical comparison of these methods for all six cases is shown in Figure~\ref{FIG759}. The conclusions are similar as before. Method~$(a)$ is most efficient, despite a relatively large number of additional stochastic noise terms is introduced, because the diagrams are averaged over space. In contrast to that method~$(b)$ and method~$(c)$ consider only a single sample.

It is interesting to note that even though in method~$(b)$ stochastic noise terms are present, it performs better than method~$(c)$. It seems that omitting a sum over space where only two quark lines end (as in method~$(b)$), is more cost effective than omitting a sum over space where four quark lines end (as in method~$(c)$).

Finally, our findings are consistent with the expectation that method~$(b_1)$ is superior to method~$(b_2)$. One reason is that the larger number of 12 inversions needed for the point-to-all propagators requires much less computing time for the heavier $s$ quark than for the lighter $u/d$ quarks. Secondly, the one-end trick has been found to be more efficient than point-to-all-propagators when applied to light quarks \cite{Wiese2012}.

% ********************

\subsubsection{$2 \times$ connected $C_{2 5}$, $C_{3 5}$, $C_{4 5}$, ($2 \equiv K \bar{K}\textrm{, point}$, $3 \equiv \eta_s \pi\textrm{, point}$, $4 \equiv Q \bar{Q}$, $5 \equiv K \bar{K} \textrm{, 2part}$)}

Similarly to $C_{1 5}$, one of the $s$ quarks propagates within a timeslice and therefore using a sequential propagator is expected to be quite efficient. One can combine the sequential propagator either with the one-end trick or with point-to-all propagators.
\begin{figure}[H]
\hspace{-6.7cm}\includegraphics[scale=.7, page=8]{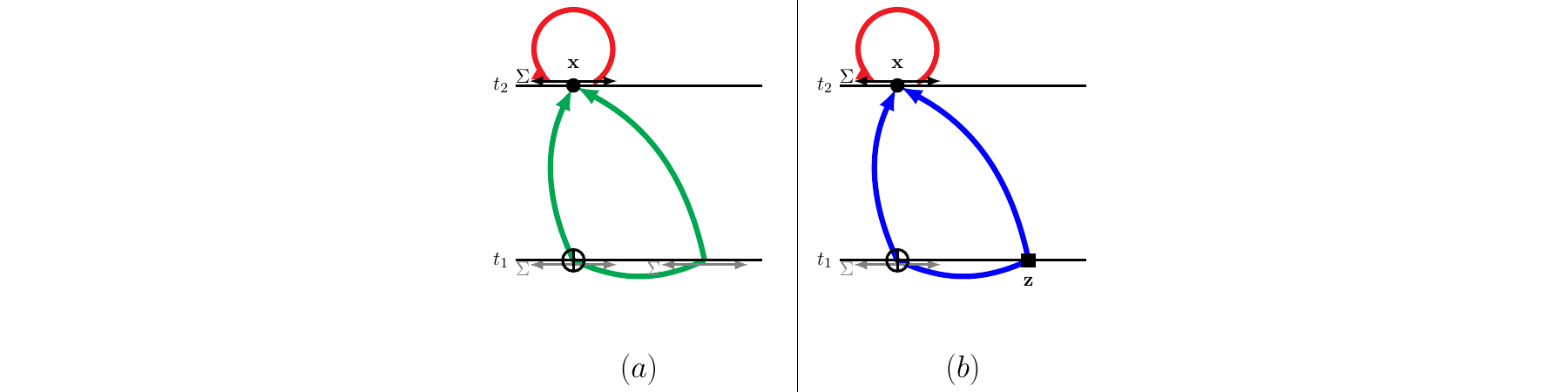}
\end{figure}
\begin{itemize}
\item[$(a)$] \textbf{one-end trick (with 1 sequential propagator)}: \\
2 $u/d$ inversions, 1 $s$ inversion \\
\textbf{stochastic timeslice-to-all propagator}: \\
$\# t$ $s$ inversions
\item[$(b)$] \textbf{2 point-to-all propagators (1 is a sequential propagator)}: \\
24 $u/d$ inversions, 12 $s$ inversions \\
\textbf{stochastic timeslice-to-all propagator}: \\
$\# t$ $s$ inversions
\end{itemize}

\begin{figure}[htb]
\begin{center}
\includegraphics[scale=0.39,page=13]{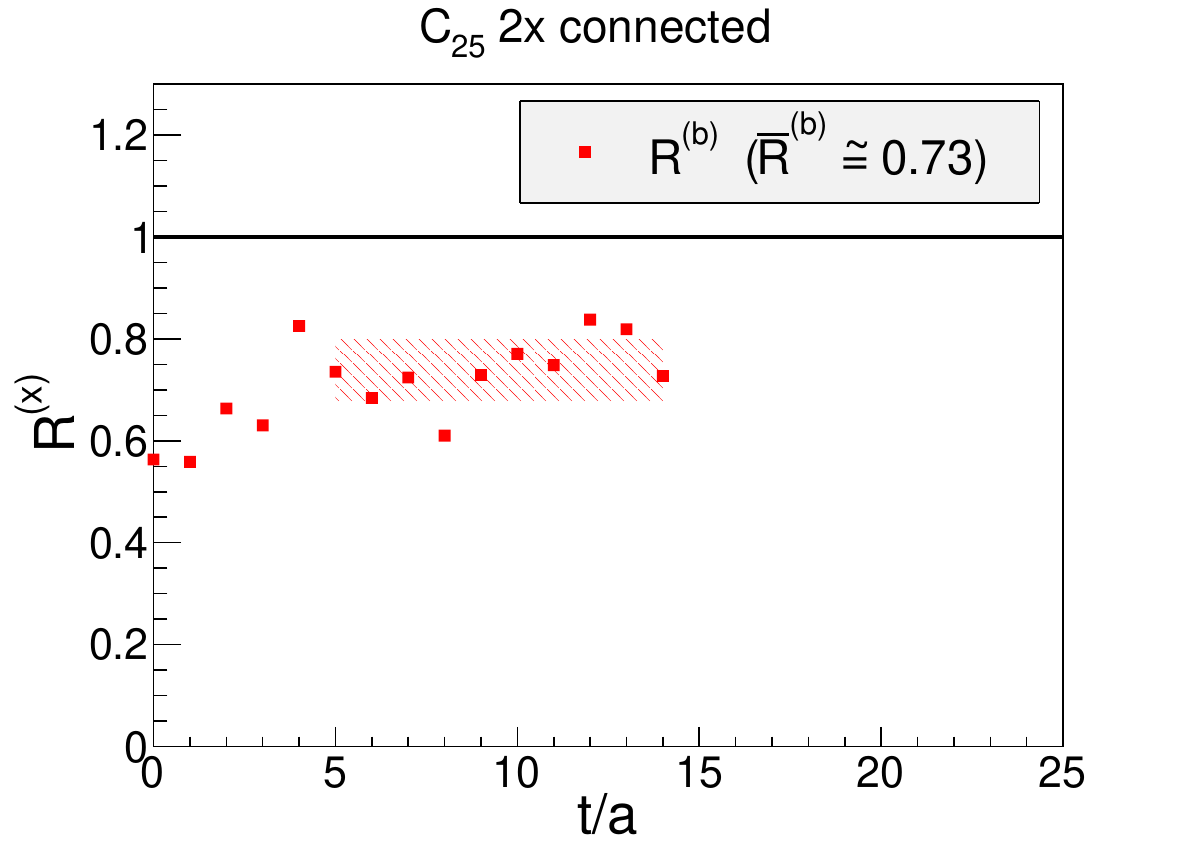}
\includegraphics[scale=0.39,page=14]{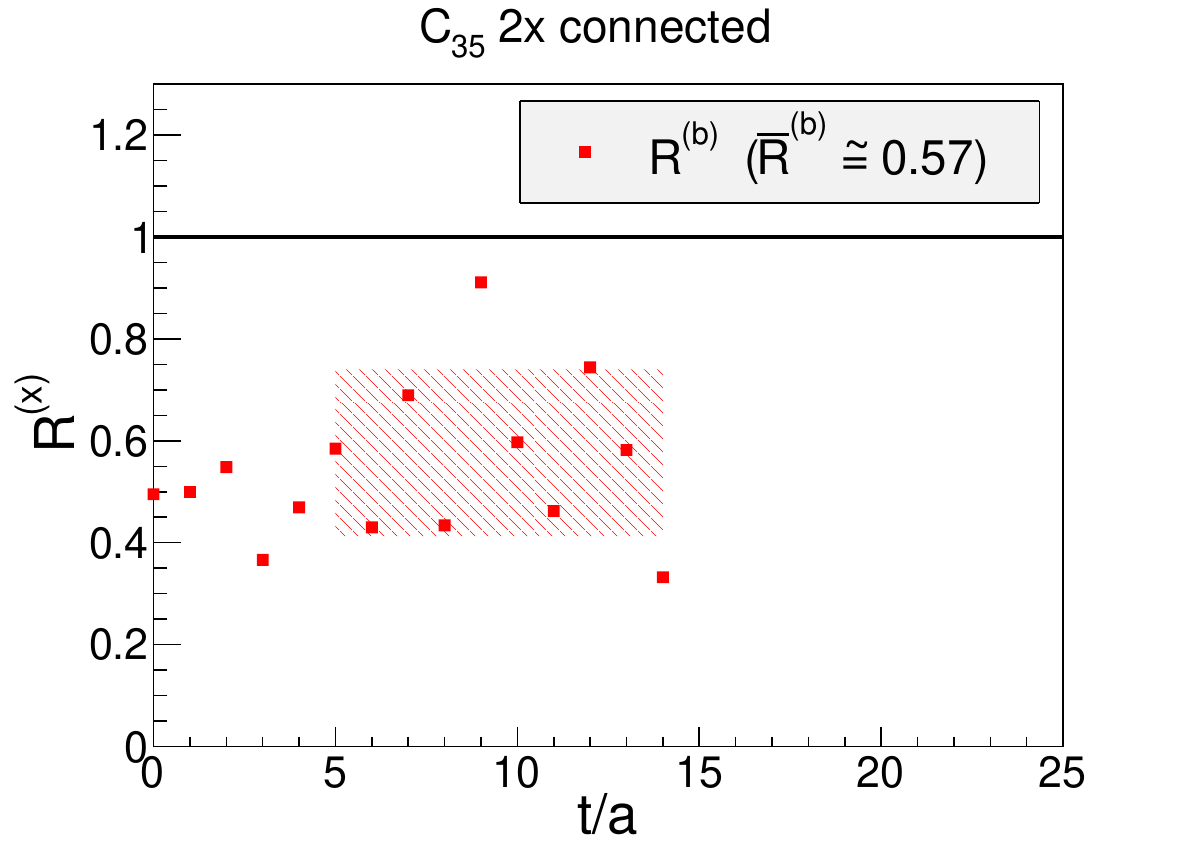}
\includegraphics[scale=0.39,page=15]{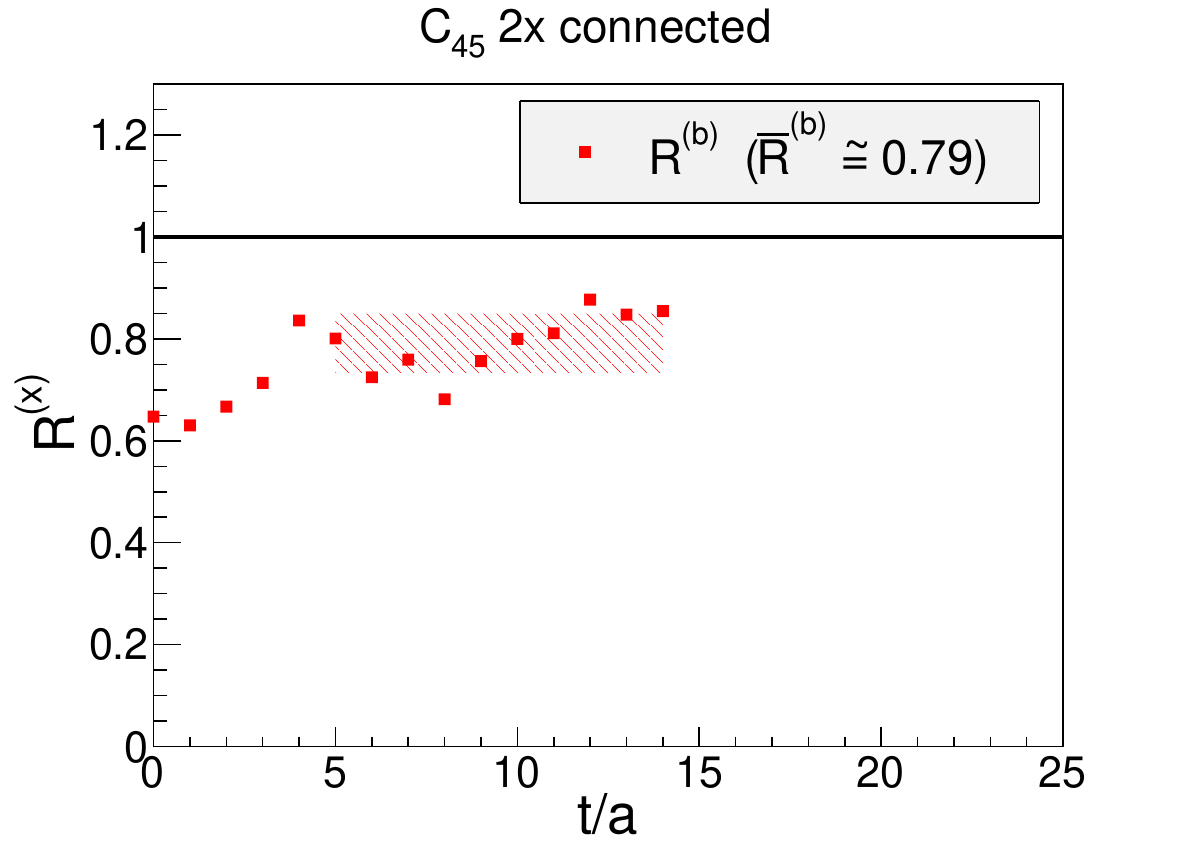}
\caption{\label{FIG760}Efficiency of different methods for $2 \times$ connected $C_{2 5}$, $C_{3 5}$, $C_{4 5}$.}
\end{center}
\end{figure}

Treating the light quarks with the one-end trick, which only introduces a moderate number of additional stochastic noise terms but averages the diagram over space, is more efficient than point-to-all propagators (cf.\ Figure~\ref{FIG760}). This is consistent with the results for matrix element $C_{1 5}$ discussed in section~\ref{SEC459}.

% ********************

\subsubsection{\label{SEC218}$2 \times$ connected $C_{2 6}$, $C_{3 6}$, $C_{4 6}$, ($2 \equiv K \bar{K}\textrm{, point}$, $3 \equiv \eta_s \pi\textrm{, point}$, $4 \equiv Q \bar{Q}$, $6 \equiv \eta_s \pi \textrm{, 2part}$)}

In this case, at least one of the closed quark loops should be computed using a stochastic timeslice-to-all propagator. Therefore, promising methods are the following.
\begin{figure}[H]
\hspace{-6.7cm}\includegraphics[scale=.7, page=9]{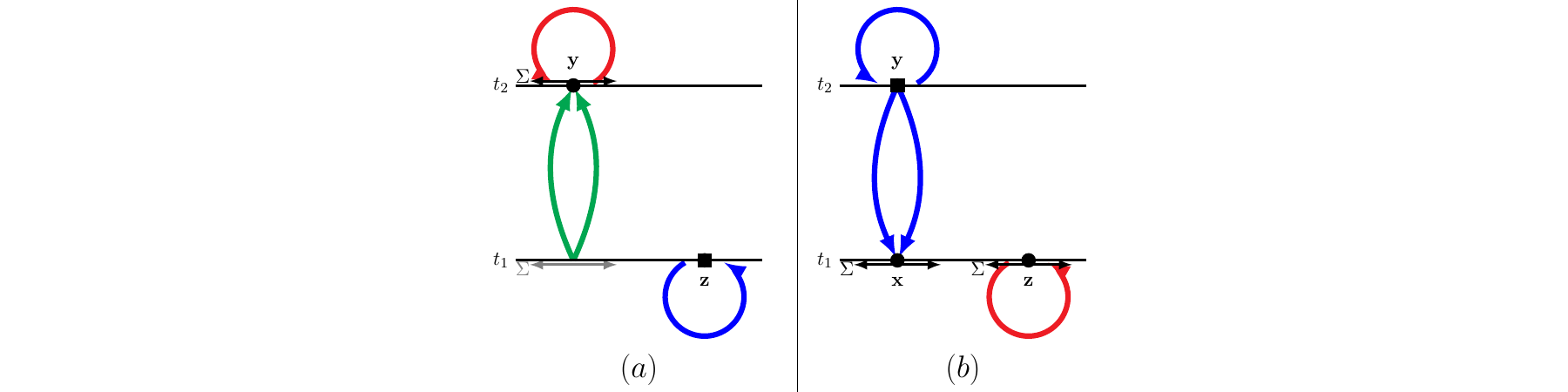}
\end{figure}
\begin{itemize}
\item[$(a)$] \textbf{one-end trick}: \\
1 $u/d$ inversion \\
\textbf{point-to-all propagator}: \\
12 $s$ inversions \\
\textbf{stochastic timeslice-to-all propagator}: \\
$\# t$ $s$ inversions
\item[$(b)$] \textbf{3 point-to-all-propagators}: \\
12 $u/d$ inversions, 12 $s$ inversions \\
\textbf{stochastic timeslice-to-all propagator}: \\
$\# t$ $s$ inversions
\end{itemize}

\begin{figure}[htb]
\begin{center}
\includegraphics[scale=0.39,page=16]{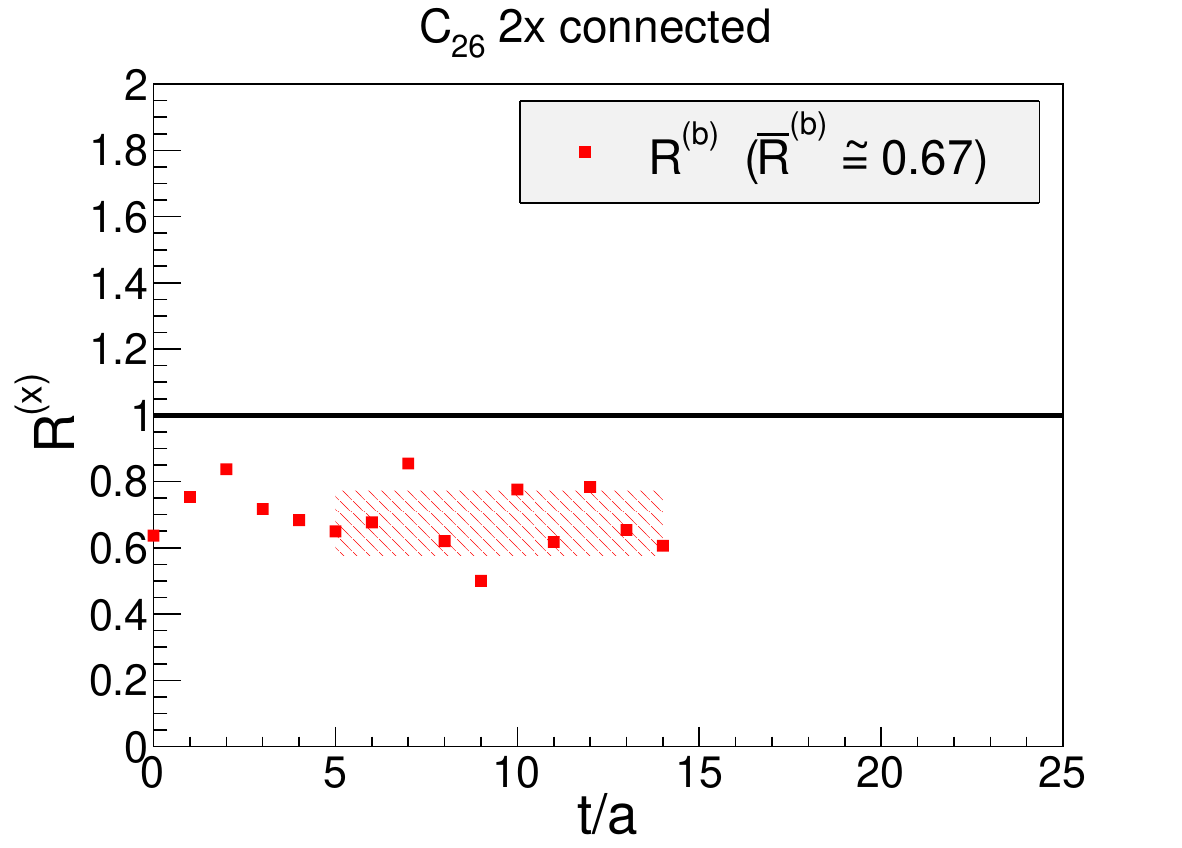}
\includegraphics[scale=0.39,page=17]{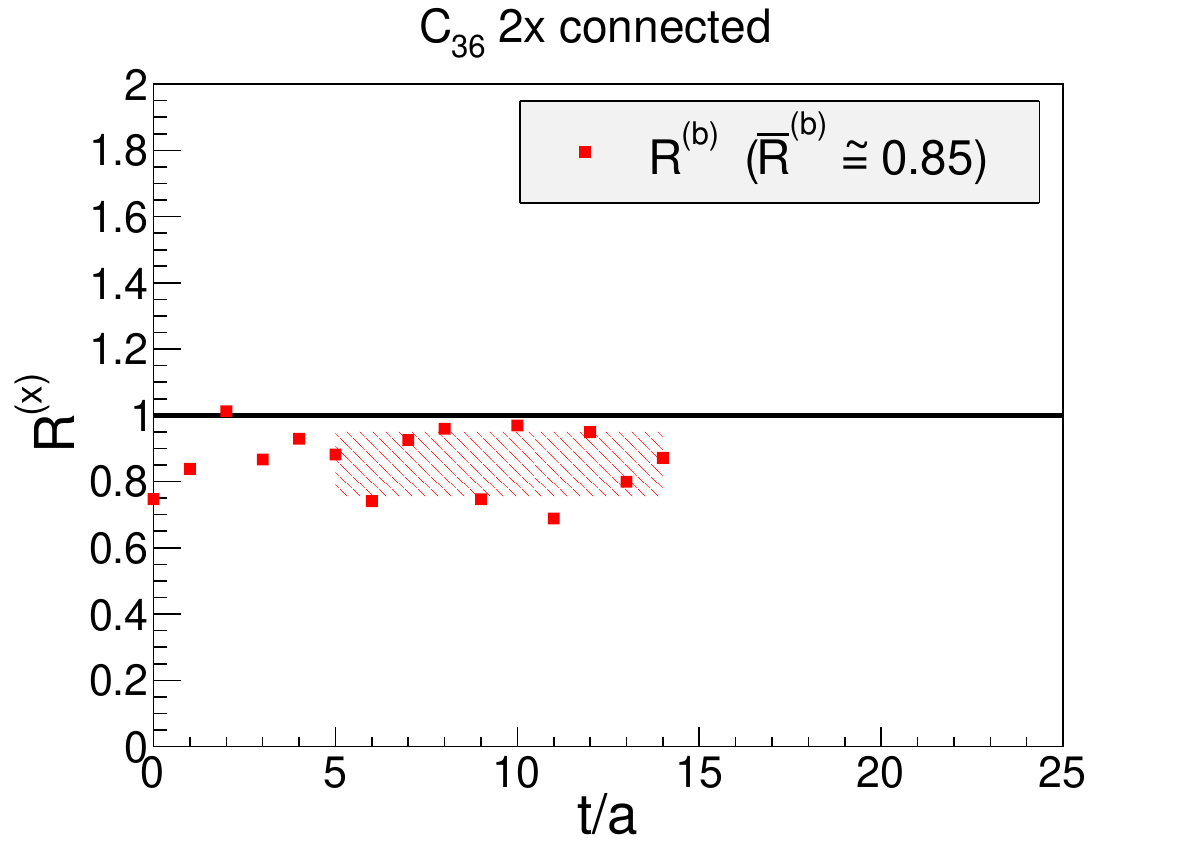}
\includegraphics[scale=0.39,page=18]{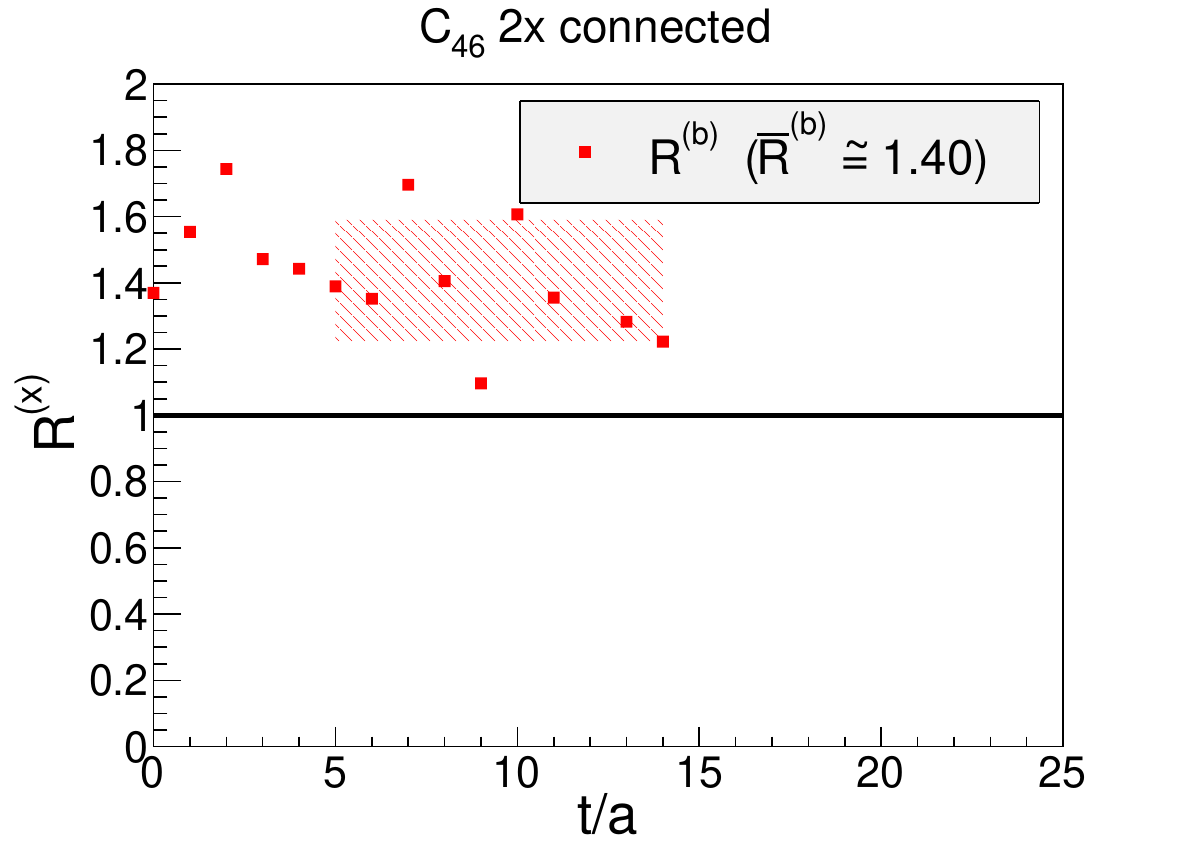}
\caption{\label{FIG761}Efficiency of different methods for $2 \times$ connected $C_{2 6}$, $C_{3 6}$, $C_{4 6}$.}
\end{center}
\end{figure}

A numerical comparison of these methods is shown in Figure~\ref{FIG761}. It is interesting to note that method~$(a)$ is superior to method~$(b)$ for $C_{2 6}$ and $C_{3 6}$, while the opposite is true for $C_{4 6}$. This is the only case we have investigated, where the spin and color structure of a diagram has a significant impact on the efficiency of the methods used for its computation.

Similarly to what has been observed for the $4 \times$ connected $C_{2 5}$, $C_{2 6}$, $C_{3 5}$, $C_{3 6}$, $C_{4 5}$ and $C_{4 6}$ (cf.\ section~\ref{SEC972}) method~(a) and method~(b) perform on a similar level, even though a significantly larger amount of noise terms is introduced by method~(a). This supports the conclusion that omitting a sum over space where only two quarks are present (as in method~$(a)$), 
is more cost effective than omitting a sum over space where four quarks are present (as in method~$(b)$).

% ********************

\subsubsection{Comparison of $4 \times$ connected and $2 \times$ connected diagrams}

A comparison of $4 \times$ connected and $2 \times$ connected diagrams for $C_{2 5}$, $C_{3 5}$, $C_{4 5}$, $C_{2 6}$, $C_{3 6}$ and $C_{4 6}$ is shown in Figure~\ref{FIG833}. Results are qualitatively very similar to those shown in section~\ref{SEC369} and the conclusions are essentially the same. Compared to their $4\times$ connected
counterparts, the statistical errors of the $2 \times$ connected diagrams increase exponentially as $e^{(2 m_K - m_\pi) t}$, when the temporal separation $t$ 
is increased. This independently of the methods used for the diagrams computation. %Hence, statistical errors will be dominated by the $2 \times$ connected diagrams and it is, thus, much more important to determine the optimal method of computation for the $2 \times$ connected diagrams than for the $4 \times$ connected diagrams.

\begin{figure}[htb]
\begin{center}
\includegraphics[scale=0.39,page=05]{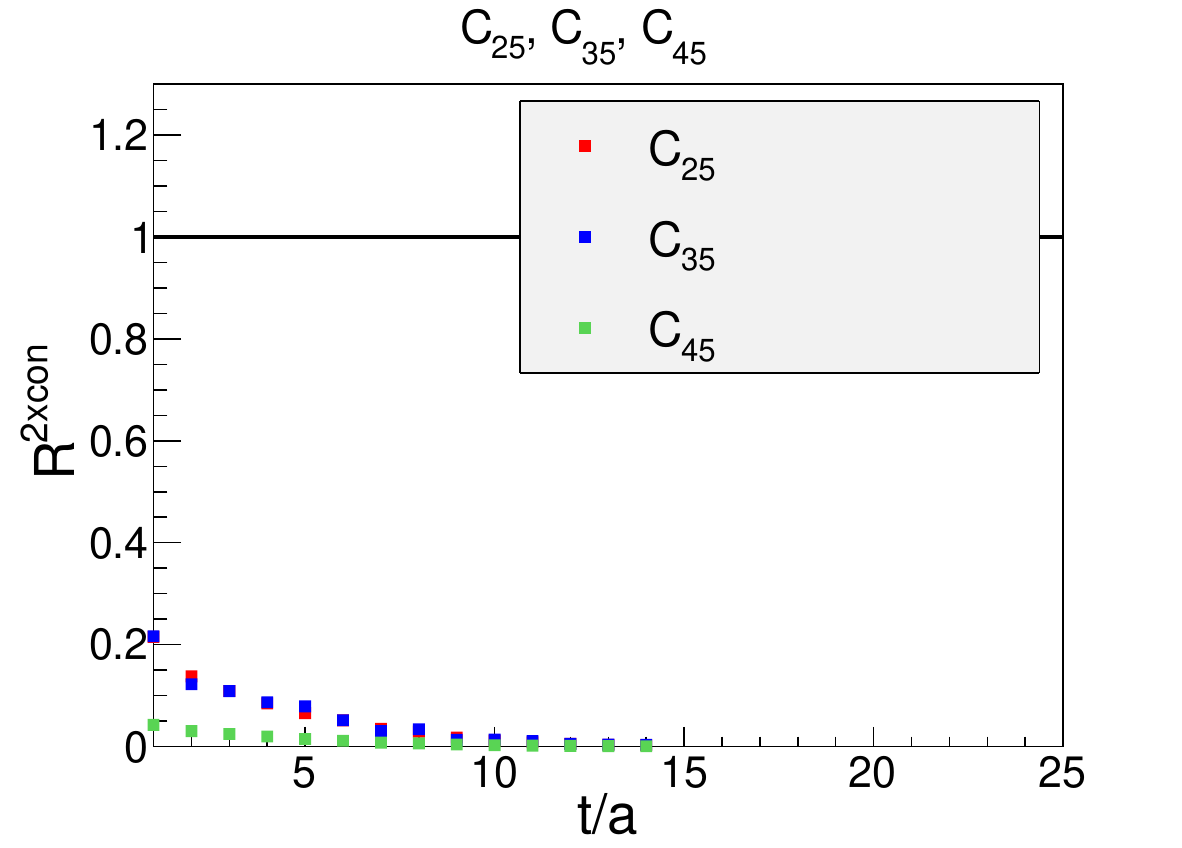}
\includegraphics[scale=0.39,page=07]{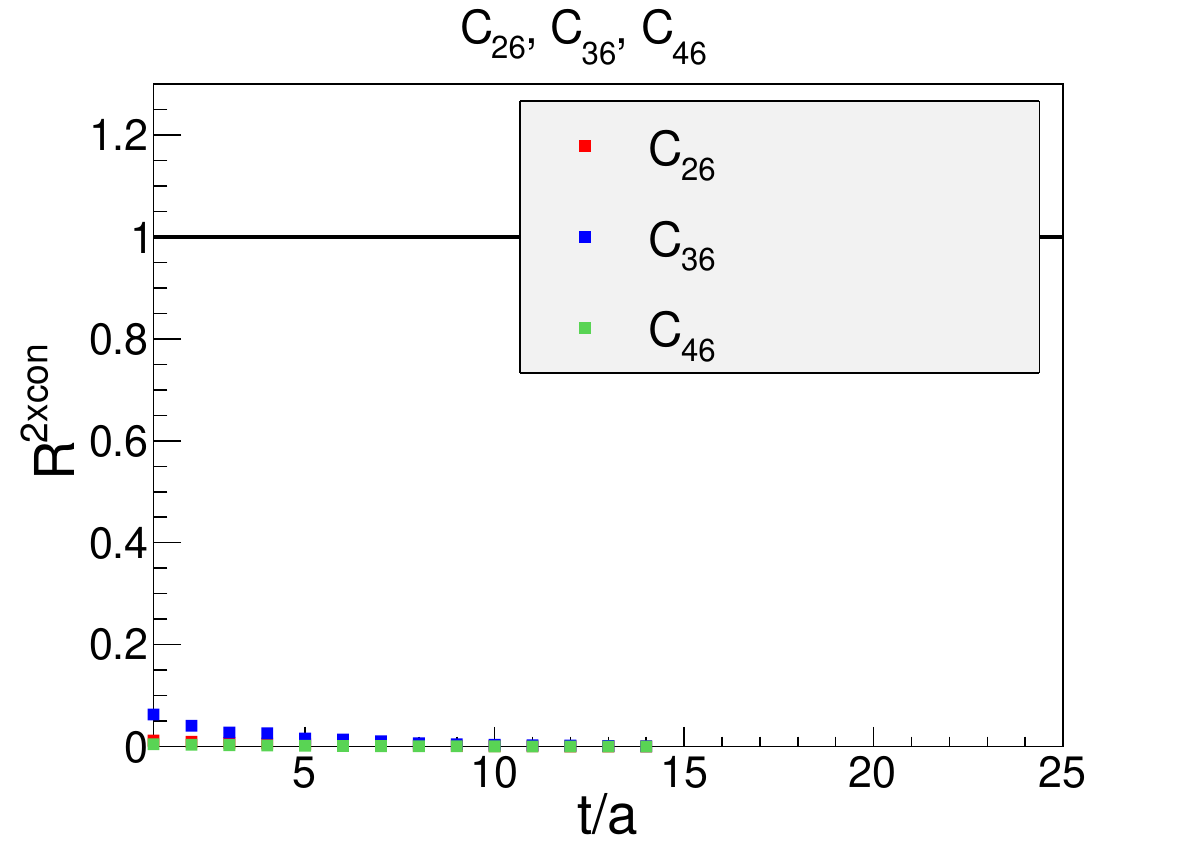}
\includegraphics[scale=0.39,page=06]{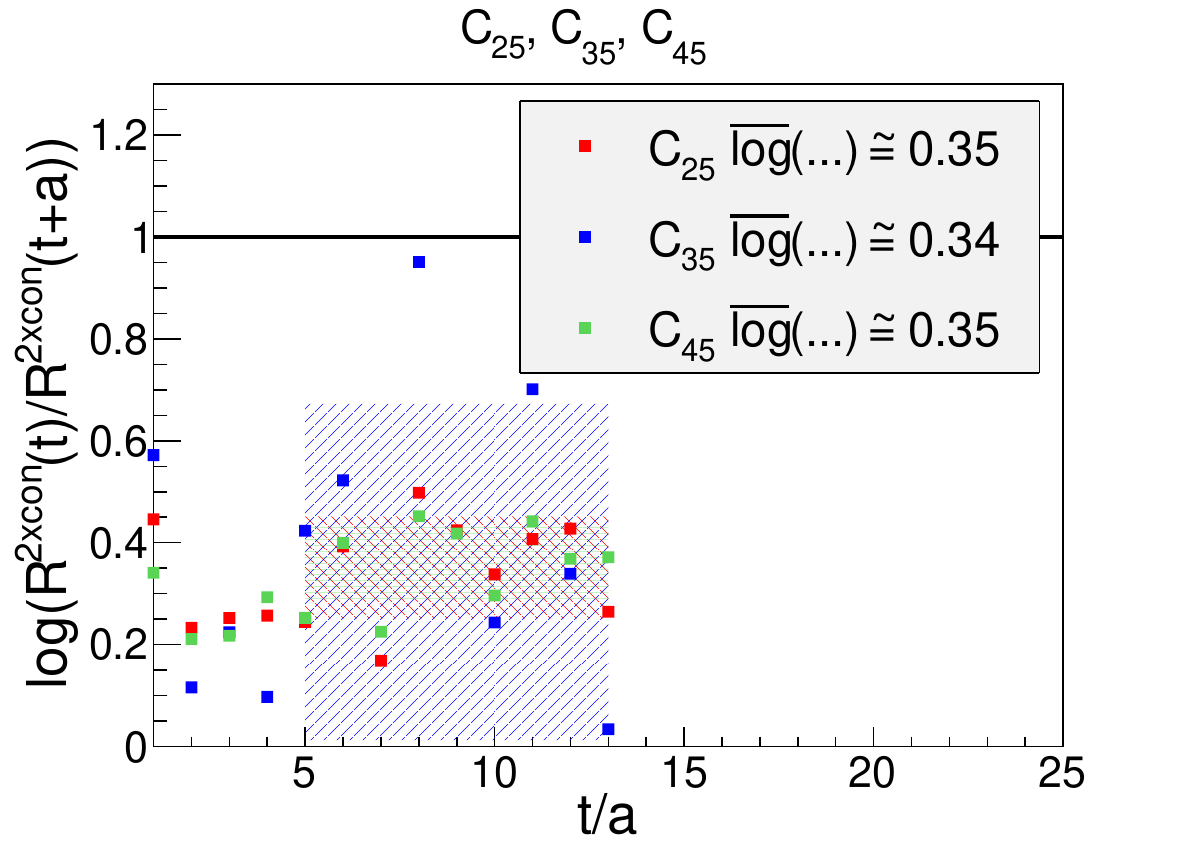}
\includegraphics[scale=0.39,page=08]{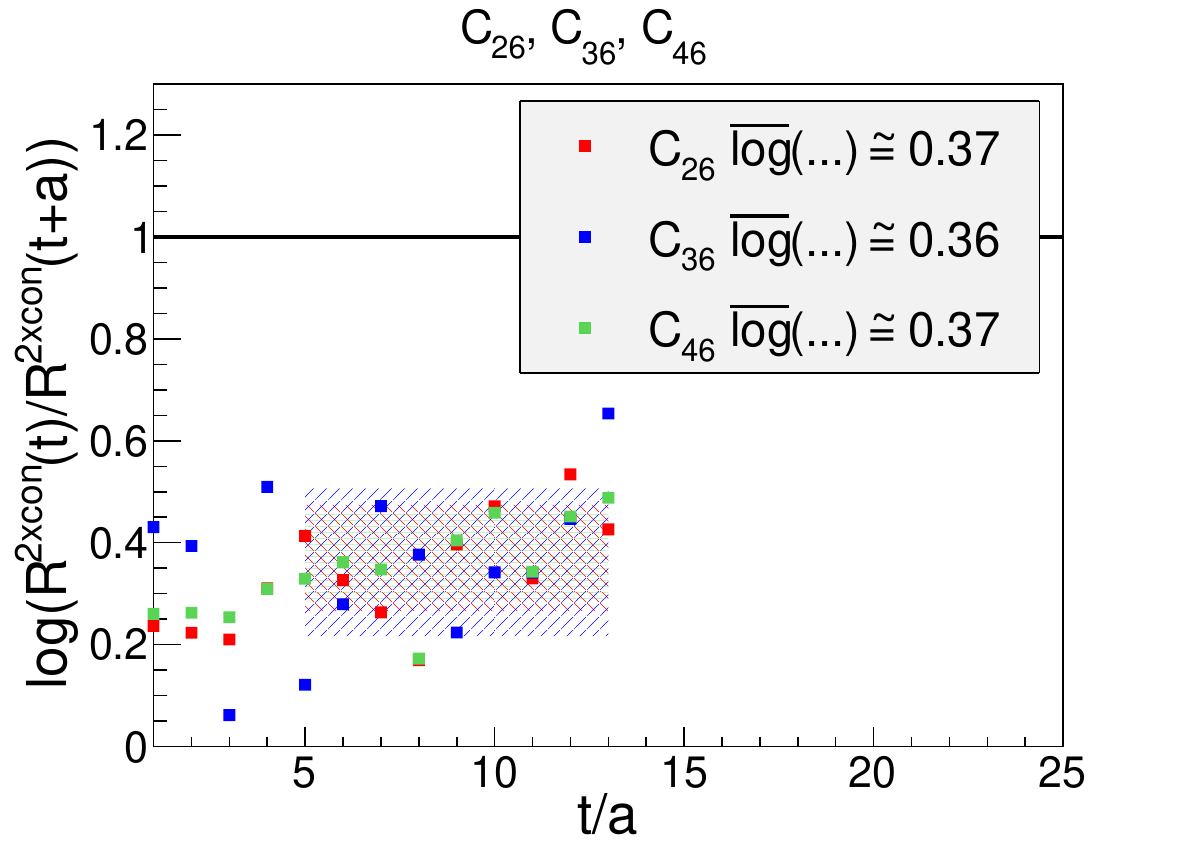}
\caption{\label{FIG833}Comparison of $4 \times$ connected and $2 \times$ connected for $C_{2 5}$, $C_{3 5}$, $C_{4 5}$, $C_{2 6}$, $C_{3 6}$, $C_{4 6}$. }
\end{center}
\end{figure}

In this respect, note that the quality ratio $R^{2 \times \textrm{con}}(t)$ is particularly small for $C_{2 6}$, $C_{3 6}$, $C_{4 6}$. This is so because even the most efficient method to compute the $2 \times$ connected diagrams (method~(a) in section~\ref{SEC218}) requires extensive use of stochastic techniques, while it still does not average the diagrams over space.

% ********************
% ********************
% ********************

\subsection{\label{sec:2m2m}Two-meson -- two-meson correlation functions}

% ********************

\subsubsection{\label{SEC973}$4 \times$ connected $C_{5 5}$, $C_{6 6}$ ($5 \equiv K \bar{K} \textrm{, 2part}$, $6 \equiv \eta_s \pi \textrm{, 2part}$)}

The two disconnected parts of these diagrams are in fact identical to $C_{1 1}$. Therefore, the methods discussed in section~\ref{SEC489} are applied to both parts.
\begin{figure}[H]
\hspace{-6.7cm}\includegraphics[scale=.7, page=10]{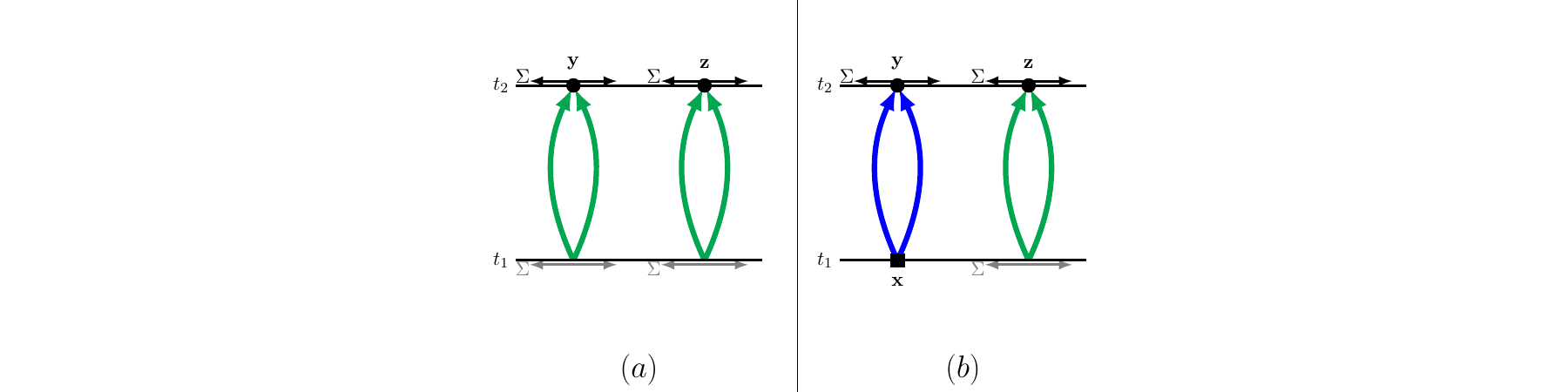}
\end{figure}
\begin{itemize}
\item $C_{5 5}$:
\begin{itemize}
\item[$(a)$] \textbf{2}$\times$\textbf{ one-end trick}: \\
2 $u/d$ inversions, 2 $s$ inversions
\item[$(b)$] \textbf{2 point-to-all-propagators}: \\
12 $u/d$ inversions, 12 $s$ inversions \\
\textbf{one-end trick}: \\
1 $u/d$ inversion, 1 $s$ inversion
\end{itemize}
Similar to $C_{2 6}$, $C_{3 6}$ and $C_{4 6}$ there are two variants of method~$(b)$ for $C_{6 6}$, either applying the one-end trick to the $\pi$ meson (method (b$_1$)) or to the $\eta_s$ meson (method (b$_2$)).
\item $C_{6 6}$:
\begin{itemize}
\item[$(a)$] \textbf{2}$\times$\textbf{ one-end trick}: \\
1 $u/d$ inversions, 1 $s$ inversions
\item[$(b_1)$] \textbf{one-end trick}: \\
1 $u/d$ inversion \\
\textbf{2 point-to-all-propagators}: \\
12 $s$ inversions
\item[$(b_2)$] \textbf{2 point-to-all-propagators}: \\
12 $u/d$ inversions \\
\textbf{one-end trick}: \\
1 $s$ inversion
\end{itemize}
\end{itemize}

\begin{figure}[htb]
\begin{center}
\includegraphics[scale=0.39,page=19]{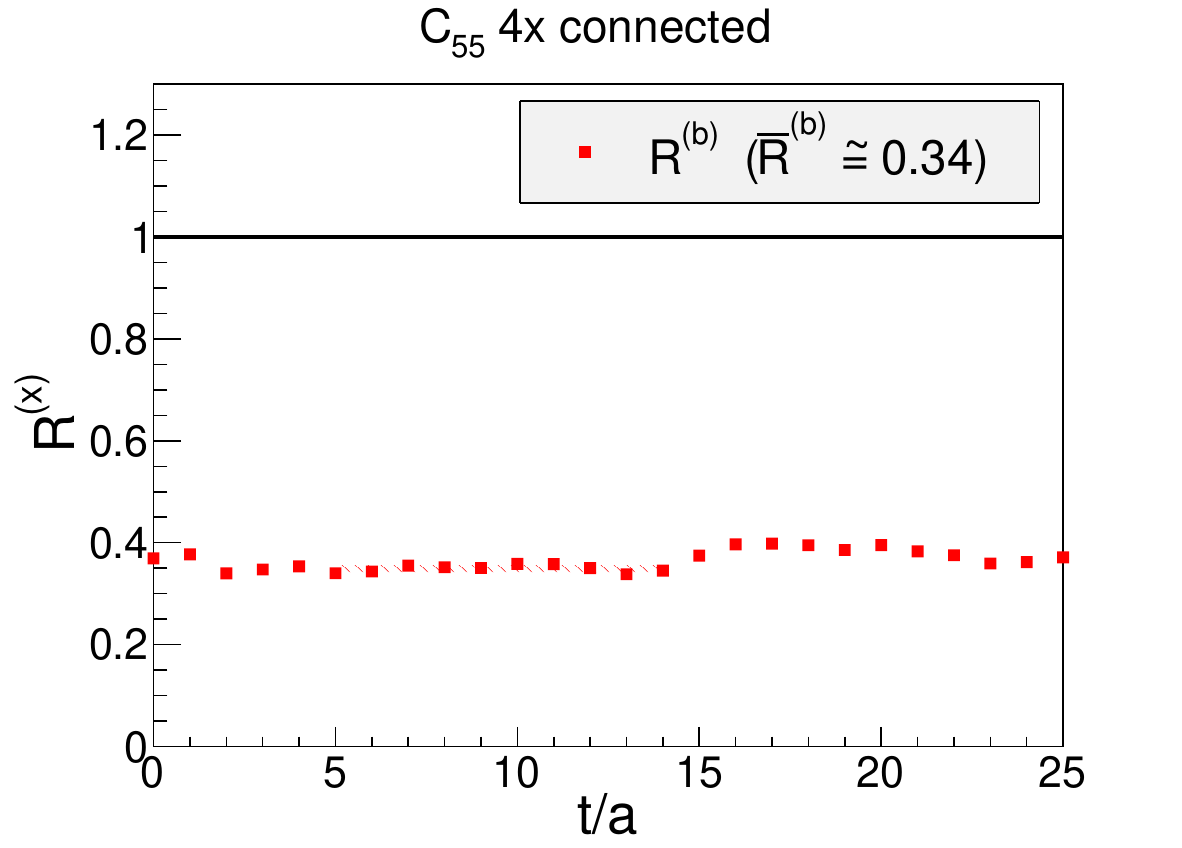}
\includegraphics[scale=0.39,page=20]{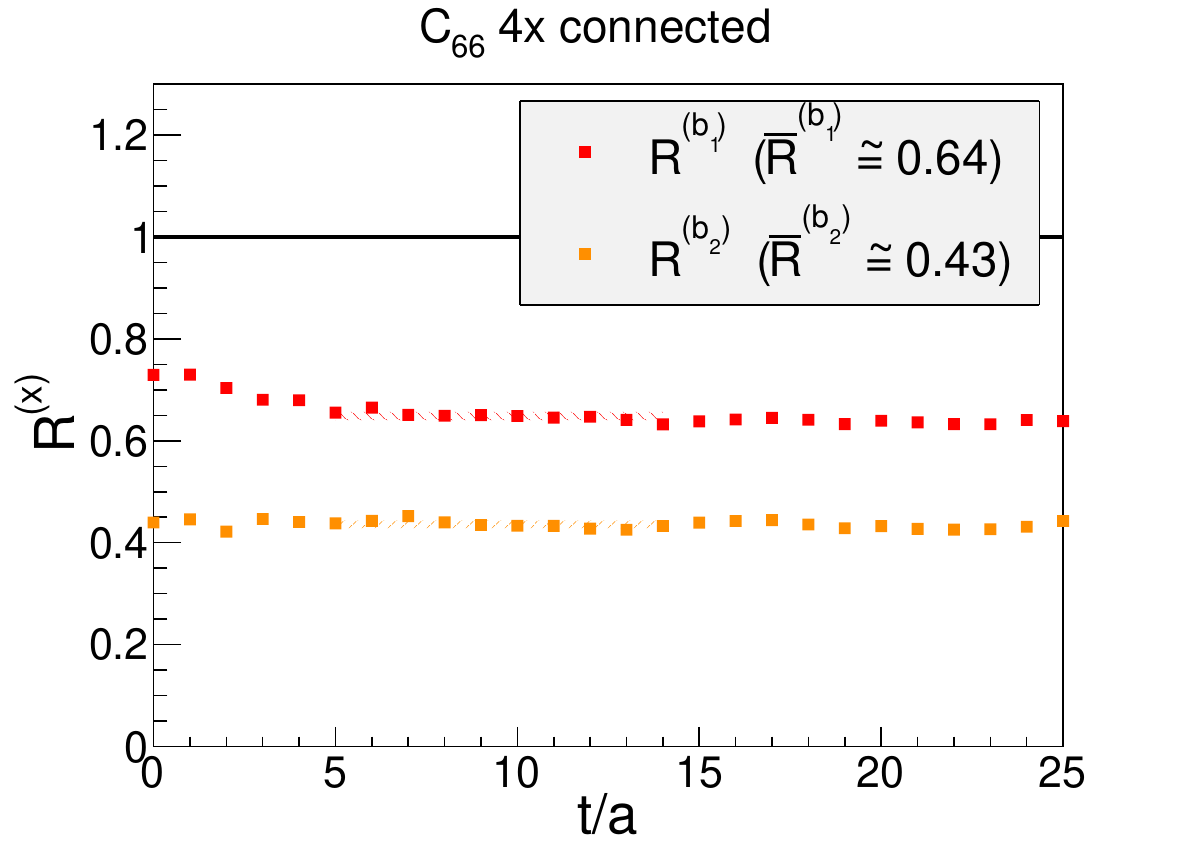}
\caption{\label{FIG762}Efficiency of different methods for $4 \times$ connected $C_{5 5}$, $C_{6 6}$.}
\end{center}
\end{figure}

A numerical comparison of these methods is shown in Figure~\ref{FIG762}. The conclusions are essentially the same as in section~\ref{SEC972} for $C_{2 5}$, $C_{2 6}$, $C_{3 5}$, $C_{3 6}$, $C_{4 5}$ and $C_{4 6}$. Method~$(a)$ is more efficient, because the diagrams are averaged over space, while method~$(b)$ considers only a single sample. Method~$(b_1)$ is superior to method~$(b_2)$, because it is cheaper to perform the larger number of 12 inversions for the heavier $s$ quark than for light $u/d$ quarks and the one-end trick has been found to be more efficient than point-to-all-propagators when applied to light quarks \cite{Wiese2012}.

% ********************

\subsubsection{$2 \times$ connected $C_{5 5}$ ($5 \equiv K \bar{K} \textrm{, 2part}$)}

Since both the $s$ quark at $t_1$ and the $s$ quark at $t_2$ propagate within the timeslice, the diagram can be computed rather efficiently using two sequential propagators combined with the one-end trick.
\begin{figure}[H]
\hspace{-8.9cm}\includegraphics[scale=.7, page=11]{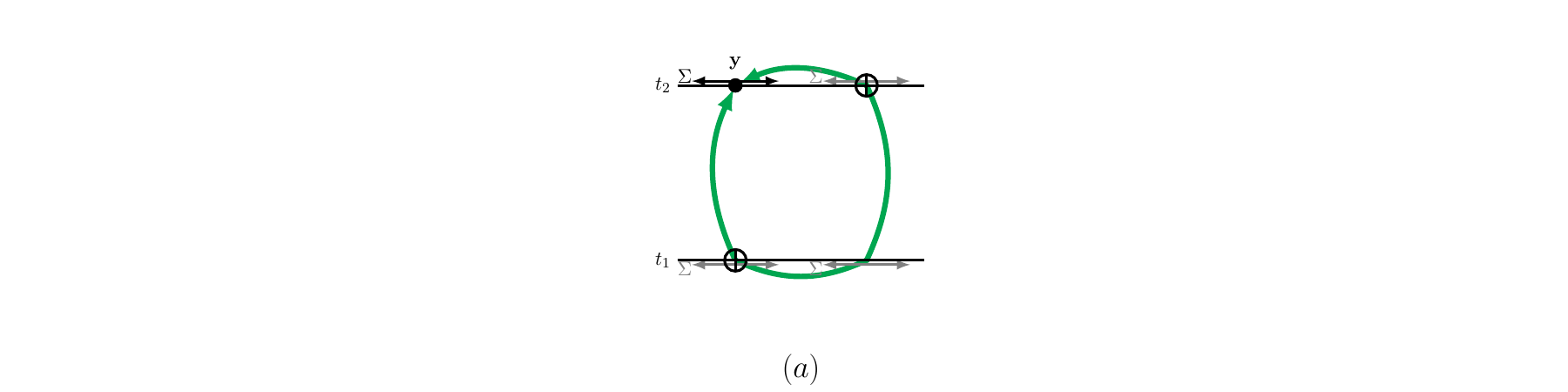}
\end{figure}
\begin{itemize}
\item[$(a)$] \textbf{2}$\times$\textbf{ one-end trick (with 2 sequential propagators)}: \\
2 $u/d$ inversions, $(\# t + 1)$ $s$ inversions
\end{itemize}

% ********************

\subsubsection{$2 \times$ connected $C_{6 6}$ ($6 \equiv \eta_s \pi \textrm{, 2part}$)}

At least one of the closed quark loops needs to be computed using a stochastic timeslice-to-all propagator. Therefore,
there is only a single possibility to compute this type of diagram efficiently.
\begin{figure}[H]
\hspace{-8.9cm}\includegraphics[scale=.7, page=14]{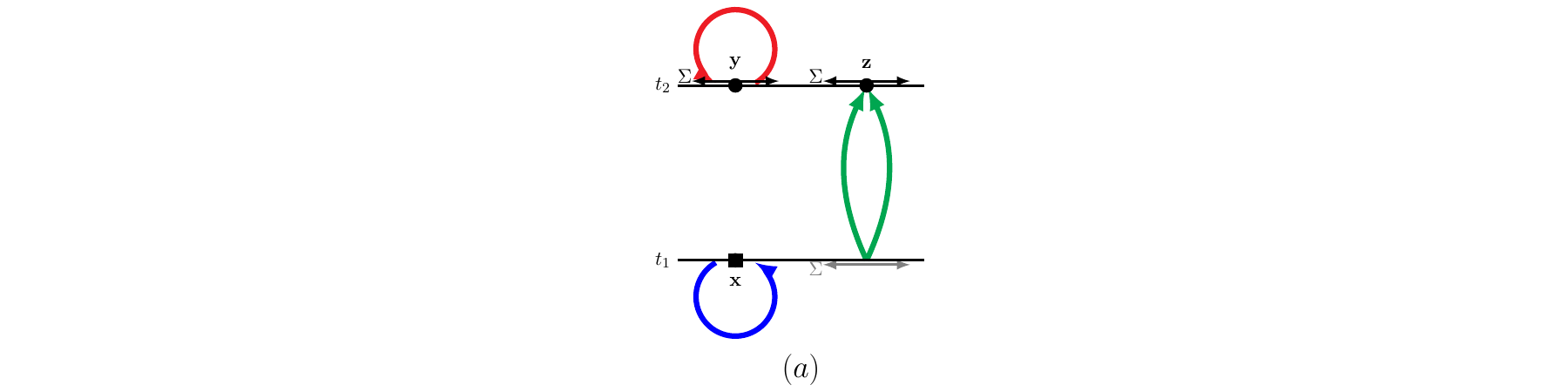}
\end{figure}
\begin{itemize}
\item[$(a)$] \textbf{one-end trick}: \\
1 $u/d$ inversions \\
\textbf{point-to-all propagator}: \\
12 $s$ inversions \\
\textbf{stochastic timeslice-to-all propagator}: \\
$\# t$ $s$ inversions
\end{itemize}

% ********************

\subsubsection{$4 \times$ connected $C_{5 6}$ ($5 \equiv K \bar{K} \textrm{, 2part}$, $6 \equiv \eta_s \pi \textrm{, 2part}$)}

Similar to $4 \times$ connected $C_{5 5}$ and $C_{6 6}$ there are two promising methods.
\begin{figure}[H]
\hspace{-2.2cm}\includegraphics[scale=.7, page=12]{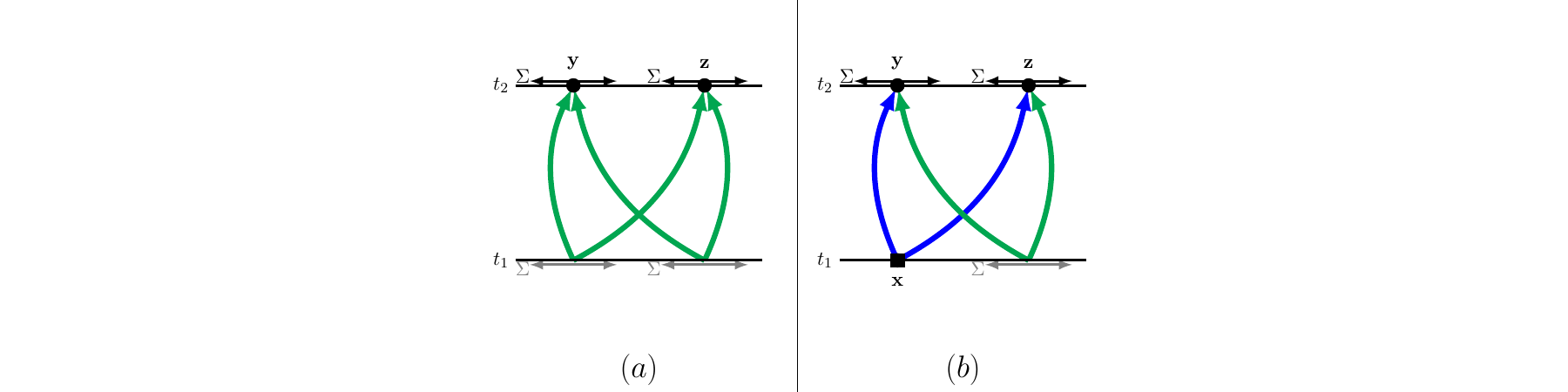}
\end{figure}
\begin{itemize}
\item[$(a)$] \textbf{2}$\times$\textbf{ one-end trick}: \\
1 $u/d$ inversions, 1 $s$ inversions
\item[$(b_1)$] \textbf{one-end trick}: \\
1 $u/d$ inversion \\
\textbf{2 point-to-all-propagators}: \\
12 $s$ inversions
\item[$(b_2)$] \textbf{2 point-to-all-propagators}: \\
12 $u/d$ inversions \\
\textbf{one-end trick}: \\
1 $s$ inversion
\end{itemize}

\begin{figure}[htb]
\begin{center}
\includegraphics[scale=0.39,page=21]{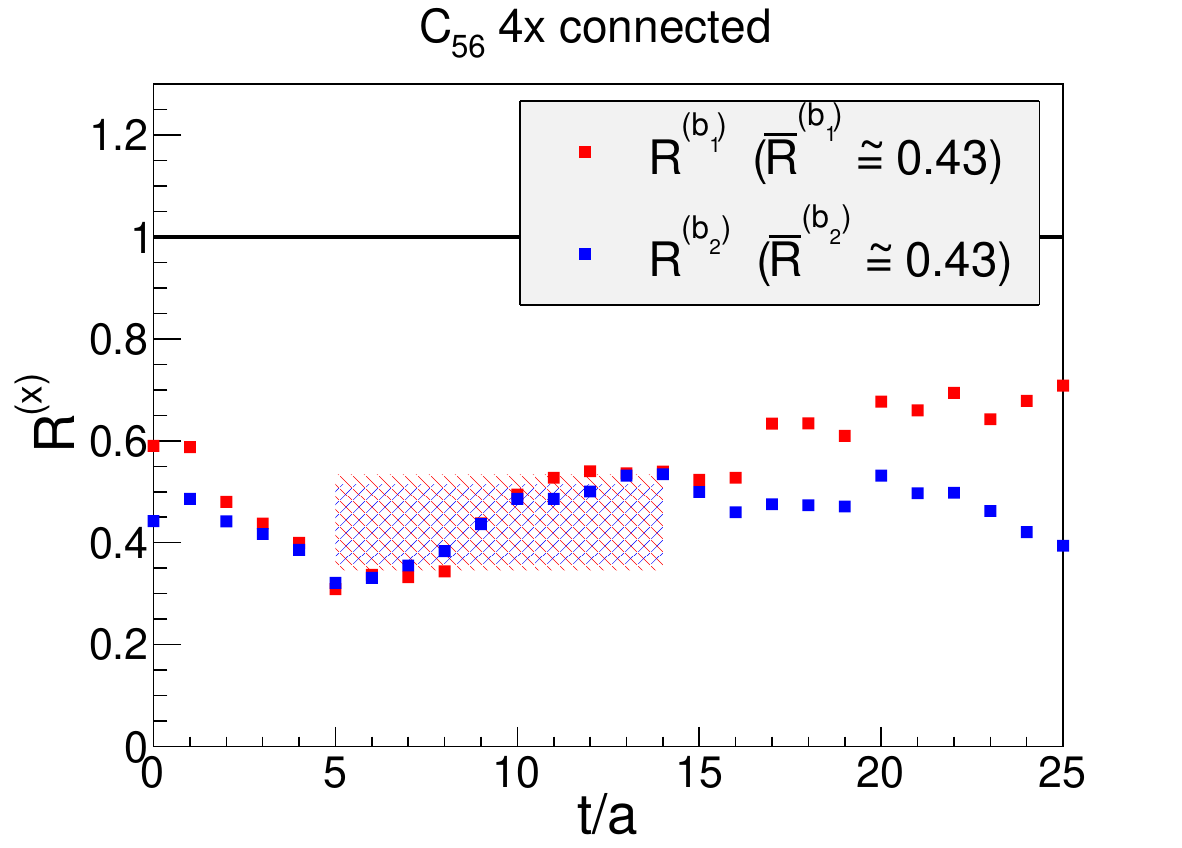}
\caption{\label{FIG764}Efficiency of different methods for $4 \times$ connected $C_{5 6}$.}
\end{center}
\end{figure}

A numerical comparison of these methods is shown in Figure~\ref{FIG764}. The conclusions are essentially the same as for the diagrams discussed in section~\ref{SEC972} and section~\ref{SEC973}. Method~$(a)$ is more efficient, because the diagrams are averaged over space, while method~$(b)$ considers only a single sample.

% ********************

\subsubsection{\label{SEC370}$2 \times$ connected $C_{5 6}$ ($5 \equiv K \bar{K} \textrm{, 2part}$, $6 \equiv \eta_s \pi \textrm{, 2part}$)}

Again the triangular loop can be computed efficiently with a sequential propagator. There are several promising methods, which combine that propagator with point-to-all propagators, stochastic timeslice-to-all propagators and the one-end trick.
\begin{figure}[H]
\hspace{-4.4cm}\includegraphics[scale=.7, page=13]{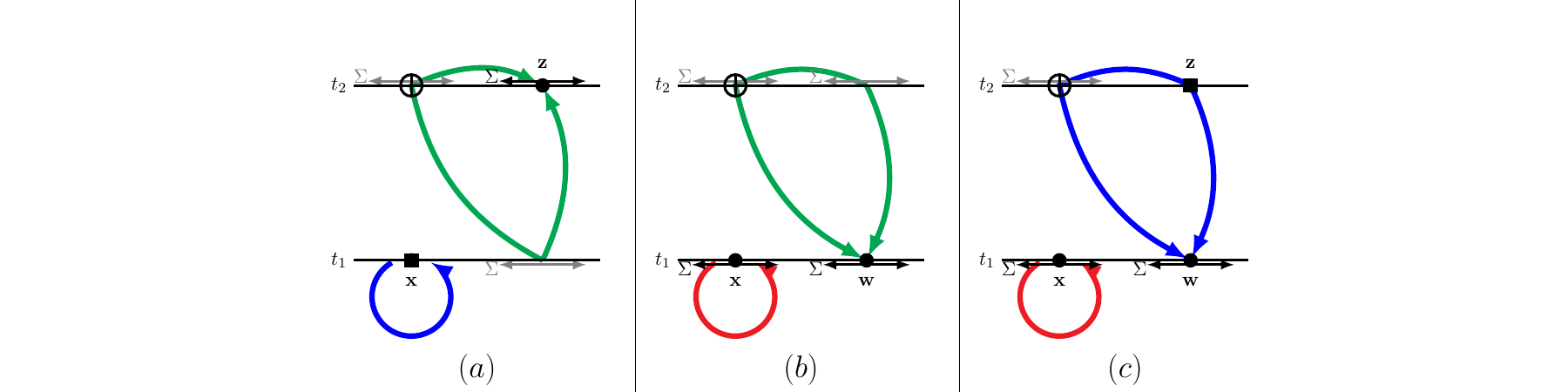}
\end{figure}
\begin{itemize}
\item[$(a)$] \textbf{one-end trick (with 1 sequential propagator)}: \\
2 $u/d$ inversion, $\# t$ $s$ inversions \\
\textbf{point-to-all propagator}: \\
12 $s$ inversions

\item[$(b)$] \textbf{one-end trick (with 1 sequential propagator)}: \\
2 $u/d$ inversion, 1 $s$ inversions \\
\textbf{stochastic timeslice-to-all propagator}: \\
$\# t$ $s$ inversion

\item[$(c)$] \textbf{2 point-to-all propagators (1 is a sequential propagator)}: \\
24 $u/d$ inversions, 12 $s$ inversions \\
\textbf{stochastic timeslice-to-all propagator}: \\
$\# t$ $s$ inversions
\end{itemize}

Note that there is another possibility very similar to method~$(b)$, which also uses the one-end trick (with 1 sequential propagator) and a stochastic timeslice-to-all propagator, and requires the same number of inversions. The only difference is that the stochastic noise introduced by the one end-trick is located on the timslice of the closed quark loop (i.e.\ also at $t_1$), while in method~$(b)$ it is located on the opposite timeslice (i.e.\ at $t_2$). We did not explore this additional method because in numerical studies of related diagrams we found that distributing the noise on the two timeslices normally results in smaller statistical errors.

\begin{figure}[htb]
\begin{center}
\includegraphics[scale=0.39,page=22]{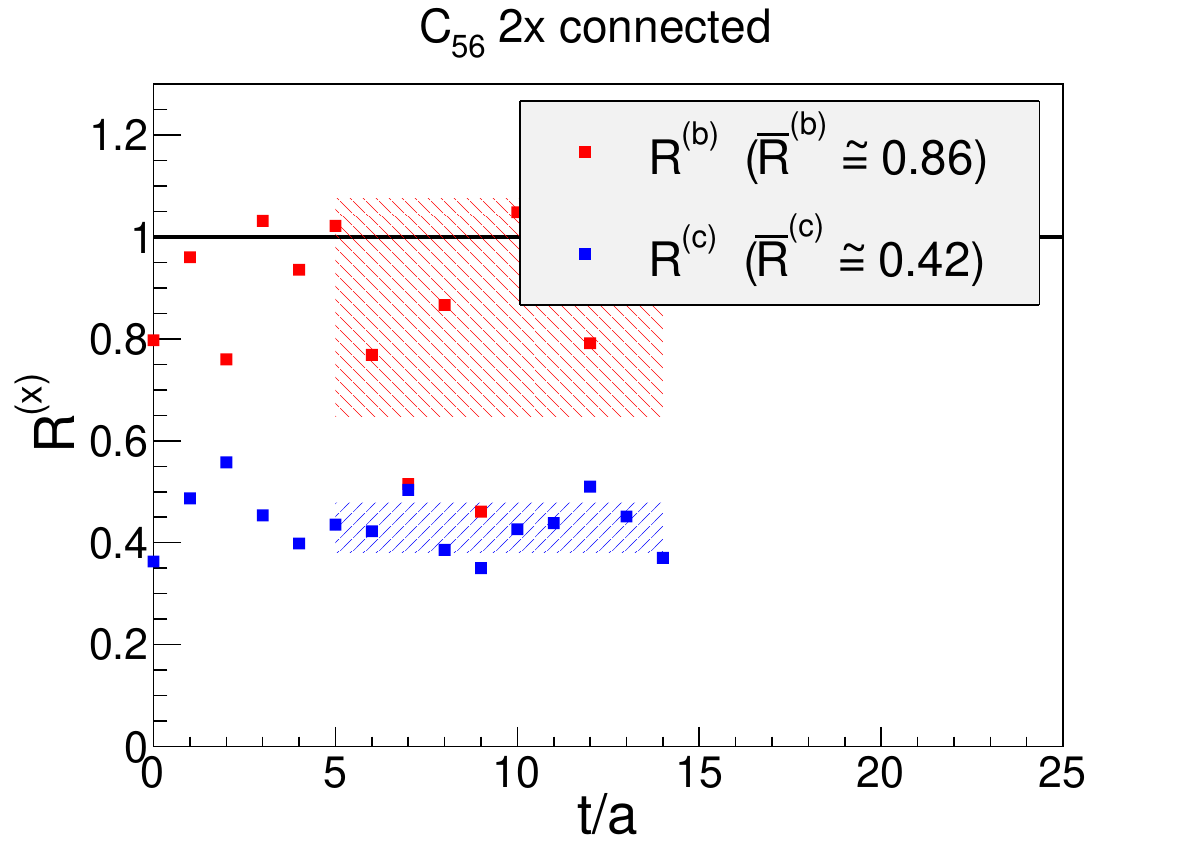}
\caption{\label{FIG765}Efficiency of different methods for $2 \times$ connected $C_{5 6}$}.
\end{center}
\end{figure}

A numerical comparison of the methods is shown in Figure~\ref{FIG765}. Method~$(a)$ and method~$(b)$ perform on a similar level. At first glance this is a bit surprising, because in previous subsections we typically observed that a method averaging a diagram over space is superior when only a moderate number of additional stochastic noise terms is introduced (cf.\ e.g.\ the correlation matrix element $C_{1 6}$ discussed in section~\ref{SEC298}, which has a very similar structure, and which can be computed by essentially the same combinations of techniques). Note, however, that this time both method~$(a)$ and method~$(b)$ require a comparably large number of $s$ inversions ($> \# t$), which is a consequence of the sequential propagator. This is not the case in many of the previously discussed cases, e.g.\ $C_{1 6}$. There the computation of the closed quark loop with a stochastic timeslice-to-all propagator requires only a single $s$ inversion, while here a point-to-all propagator needs to be computed, which requires a significantly larger amount of $s$ inversions i.e. 12 inversions.

% ********************

\subsubsection{Comparison of $4 \times$ connected and $2 \times$ connected diagrams}

A comparison of $4 \times$ connected and $2 \times$ connected diagrams for $C_{5 5}$, $C_{5 6}$ and $C_{6 6}$ is shown in Figure~\ref{FIG833}. Results and conclusions are again very similar to previous comparisons. The statistical errors of the $2 \times$ connected diagrams increase exponentially as $e^{(2 m_K - m_\pi) t}$ with respect to the temporal separation $t$ compared to their $4 \times$ connected counterparts independent of the methods employed. %{\color{red}\sout{Hence, statistical errors will be dominated by the $2 \times$ connected diagrams and it is, thus, much more important to determine the optimal method of computation for the $2 \times$ connected diagrams than for the $4 \times$ connected diagrams.}}
\begin{figure}[htb]
\begin{center}
\includegraphics[scale=0.39,page=09]{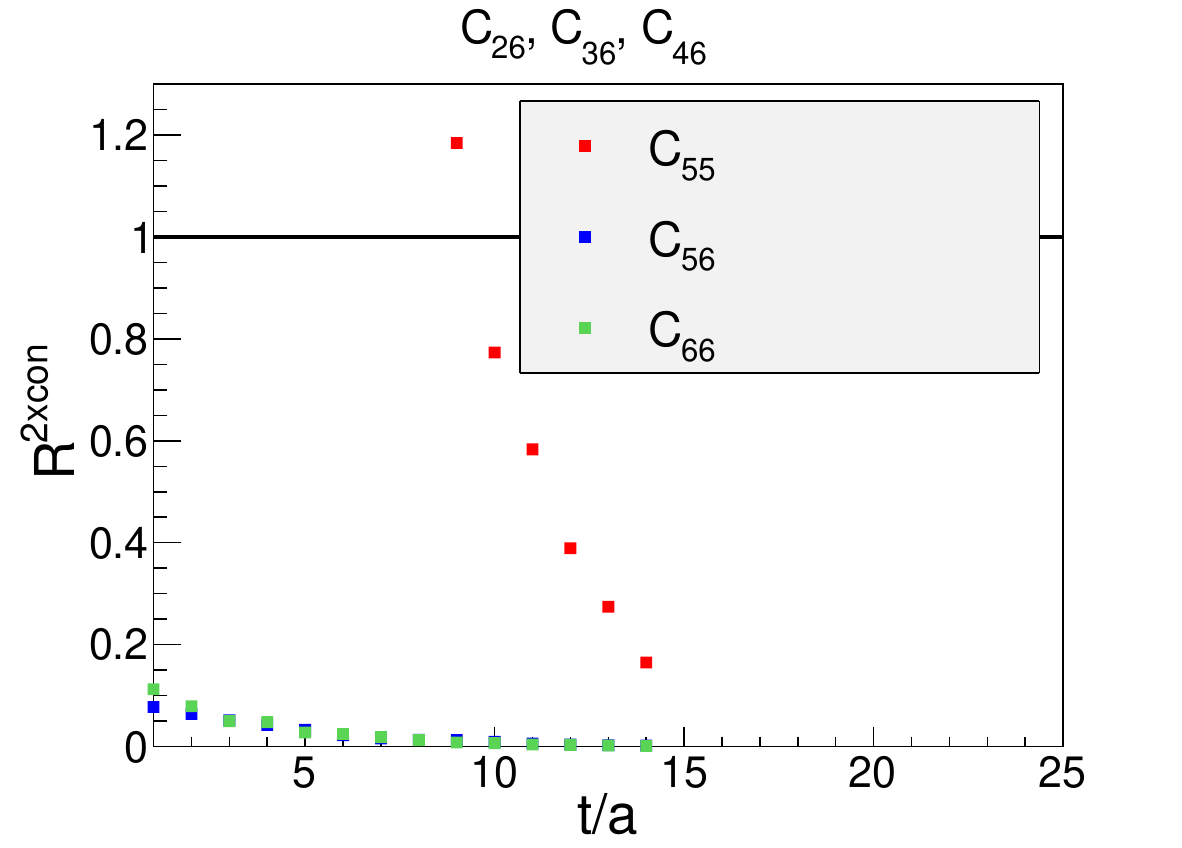}
\includegraphics[scale=0.39,page=10]{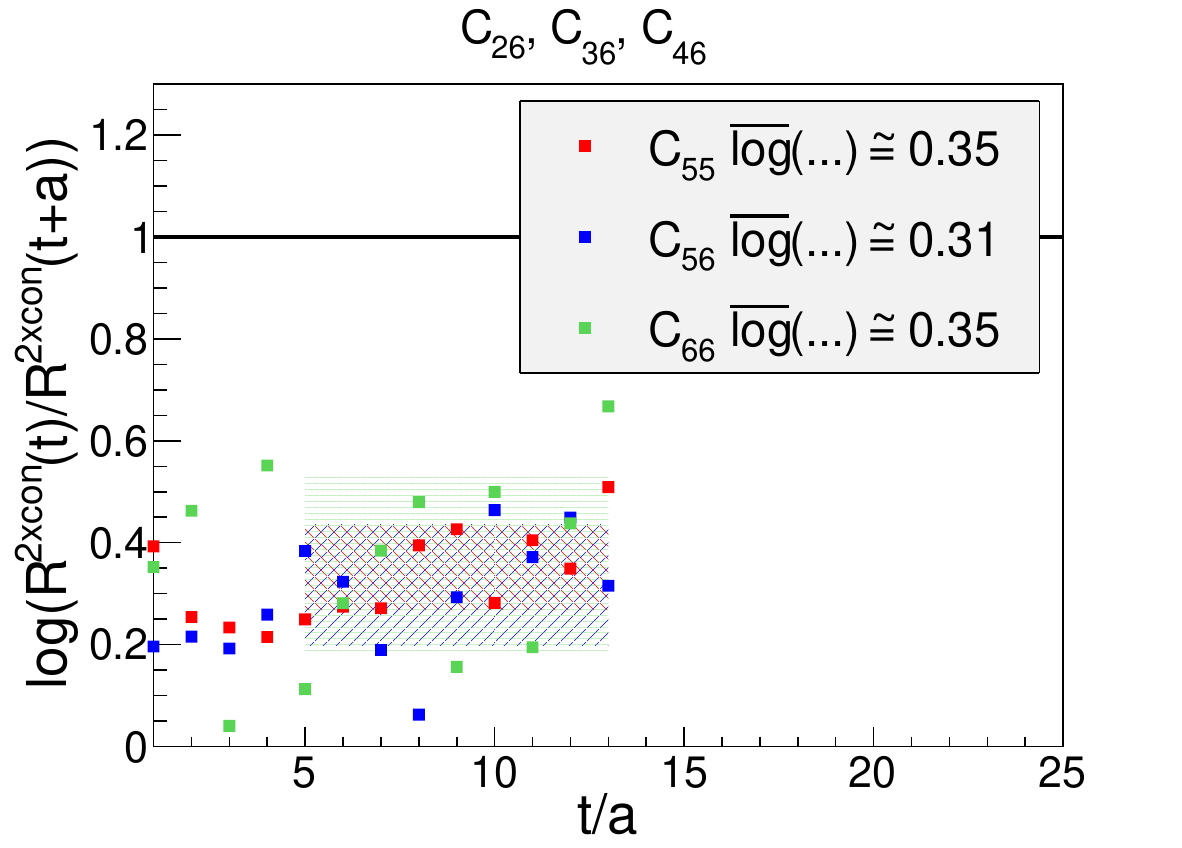}
\caption{\label{FIG833}Comparison of $4 \times$ connected and $2 \times$ connected for $C_{5 5}$, $C_{5 6}$, $C_{6 6}$.}
\end{center}
\end{figure}

On the other hand, it is interesting to note that the quality ratio $R^{2 \times \textrm{con}}(t)$ for $C_{5 5}$ is orders of magnitude larger than for all other diagrams. This seems to be the case due to a combination of the following two reasons. (1)~The $2 \times$ connected $C_{5 5}$ diagram is a single connected loop, i.e.\ it is not a combination of several disconnected pieces as e.g.\ for $C_{5 6}$ or $C_{6 6}$. (2)~The method of computation used for $C_{5 5}$ introduces only a moderate number of noise terms (only 1 application of the one-end trick), but still averages the diagram over space, i.e.\ it is very efficient. Consequently, up to temporal separation $t/a < 10$ the statistical error of the $2 \times$ connected diagram is smaller than that of the $4 \times$ connected diagram.

% ********************
% ********************
% ********************
% ********************
% ********************

\newpage

\section{\label{SEC588}Conclusions}

We have explored various methods to compute correlation functions of two- and four-quark interpolating operators of different structures including quark-antiquark pairs, mesonic molecules, diquark-antidiquark pairs and two independent mesons. Computing such correlation functions in an efficient way and with small statistical errors is essential to study tetraquark candidates or meson-meson scattering with lattice QCD.

The investigations have been performed in the context of a long-term project focused on the $a_0(980)$ meson, which might have a significant tetraquark component. We expect, however, that our conclusions hold at least qualitatively also for other tetraquark candidates and four-quark systems, e.g.\ for the positive parity $D_s$ mesons $D_{s0}^\ast(2317)$ and $D_{s1}(2460)$ or for certain $Z_c$ systems. Of course, the quality ratios presented and discussed in the previous section, might be somewhat different, when changing quark masses or quark flavors, when studying different lattice spacings or spacetime volumes, or when using another lattice QCD quark action. The general tendencies we have observed should, however, be the same also for other lattice QCD setups and other four-quark systems. Therefore, the investigations reported in this paper might provide a comprehensive overview of existing methods and helpful guidelines, which methods to choose for which kind of diagrams.

We have considered combinations of several techniques for quark propagator and correlation function computation: (A) point-to-all propagators, (B) stochastic timslice-to-all propagators, (C) the one-end trick and (D) sequential propagators. 
%We did not study the distillation method \cite{Peardon:2009gh}, a more recent alternative to compute all-to-all propagators, which is also a very promising technique to compute correlation functions of two- and four-quark interpolating operators, but which is somewhat different from a technical point-of-view. 
For applications of the distillation method to the computation of four-quark correlation functions we refer to \cite{Lang:2015sba,Dudek:2016cru,Moir:2016srx}.

For each diagram of the $6 \times 6$ correlation matrix (\ref{eq:corrmatrix}) we have implemented up to four promising methods, i.e.\ combinations of the above mentioned techniques (A) to (D). While details can be found in section~\ref{seq:compofdiagrams}, there are a few general observations which seem to hold for most diagrams:

\begin{itemize}
\item Methods, which average a diagram over space (i.e.\ which are not using any point-to-all propagators), are usually quite efficient. Exceptions are methods which extensively use stochastic methods ($M \geq 2$, where $M$ counts the number of stochastic timeslice-to-all propagators, and applications of the one-end trick, i.e.\ the number of stochastic sources) for diagrams, which can also be computed without introducing any stochastic noise (and without averaging over space). An example of the latter is method~$(a)$ in section~\ref{SEC490}. A rule of thumb, which is fulfilled by the majority of the diagrams and methods explored in this work, is the following. \vspace{0.2cm}
\\
\textbf{Method~$(a)$ averaging a diagram over space is more efficient than method~$(b)$ not averaging a diagram over space, if} $M^{(a)} \leq M^{(b)} + 1$\textbf{.} \vspace{0.2cm}
\\
Exceptions have been found in section~\ref{SEC972}, method~$(b)$ versus method~$(c)$ and section~\ref{SEC218}, method~$(a)$ versus method~$(b)$ (cf.\ the next item) and section~\ref{SEC370}, method~(a) versus method~$(b)$ (specific reasons are discussed in the section).

\item When using one or more point-to-all propagators a spatial sum has to be omitted in the correlation function, i.e.\ the diagram is not averaged over space. \vspace{0.2cm} \\
\textbf{Omitting spatial sums over only two quark field operators, seems to result in much smaller statistical errors than omitting a sum over four quark field operators.} \vspace{0.2cm} \\
Examples are presented in section~\ref{SEC972}, method~$(b)$ versus method~$(c)$ and section~\ref{SEC218}, method~$(a)$ versus method~$(b)$.

\item Sequential propagators are particularly useful for $2 \times$ connected diagrams, where at least one operator $\mathcal{O}^{\Op{5}}$ is involved, i.e.\ to compute the strange quark propagating between different points in space within a timeslice.

\item Stochastic methods are more efficient for lighter quarks (where the propagators exhibit rather strong statistical fluctuations due to the gauge links) and less efficient for heavy quarks (where the opposite is the case). An example is presented in section~\ref{SEC370}, method~$(b_1)$ versus method~$(b_2)$ (cf.\ also \cite{Wiese2012}).
\end{itemize}

A severe problem when computing a correlation matrix like (\ref{eq:corrmatrix}), is that statistical errors of $2 \times$ connected diagrams are increasing exponentially with respect to the temporal separation compared to their $4 \times$ connected counterparts. This is independent of the method of computation and inherent to these diagrams as explained in detail in section~\ref{SEC369}. Their computation with sufficient statistical precision is, therefore, extremely challenging and the statistical errors of the correlation matrix elements will be dominated by the $2 \times$ connected contributions.

\begin{figure}[htb]
\begin{center}
\includegraphics[scale=0.39,page=01]{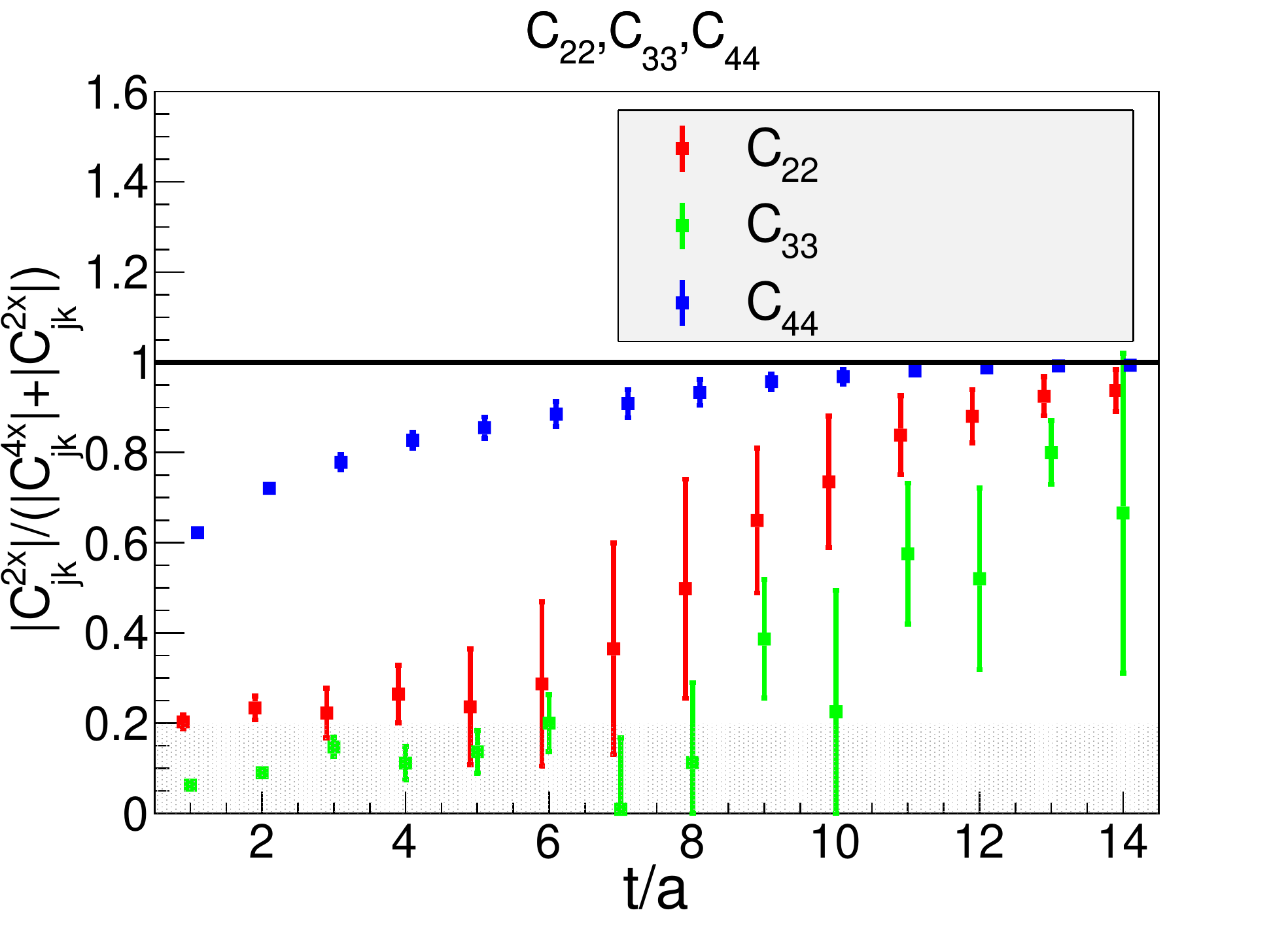}
\includegraphics[scale=0.39,page=02]{2xconn_rel.pdf}
\caption{\label{FIG906}Relative importance of the $2 \times$ connected diagrams for $C_{2 2}$, $C_{3 3}$, $C_{4 4}$ and $C_{5 6}$.}
\end{center}
\end{figure}

In the past several studies have neglected $2 \times$ connected diagrams (cf.\ e.g.\ \cite{Prelovsek:2010kg,Alexandrou:2012rm}). While in specific cases this might be justified, it is certainly not a general rule that these diagrams are negligible compared to their $4 \times$ connected counterparts. A recent study \cite{Wakayama:2014gpa} has stressed this and we moreover have found that the contribution of 2$\times$ connected diagrams is in many cases sizable. For illustration we show a few cases in Figure~\ref{FIG906}, where the ratio $|C_{j k}^{2 \times \textrm{con}}| / (|C_{j k}^{4 \times \textrm{con}}| + |C_{j k}^{2 \times \textrm{con}}|)$ is plotted, i.e.\ the percentage of contribution of the $2 \times$ connected diagram. For $C_{4 4}$ the $2 \times$ connected diagram contributes more than the corresponding $4 \times$ connected diagram to the whole correlation function, but also for the other examples, $C_{2 2}$, $C_{3 3}$ and $C_{5 6}$ both contributions are at least of similar importance.

On the one hand $2 \times$ connected diagrams may not be neglected, on the other hand they are the reason why certain elements of the correlation matrix are very noisy. Therefore, performing a standard analysis by solving a generalized eigenvalue problem (cf.\ \cite{Blossier:2009kd} and references therein) might not be the most successful strategy. Clearly, the numerical solution of a generalized eigenvalue problem requires the knowledge of all elements of a correlation matrix and, hence, the quality of its result will be dictated by the statistical errors of the most imprecise elements. A more promising strategy might be to fit a sum of decaying exponential functions, where the statistical error is rather related to the most precise elements of the correlation matrix, and, where in contrast to the generalized eigenvalue problem, strongly fluctuating matrix elements can be excluded from the analysis or will not affect the fitting result in a significant way. An advanced related variant, which we have recently explored in the context of this work is the AMIAS method \cite{FINKENRATH2016}.

After significantly increasing the statistical accuracy using the best method for each diagram as detailed in the main section of this paper, we plan to carry out a physics analysis using exponential fitting and the AMIAS method in the near future. The corresponding results regarding $a_0(980)$ meson will be part of an upcoming publication.

% ********************
% ********************
% ********************
% ********************
% ********************

\newpage

\appendix

\section{\label{SEC876}Lattice setup}

The numerical investigations presented in this paper are based on gauge link configurations generated by the PACS-CS collaboration \cite{Aoki:2008sm}. The gluonic action is the Iwasaki gauge action \cite{Iwasaki:1985we} and the quark action for 2+1 flavors is a non-perturbatively $\mathcal{O}(a)$ improved Wilson quark action,
\begin{align}
\nonumber & S_\textrm{quark}[q,\bar{q},U] = \sum_{q=u,d,s} \bigg(\sum_x \bar{q}_x q_x - \kappa_q c_\textrm{SW} \sum_{x,\mu,\nu} \frac{i}{2} \bar{q}_x \sigma_{\mu \nu} F_{x,\mu \nu} q_x \\
 \label{eq:fermionaction} & \hspace{0.7cm} - \kappa_q \sum_{x,\mu} \Big(\bar{q}_x (1-\gamma_\mu) U_{x,\mu} q_{x+\hat{\mu}} + \bar{q}_x (1+\gamma_\mu) U_{x-\hat{\mu},\mu}^\dagger q_{x-\hat{\mu}}\Big)\bigg) ,
\end{align}
with the field strength
\begin{align}
F_{x,\mu \nu} = \frac{1}{4} \sum_{j=1}^4 \frac{1}{2 i} \Big(U_{x,\mu \nu}(j) - U_{x,\mu \nu}^\dagger(j)\Big) ,
\end{align}
where $U_{x,\mu \nu}(j)$, $j=1,\ldots,4$ denotes the four plaquettes in the $\mu\nu$-plane attached to $x$.

We consider a single ensemble with gauge coupling $\beta = 1.90$, corresponding lattice spacing $a \approx 0.091 \, \textrm{fm}$ and pion mass $m_\pi \approx 300 \, \textrm{MeV}$. Further details are listed in Table~\ref{table:ensemble}.

\begin{table}[htb]
\begin{center}
\begin{tabular}{|c|c|c|c|c||c|c||c|}
\hline
& & & & & & & \vspace{-0.40cm} \\
$\beta$ & $(L/a)^3 \times T/a$ & $\kappa_{u,d}$ & $\kappa_s$ & $c_\textrm{SW}$ & $a$ [$\textrm{fm}$] & $m_\pi$ [$\textrm{MeV}$] & \# configurations \\
& & & & & & & \vspace{-0.40cm} \\
\hline
& & & & & & & \vspace{-0.40cm} \\
1.90 & $32^3 \times 64$ & $0.1370$ & $0.1364$ & $1.715$ & $0.091$ & $300$ & $500$ \vspace{-0.40cm} \\
& & & & & & & \\
\hline
\end{tabular}
\caption{\label{table:ensemble}PACS-CS gauge link ensemble used (cf.\ also \cite{Aoki:2008sm}).}
\end{center}
\end{table}

To enhance the ground state overlap of trial states $\mathcal{O}^j | \Omega \rangle$ generated by our interpolating operators (\ref{eq:operatorone}) to (\ref{eq:operatorsix}), we apply standard smearing techniques. For the quark field operators $u_\mathbf{x}$, $d_\mathbf{x}$ and $s_\mathbf{x}$ we use Gaussian smearing with APE smeared spatial links (parameters $N_\textrm{Gauss} = 50$, $\kappa_\textrm{Gauss} = 0.5$, $N_\textrm{APE} = 20$, $\alpha_\textrm{APE} = 0.45$; cf.\ \cite{Jansen:2008si} for detailed equations). The widths of these smeared quark field operators can be estimated, $D \approx \sqrt{2 N_\textrm{Gauss} \kappa_\textrm{Gauss} / (1 + 6 \kappa_\textrm{Gauss})} a \approx 0.3 \, \textrm{fm}$ (eq.\ (27) in \cite{Jansen:2008si}), i.e.\ they excite the corresponding quark fields in spherically extended regions with radius size $D$ centered around $\mathbf{x}$.

% ********************
% ********************
% ********************
% ********************
% ********************

\newpage

\section{\label{APP002}Computation of the correlation matrix: technical aspects}

% ********************

\subsection{\label{APP003}$C_{1 2}$, $C_{1 3}$, $C_{1 4}$ ($1 \equiv q \bar{q}$, $2 \equiv K \bar{K}\textrm{, point}$, $3 \equiv \eta_s \pi\textrm{, point}$, $4 \equiv Q \bar{Q}$)}

Even though the diagrams for $C_{1 2}$, $C_{1 3}$ and $C_{1 4}$ are identical (cf.\ Figure~\ref{fig:diagrammatrix}), the corresponding mathematical expressions differ due to a different spin and color structure of the interpolating operators (\ref{eq:operatortwo}), (\ref{eq:operatorthree}) and (\ref{eq:operatorfour}),
\begin{align}
\label{EQN806} & C_{1 2}(t) = -\frac{1}{\sptV} \sum_{{\bf x},{\bf y}} \Big\langle \textrm{Tr}\Big(\D{(d)}{x}{t_1}{y}{t_2} \D{(u)}{y}{t_2}{x}{t_1} \gamma_5 \D{(s)}{x}{t_1}{x}{t_1} \gamma_5\Big) \Big\rangle_U \\
 & C_{1 3}(t) = +\frac{1}{\sptV} \sum_{{\bf x},{\bf y}} \Big\langle \textrm{Tr}\Big(\D{(d)}{x}{t_1}{y}{t_2} \D{(u)}{y}{t_2}{x}{t_1} \gamma_5 \Big) \textrm{Tr}\Big(\D{(s)}{x}{t_1}{x}{t_1} \gamma_5\Big) \Big\rangle_U \\
\nonumber & C_{1 4}(t) = +\frac{1}{\sptV} \sum_{{\bf x},{\bf y}} \epsilon_{a b c} \epsilon_{a d e} \Big\langle \textrm{Tr}_\textrm{spin}\Big(G^{(d)}_{c f}({\bf x},t_1;{\bf y},t_2) G^{(u)}_{f d}({\bf y},t_2;{\bf x},t_1) (C \gamma_5)^\ast \\
 & \hspace{0.7cm} \Big(G^{(s)}_{b e}({\bf x},t_1;{\bf x},t_1)\Big)^T (C \gamma_5)^\ast\Big) \Big\rangle_U .
\end{align}

The three combinations of techniques investigated in section~\ref{SEC490} are rather independent of the spin and color structure introduced by either $\mathcal{O}^{\Op{2}}$, $\mathcal{O}^{\Op{3}}$ or $\mathcal{O}^{\Op{4}}$ and, hence, can be applied in essentially the same way to $C_{1 2}$, $C_{1 3}$ and $C_{1 4}$. Therefore, we present equations only for $C_{1 2}$.
\begin{itemize}
\item Method~$(a)$:
\begin{align}
\nonumber & C_{1 2}(t) = -\bigg\langle \bigg(\sum_{\bf y} \phi_\textrm{pnt}^{(d)}({\bf y},t_2)[a,A,{\bf x},t_1]^\dag \gamma_5 \phi_{\textrm{pnt}}^{(u)}({\bf y},t_2)[b,B,{\bf x},t_1]\bigg) \\
\label{EQN610} & \hspace{0.7cm} (\gamma_5)_{B;C} \phi^{(s)}_{\textrm{pnt}\, b,C}({\bf x},t_1)[a,A,{\bf x},t_1] \bigg\rangle_U .
\end{align}

\item Method~$(b)$:
\begin{align}
\nonumber & C_{1 2}(t) = -\frac{1}{N_\textrm{one}} \sum_{n=1}^{N_\textrm{one}} \frac{1}{N_\textrm{sto}} \sum_{n'=1}^{N_\textrm{sto}} \frac{1}{\sptV} \bigg\langle \sum_{\bf x} \Big(\tilde{\phi}_{\textrm{one}}^{(u)}({\bf x},t_1)[t_2,\mathbbm{1},n]^\dag \gamma_5 \phi_{\textrm{sto}}^{(s)}({\bf x},t_1)[t_1,n']\Big) \\
\label{EQN608} & \hspace{0.7cm} \Big(\xi_{\textrm{sto}}({\bf x},t_1)[t_1,n']^\dag \phi_{\textrm{one}}^{(d)}({\bf x},t_1)[t_2,n]\Big) \bigg\rangle_U .
\end{align}

\item Method~$(c)$:
\begin{align}
\nonumber & C_{1 2}(t) = -\frac{1}{N_\textrm{sto}} \sum_{n=1}^{N_\textrm{sto}} \bigg\langle (\gamma_5)_{B;A} \sum_{\bf x}  \Big(\phi_{\textrm{pnt}}^{(u)}({\bf x},t_1)[a,A,{\bf y},t_2]^\dag \phi_{\textrm{sto}}^{(s)}({\bf x},t_1)[t_1,n] \Big) \\
\label{EQN609} & \hspace{0.7cm} \Big(\xi_{\textrm{sto}}({\bf x},t_1)[t_1,n]^\dag \gamma_5 \phi_{\textrm{pnt}}^{(d)}({\bf x},t_1)[b,B,{\bf y},t_2] \Big) \bigg\rangle_U .
\end{align}
\end{itemize}

% ********************

\subsection{\label{APP004}$C_{1 5}$ ($1 \equiv q \bar{q}$, $5 \equiv K \bar{K} \textrm{, 2part}$)}

With respect to spin and color $C_{1 5}$ is identical to $C_{1 2}$ (cf. eq.\ (\ref{EQN806})). The crucial difference is that the $K$ and the $\bar{K}$ meson are not created at the same point in space, but at different points ${\bf x}$ and ${\bf z}$,
\begin{align}
C_{1 5}(t) = -\frac{1}{\sptV^{3/2}} \sum_{{\bf x},{\bf y},{\bf z}} \Big\langle \textrm{Tr}\Big(\D{(d)}{z}{t_1}{y}{t_2} \D{(u)}{y}{t_2}{x}{t_1} \gamma_5 \D{(s)}{x}{t_1}{z}{t_1} \gamma_5\Big) \Big\rangle_U .
\end{align}

Equations corresponding to the two combinations of techniques investigated in section~\ref{SEC459} are the following.
\begin{itemize}
\item Method~$(a)$:
\begin{align}
\label{EQN611} C_{1 5}(t) = -\frac{1}{N} \sum_{n=1}^N \frac{1}{\sptV^{3/2}} \bigg\langle \sum_{\bf y} \Big(\tilde{\phi}^{(d)}({\bf y},t_2)[t_1,\gamma_5,n]^\dagger \gamma_5 \psi^{(u;s)}({\bf y},t_2)[t_1,\gamma_5,n]\Big) \bigg\rangle_U ,
\end{align}
where
\begin{align}
 & \sum_y D_{a,A;b,B}^{(s)}(x;y) \phi_{b,B}^{(s)}(y)[t_1,n] = \xi_{a,A}(x)[t_1,n] \\
 & \sum_x D_{a,A;b,B}^{(u)}({\bf x},x_0;y) \psi_{b,B}^{(u;s)}(y)[t_1,\gamma_5;t_1,n] = (\gamma_5 \phi)_{a,A}^{(s)}({\bf x},x_0)[t_1,n] \delta(x_0;t_1)
\end{align} 
are the analogs of (\ref{EQN313}) and (\ref{EQN311}) for combining the one-end trick with a sequential propagator, which are straightforward to derive.

\item Method~$(b)$:
\begin{align}
\label{EQN612} C_{1 5}(t) = -\frac{1}{\sptV^{1/2}} \bigg\langle \sum_{\bf y} \Big(\phi^{(d)}({\bf y},t_2)[a,A,{\bf z},t_1]^\dagger \gamma_5 \psi^{(u;s)}({\bf y},t_2)[t_1,\gamma_5;a,A,{\bf z},t_1]\Big) \bigg\rangle_U ,
\end{align}
where $\psi^{(u;s)}$ is the sequential propagator $G^{(u)} \gamma_5 G^{(s)}$ according to (\ref{EQN866}).
\end{itemize}

% ********************
% ********************
% ********************
% ********************
% ********************

\newpage

\section*{Acknowledgments}

We thank Mario Gravina for his contributions in an early stage of this project.
We acknowledge useful conversations with Stefan Meinel and Christian Wiese. 

J.B.\ and M.W.\ acknowledge support by the Emmy Noether Programme of the DFG (German Research Foundation), grant WA 3000/1-1.
M.D.B. is grateful to the Theoretical Physics Department at CERN for the hospitality and support.

This work was supported in part by the Helmholtz International Center for FAIR within the framework of the LOEWE program launched by the State of Hesse.

This work was cofunded by the European Regional Development Fund and the Republic of Cyprus through the Research Promotion Foundation (Project Cy-Tera ΝΕΑ
$\Upsilon\Pi\text{O}\Delta\text{OMH} / \Sigma\text{TPATH}/0308/31$
%ΥΠΟΔΟΜΗ/ΣΤΡΑΤΗ/0308/31 
) by the grand cypro914.

Calculations on the LOEWE-CSC and on the on the FUCHS-CSC high-performance computer of the Frankfurt University were conducted for this research. We would like to thank HPC-Hessen, funded by the State Ministry of Higher Education, Research and the Arts, for programming advice.

Computations have been performed using the Chroma software library \cite{Edwards:2004sx}.

% ********************
% ********************
% ********************
% ********************
% ********************

% ********************
% ********************
% ********************
% ********************
% ********************

\end{document}